%% file: HEGRA_tsukuba.tex
\documentclass[12pt]{book}
\usepackage{makeidx,HEGRA_tsukuba}%
\usepackage[dvips]{graphicx,color}
\makeauthorindex
\makeindex
\printAuthorIndex

\title{HEGRA Contributions  to the \\
\textbf{28th International Cosmic Ray Conference}\\
 Tsukuba, Japan}
\date{July 31 - August 7, 2003}
\begin{document}
\maketitle
\BookTitle{\itshape HEGRA Contributions to the 28th
International Cosmic Ray Conference}
\CopyRight{\copyright 2003 by Universal Academy Press, Inc.}
\parbox{\linewidth}{\rule{0pt}{29cm}}
\input{toc}
\parbox{\linewidth}{\rule{0pt}{29cm}}
{\input{my_008809-1}}  
{\input{my_icrc2003.hegra.highlight.tex}}  
{\input{my_008827_1}}  
{\input{my_008277-1}}  
{\input{my_008277-2}}  
{\input{my_009254-1}}  
{\input{my_009018-2}}  
{\input{my_009018-1}}  
{\input{my_009463-1}}  
{\input{my_009463-2}}  
{\input{tluczykont_agn.tex}}  
{\input{my_009271-1}}  
{\input{my_009109-1}}  

\end{document}

%% file: toc.tex
{\Huge \bf Contents}\\
\begin{enumerate}
\bf
\item The Giant Radio Galaxy M\,87 as a TeV $\gamma $-Ray Emitter observed with the HEGRA Cherenkov Telescopes \hfill 5
\item Highlights from 6 years of TeV gamma-ray astrophysics with the HEGRA imaging Cherenkov telescopes \hfill 9
\item Search for TeV Gamma-Rays from the Andromeda Galaxy and for Supersymmetric Dark Matter in the Core of M31 \hfill 13
\item Studies of the Crab Nebula based upon 400\nobreakspace {}hours of Observations with the HEGRA System of Cherenkov Telescopes \hfill 17
\item High Energy Emission from H1426+428 and Absorption on the Extragalactic Background Light \hfill 21
\item An new method to determine the arrival direction of individual air showers with a single Air Cherenkov Telescope \hfill 25
\item Scans of the TeV Gamma-Ray Sky with the HEGRA System of Cherenkov Telescopes  \hfill 29
\item The Technical Performance of the HEGRA IACT System \hfill 33
\item The New Unidentified TeV Source in Cygnus (TeV\nobreakspace {}J2032+4130): HEGRA IACT-System Results \hfill 37
\item TeV Observations of Selected GeV Sources with the HEGRA IACT-System \hfill 41
\item Observations of 54 Active Galactic Nuclei with the \\ HEGRA Cherenkov Telescopes \hfill 45
\item Study of the VHE Gamma Ray Emission from the AGN \\1ES1959+650 with the HEGRA Cherenkov Telescope CT1 \hfill 49
\item Observation of VHE Gamma Rays from the Remnant of SN 1006 with HEGRA CT1 \hfill 53
\end{enumerate}

%% file: my_008809-1.tex
\boldmath
\chapter{
  The Giant Radio Galaxy M\,87 as a TeV $\gamma$-Ray 
  Emitter
  observed with the HEGRA Cherenkov Telescopes
  }
\unboldmath

\author{%
%
%
Niels G\"otting,  
Martin Tluczykont 
and
G\"otz Heinzelmann 
\\
for the HEGRA Collaboration
\\
{\it Institut f\"ur Experimentalphysik, Universit\"at Hamburg,
  Luruper Chaussee~149, 
  \\
  D-22761 Hamburg, Germany
  }
}


\section*{Abstract}

For the first time an excess of photons above an energy
threshold of 730\,GeV from the giant radio galaxy M\,87 has
been measured with a significance of 4.1\,$\sigma$. The
data have been taken during the years 1998 and 1999 with
the HEGRA stereoscopic system of 5 imaging atmospheric
Cherenkov telescopes. The excess corresponds to an
integral flux of 3.3\,\% of the Crab flux. Making use of
the imaging atmospheric Cherenkov technique, this is the
first object of the AGN class observed in this
energy range not belonging to the BL Lac type objects.


\section{Introduction}

Extragalactic TeV $\gamma$-ray emission 
has been observed with the imaging atmospheric Cherenkov technique
so far from AGN only of the BL Lac type, i.\,e.~objects 
ejecting matter in a relativistic jet oriented very close to the observer's line of sight. In BL 
Lacs, TeV photons are commonly believed to originate in the 
jets, most popularly due to inverse Compton scattering.
The well studied objects Mkn\,421 ($z =$~0.030), 
Mkn\,501 ($z =$~0.034), 
1ES\,1959+650 ($z =$~0.047) 
and H\,1426+428 ($z =$~0.129) 
belong to this type of TeV $\gamma$-ray emitters.
However, other types of AGN, e.\,g.~giant radio galaxies,
also show jets,
though, in contrast to BL Lac type objects, under large viewing
angles. Amongst these the nearby radio galaxy M\,87 
($z = 0.00436$) 
-- containing a supermassive black hole with
$M_{\mbox{\tiny BH}} \approx 2 \mbox{--} 3 \times 10^9\,\mbox{M}_\odot$
[11] --  
has been speculated to be a powerful accelerator of cosmic
rays, including the highest energy particles observed in the universe,
see e.\,g.~[10, 8].  
The angle of the M\,87 jet axis
to the line of sight was determined to be $30^\circ$ -- $35^\circ$ 
[7].  

The HEGRA collaboration has extensively observed M\,87
in 1998 and 1999 with the stereoscopic system 
of 5~imaging atmospheric Cherenkov telescopes (IACT system) 
[9]  
as one of the prime candidates for TeV $\gamma$-ray emission from the class
of radio galaxies.
In this paper the encouraging results of these observations
are reported, applying a very sensitive analysis method (see also 
[3]).  
Astrophysical conclusions concerning the nature of the observed
excess are discussed.


\section{Observations and results of analysis}

M\,87 was observed in 1998 and 1999
with the HEGRA IACT system for a total of 83.4\,h~[3]  
above a mean energy threshold 
of 730\,GeV for a Crab-like spectrum~[14].  
Only data of good quality were considered for the analysis.

All M\,87 observations were performed in the 
{\em wobble} mode 
allowing for simultaneous estimation of the 
background (``OFF'') rate induced by charged cosmic rays
[1].  
This analysis uses a ring segment 
as extended OFF-region reducing the statistical error on the number of 
background events [3].  
The radius and width of the ring are set according to the position and
size of the ON-source area in order to 
provide the same acceptance for ON- resp.~OFF-source events. 
The event reconstruction (described elsewhere, e.\,g.~[2])  
makes use of algorithm~\#3 for the reconstruction of the air shower
direction [13].  
A ``tight shape cut'' 
(parameter {\em mscw}~$< 1.1$) [14]  
is applied for an effective $\gamma$-hadron separation.
The optimum angular cut was derived using $\gamma$-ray
events from the Crab nebula on the basis of a nearly contemporaneous data set
at similar zenith angles.

\begin{figure*}[t]
  \centering
  \includegraphics[width=0.49\textwidth]{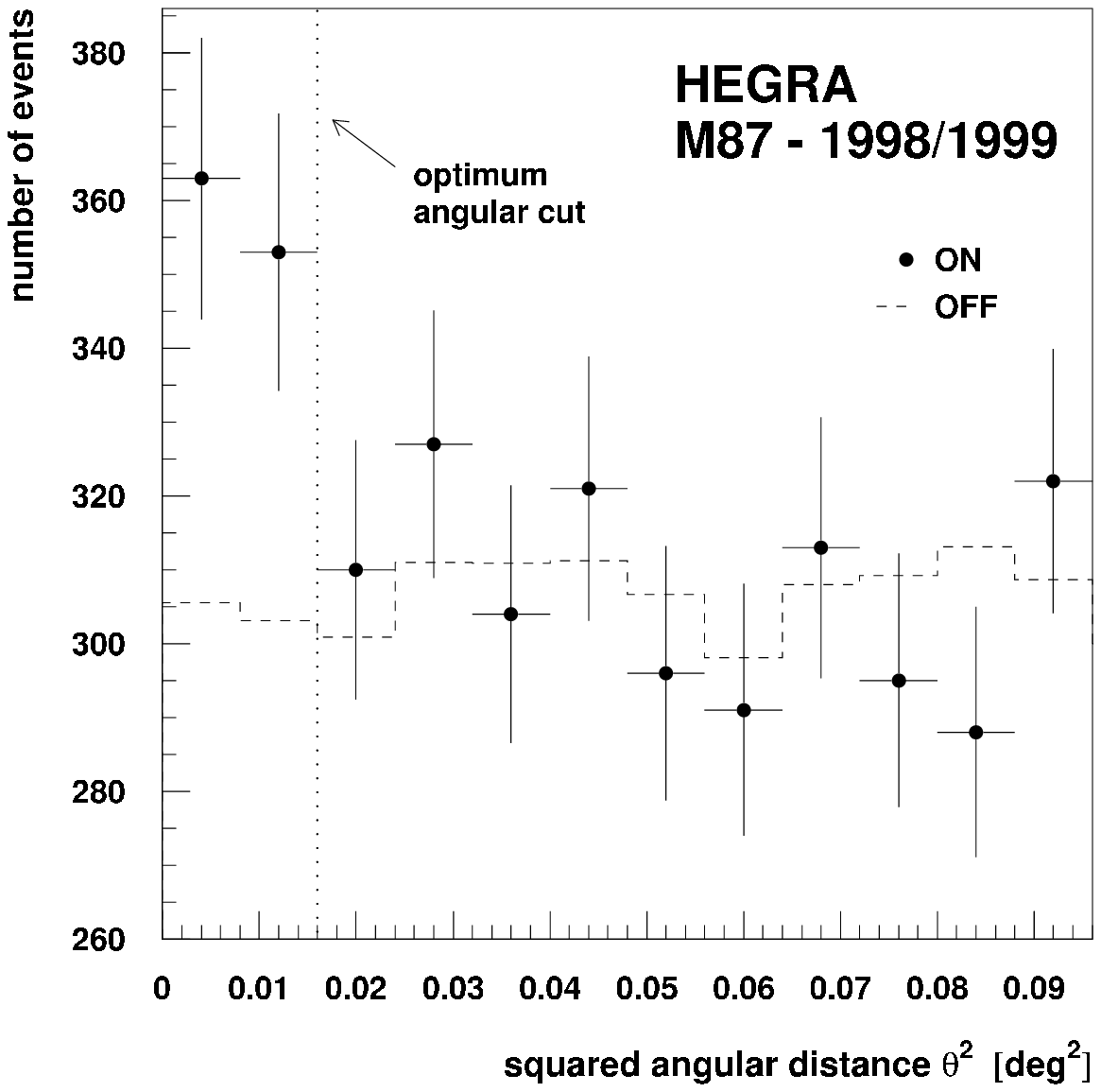}
  \hfill
  \begin{minipage}[b]{0.49\textwidth}
    \includegraphics[width=\textwidth]{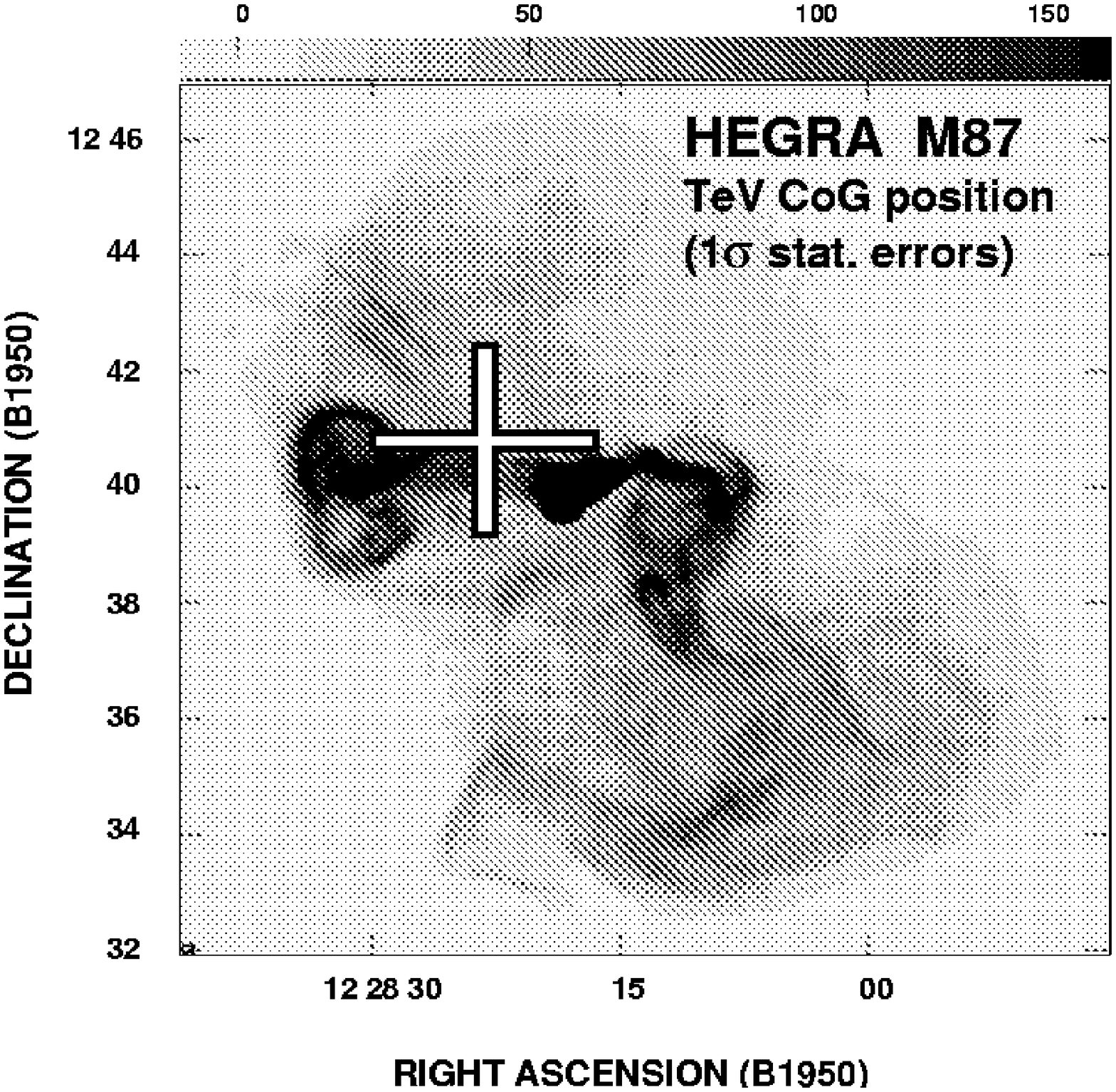}
    \vspace*{-0.2cm}
  \end{minipage}
  \caption{
    Left: Number of events vs.~squared angular distance 
    to M\,87 as observed in 
    1998 and 1999 with the HEGRA IACT system
    (dots: ON-source events, dashed histogram:
    background).
    The significance of the excess 
    amounts to 4.1\,$\sigma$.
    Right: The center of gravity position of the HEGRA M\,87 TeV $\gamma$-ray excess
    is marked by the cross indicating the statistical 1\,$\sigma$ errors 
    (radio image adapted from [17]).  
    }
\end{figure*}

Figure~1 (left panel) shows the event distribution
for the ON-source and the OFF-source regions as a function of the squared
angular distance of the 
shower direction to the source position.
The statistical significance of the observed excess from M\,87 
is~4.1\,$\sigma$, calculated using formula (17) from [16].  
On the basis of the limited event statistics the excess is compatible 
with a point-like source for the HEGRA IACT system at the position of M\,87, 
although a slightly extended emission region cannot be excluded. 

The event distribution in the field of view was used 
to determine the center of gravity position (CoG) of the 
TeV $\gamma$-ray excess at 
$\alpha_{\mbox{\tiny J2000.0}} = 12^{\mbox{\tiny hr}}30^{\mbox{\tiny m}}54.4^{\mbox{\tiny s}}
  \pm 6.9^{\mbox{\tiny s}}_{\mbox{\tiny stat}} \pm
  1.7^{\mbox{\tiny s}}_{\mbox{\tiny syst}}$,
$\delta_{\mbox{\tiny J2000.0}} = 12^\circ24^{\prime}17^{\prime\prime}
  \pm 1.7^{\prime}_{\mbox{\tiny stat}} \pm 0.4^{\prime}_{\mbox{\tiny syst}}$
(see Figure~1, right panel).
The accuracy of the CoG determination is limited by a systematic
pointing error of about 25\,$^{\prime\prime}$ [18].  
Within the large statistical errors, the CoG is consistent with the M\,87 position.
Currently, it is not possible to localize a candidate TeV
$\gamma$-ray production site to particular inner radio structures of M\,87.

The observed excess can be converted into an integral flux 
of ($3.3 \pm 0.8)$\,\% of the Crab nebula flux.
A conversion into absolute flux units 
results in an M\,87 $\gamma$-ray flux of
$
N_\gamma(E > 730\,\mbox{GeV}) = (0.96 \pm 0.23) \times 10^{-12}\,\mbox{phot.\,cm}^{-2}\,\mbox{s}^{-1}\mbox{.}
$
A spectral analysis of the data of the M\,87 data has been performed using
the analysis technique described in [4].  
The data can be well described with a power law 
$\mbox{d}N/\mbox{d}E \sim E^{-\alpha}$ with
$\alpha = 2.9 \pm 0.8_{\mbox{\tiny stat}} \pm 0.08_{\mbox{\tiny syst}}$.


\section{Summary and Conclusions}

The radio galaxy M\,87 has been observed with the HEGRA IACT system for a total of~83.4\,h. 
For the first time a significant excess of 
4.1\,$\sigma$ has been detected at energies above a mean energy threshold of 730\,GeV
from a member of this class of objects using the imaging atmospheric Cherenkov
technique. 

Due to the limited number of excess events detected so far
it is difficult to draw 
a conclusion about the origin of
the TeV $\gamma$-radiation. Assuming a spectral shape following a power law 
d$N$/d$E \propto E^{-2.9}$ 
the integral photon flux 
of ($3.3 \pm 0.8)$\,\% of
the Crab nebula flux converts into an energy flux of
$ F_\gamma(E > 730\,\mbox{GeV}) = (4.3 \pm 1.0) \times
10^{-12}\,\mbox{erg\,cm}^{-2}\,\mbox{s}^{-1}\mbox{.} $
Given the distance to M\,87 of about 16\,Mpc, this
corresponds to a $\gamma$-ray luminosity above 730\,GeV 
of about $10^{41}\,\mbox{erg\,s}^{-1}$
under the assumption of isotropic emission.
The integral flux observed by HEGRA is not in conflict
with an upper limit of
$2.2 \times 10^{-11}$ phot.\,cm$^{-2}$s$^{-1}$ 
above 250\,GeV reported by the VERITAS collaboration
from data collected in 2000 and 2001 
[15].  

Several different possibilites for the origin of GeV/TeV $\gamma$-radiation 
are conceivable.
M\,87 with its pc scale jet has recently been modeled
within the Synchrotron Self Compton scenario
as a BL Lac object seen at a large angle to its jet axis 
[5].  
Note, that a recent Chandra monitoring of the optical knot HST-1 in the M\,87
jet (located only 0.8$^{\prime\prime}$ away from the core) revealed a strong hint for
a synchrotron origin of the observed X-ray emission 
[12].  
The TeV $\gamma$-radiation of M\,87 has also been modeled using the so-called
Synchrotron Proton Blazar model~[19].  
In both models, the flux observed by HEGRA can be accommodated.
The large scale (kpc) jets with several knots detected at radio to X-ray 
frequencies 
is also a possible $\gamma$-ray production site in M\,87.
Moreover, $\gamma$-rays could be produced in the interstellar medium at larger 
distances from the center of M\,87.
It should be noted that M\,87 is also considered as a possible source of TeV $\gamma$-rays from
the hypothetical neutralino annihilation process 
[6].  

A weak signal at the centi-Crab level is at the sensitivity threshold
for the HEGRA IACT system. Deep observations
of M\,87 with next generation Cherenkov telescopes like
H$\cdot$E$\cdot$S$\cdot$S, MAGIC and VERITAS 
will provide a high sensitivity together with a low energy threshold.
Due to the proximity to M\,87 (16\,Mpc compared to 110\,Mpc for the
closest TeV BL Lac Mkn\,421) these measurements will allow 
an accurate location and spectral analysis of the $\gamma$-ray emission site in M\,87
thus greatly advancing our understanding of its TeV $\gamma$-radiation.
%


\section*{Acknowledgements}

  The support of the German Federal Ministry for Research and Technology BMBF and
  of the Spanish Research Council CICYT is gratefully acknowledged. 
  G.~Rowell acknowledges receipt of a von Humboldt fellowship.
  We thank the Instituto de Astrof\'{\i}sica de Canarias (IAC)
  for the use of the HEGRA site at the Observatorio del Roque de los 
  Muchachos (ORM) and for supplying excellent working conditions on La
  Palma.


\section*{References}


  \re
  1.\ Aharonian, F.\,A.,
  et al.~1997, 
  A\,\&\,Ap 327, L5
  \re
  2.\ Aharonian, F.\,A., et al.~1999, A\,\&\,Ap 342, 69
  \re
  3.\ Aharonian, F.\,A., 
  et al.~2003a,
  A\,\&\,Ap 403, L1
  %
  \re
  4.\ Aharonian, F.\,A., et al.~2003b, A\,\&\,Ap 403, 523
  %
  \re
  5.\
  Bai, J.\,M., \& Lee, M.\,G., 2001,
  ApJ 549, L173
  \re
  6.\
  Baltz, E.\,A., 
  et al.~2000,
  Phys.~Rev.~D 61 023514
  \re
  7.\
  Bicknell, G.\,V., \& Begelman, M.\,C., 1996,
  ApJ 467, 597
  \re
  8.\
  Biermann, P.\,L., 
  et al.~2000,
  Nucl.~Phys.~B (Proc.~Suppl.) 87, 417
  \re
  9.\
  Daum, A.,
  et al.~1997,
  Astropart.~Physics~8, 1
  \re
  10.\
  Ginzburg, V.\,L. \& Syrovatskii, S.\,I., 1964,
  Pergamon Press, New York
  \re
  11.\
  Harms, R.\,J., 
  et al.~1994,
  ApJ 435, L35
  \re
  12.\
  Harris, D.\,E.,
  et al.~2003,
  ApJ 586, L41
  \re
  13.\
  Hofmann, W., 
  et al.~1999, Astropart.~Physics~12, 135
  \re
  14.\
  Konopelko, A., 
  et al.~1999, 
  Astropart.~Physics 10, 275 ff.
  \re
  15.\
  Lebohec, S., et al.~2001,
  Proc.~of the 27th ICRC,
  OG~2.3.191, Vol.~7, 2643
  \re
  16.\
  Li, T.-P., \& Ma, Y.-Q., 1983,
  ApJ 272, 317
  \re
  17.\
  Owen, F.\,N., 
  et al.~2000,
  ASP Conf.~Series, Vol.~199 (astro-ph/0006152)
  \re
  18.\
  P\"uhlhofer, G.,
  et al.~1997, 
  Astropart.~Physics 8, 101
  \re
  19.\
  Protheroe, R.\,J.,
  et al.~2003,
  astro-ph/0303522

\endofpaper

%% file: my_icrc2003.hegra.highlight.tex
\chapter{Highlights from 6 years of TeV gamma-ray astrophysics with the HEGRA
  imaging Cherenkov telescopes }

\author{%
G\"otz Heinzelmann$^1$,
for the HEGRA Collaboration$^2$\\
{\it (1)  Universit\"at Hamburg, Institut f\"ur Experimentalphysik,
Luruper Chaussee 149, D-22761 Hamburg, Germany \\
(2) {\tt http://www-hegra.desy.de/hegra/}}
}
\section*{Abstract} 

\vspace{-0.5pc}

\begin{small}
  The HEGRA (High Energy Gamma Ray Astronomy) experiment achieved
  outstanding results during the operation of the six Cherenkov
  telescopes (end of 1996 - end of 2002), pioneering with 5 telescopes
  the stereoscopic observation mode. Concerning the Galactic sources
  these include the detection of Cassiopeia A, being the only shell
  type supernova remnant at TeV energies seen up to now in the
  northern sky and recently the observation of a yet unidentified TeV
  gamma-ray source \mbox{TeV J2032+4130} in the Cygnus region.  Also a
  scan of a large fraction of the Galactic plane has been achieved.
  Concerning the extragalactic sources most precise spectra at the
  highest energies have been obtained from the well studied blazars
  Mkn~501 and Mkn~421 as well as from H1426+428 and 1ES1959+650
  established in the last two years only.  Also extensive
  multi-wavelength campaigns have been successfully performed.
  Recently strong evidence for the nearby giant radio galaxy M87 being
  a TeV $\gamma$-ray source has been obtained. Some of the highlight
  results are addressed below.
\end{small}

\section{Introduction and experimental method}

\vspace{-0.5pc}
The imaging atmospheric Cherenkov telescopes (Table 1) of HEGRA
(Canary island La Palma, 28.75$^0$ N, 17.89$^0$ W, 2200m a.s.l.)
started with a prototype telescope in 1992 and consisted in fall 1996
of a stand alone telescope (CT1) [1] pioneering observations during
partial moon time and 5 telescopes (CT2-CT6) [2] introducing the very
successful stereoscopic observation mode adopted by most of the next
generation experiments.  The stereoscopic observation of an air shower
i.e. the simultaneous measurement of an air shower with several
telescopes, under different viewing angles, allows to reconstruct
unambiguously the shower direction and impact point and also the
shower height on an event by event basis.  This leads to an improved
angular and energy resolution, an improved gamma/hadron separation and
suppression of background from night sky light and muons (due to the
coincidence method). The stereoscopy also allows to record
simultaneously events from well defined background regions giving the
quoted errors a high credibility and to perform sky searches in the
field of view of the camera. The flux sensitivity achieved is a 10
$\sigma$ detection within 1 hour for a source with a flux of 1 Crab.
The telescopes CT2-CT6 ceased operation end of 2002.

\begin{table}
\caption{ \small Properties of the HEGRA Cherenkov telescopes (end of 2002) } 

\begin{center}
\begin{small}
\begin{tabular}{ ll  c             c              l       c            c               c }
\hline
Telescope  & Observ. & Mirror             & PM-Camera             & FoV   & E$_{thres}$ &$\delta \theta$ & $\delta E/E$ \\
           & mode    & area [m$^2$]       & \#pixels              & [deg] & [GeV]       &    [deg]       &  [\%]       \\
\hline
CT1 &       alone    & 5 $\rightarrow $10 & 37 $\rightarrow $ 127 & 3     & 700-900     &  $<$    0.2    &   $\sim$25  \\
CT2-CT6 & stereo     & 5 x 8.5            &  5 x  271             & 4.3   & 500-600     &  $ <$ 0.1      &   10-20   \\
\hline
\end{tabular}

\vspace{-1.5pc}               
\end{small}
\end{center}
\end{table}

\section{Galactic sources and Galactic accelerators of nuclei}                          

It is one of the main goals of the TeV $\gamma$-astronomy to find the
sources of the Galactic cosmic rays commonly believed to be the shell
type supernova remnants. In fact HEGRA detected [3] with Cassiopeia A
the only shell type supernova remnant seen so far at TeV energies in
the northern sky after a long observation time of 210 h (in 3 years)
resulting in a flux of $\sim$ 3 \% of the Crab nebula.  Although not
yet unequivocally proven it is very likely [4] that Cas A is one of
the long sought hadronic accelerators.  A scan of 1/4 of the Galactic
plane (longitude l= -2$^o$ to 85$^o$) yielded upper limits in the
range of 0.15-0.5 Crab flux units for 63 SNR's, 86 pulsars and 9
unidentified EGRET sources [5].  Recently HEGRA detected [6] with TeV
J2032+4130 (Fig.~1) a new source up to now unidentified (i.e. no
counterpart at radio, optical and X-ray energies) in a direction about
0.5 degrees north of \mbox{Cygnus X-3}.  The source is detected
meanwhile with a significance above 7$\sigma$ [6], exhibits a hard
spectrum (index $ \alpha $ $\sim$ -1.9) and is possibly extended (on a
3$\sigma$ level). Several source mechanisms are conceivable and it may
turn out that this source plays an important role in the context of
the Galactic accelerators.
\begin{small}
\begin{table}[h!]
\caption{Galactic and extragalactic sources seen by HEGRA. The fluxes are given  in units of the Crab-flux
[$ \rm dJ_{\gamma}/dE = (2.79 \pm 0.02 \pm 0.5)\cdot 10^{-7}(E/1TeV)^{-2.59\pm 0.03 \pm 0.05}$
photons m$^{-2}$ s$^{-1}$ TeV$^{-1}$]. (C=CAT, H=HEGRA, V=VERITAS, T=7Tel. Array)}
\vspace{-1.5pc}               
\begin{center}
\begin{small}
\begin{tabular}{l  l        l                    l             l           l       }
\hline
                 & Type    & distance          & significance& Flux       & Experim.   \\
                 &         &                   & [$\sigma$] (HEGRA)  & [Crab flux]&       \\
\hline  
Crab nebula      & Plerion & 1.6 kpc           & $>$ 10      & 1.0        & many   \\
Cas A            & SNR  (shell) & 3.4 kpc      & $\sim$ 6    & 0.03       & H    \\
TeV J2032+4130   & unknown & unknown           & $\sim$ 7    & 0.03       & H    \\
\hline
Mkn 421          & BL Lac  &(z = 0.030)        & $>$ 10      &  0.04-7.40 & many \\
Mkn 501          & BL Lac  &(z = 0.034)        & $>$ 10      &  0.33-6.00 & many  \\
1ES 1959+650     & BL Lac  &(z = 0.047)        & $>$ 10      &  0.05-2.20 & H,V,C,T \\
H1426+428        & BL Lac  &(z = 0.129)        & $\sim$ 7    &  0.03-0.08 & H,V,C \\
1ES 2344+514     & BL Lac  &(z = 0.044)        & $\sim$ 4    &  0.03      & H,V  \\
M87              & radio gal. &(z = 0.0044)    & $\sim$ 4    &  0.03      & H  \\
\hline
\end{tabular}
\end{small}
\end{center}
\end{table}
\end{small}

\begin{figure}[t]
\begin{minipage}[t]{0.49\linewidth}
 \begin{center}
 \includegraphics[width=0.99\linewidth]{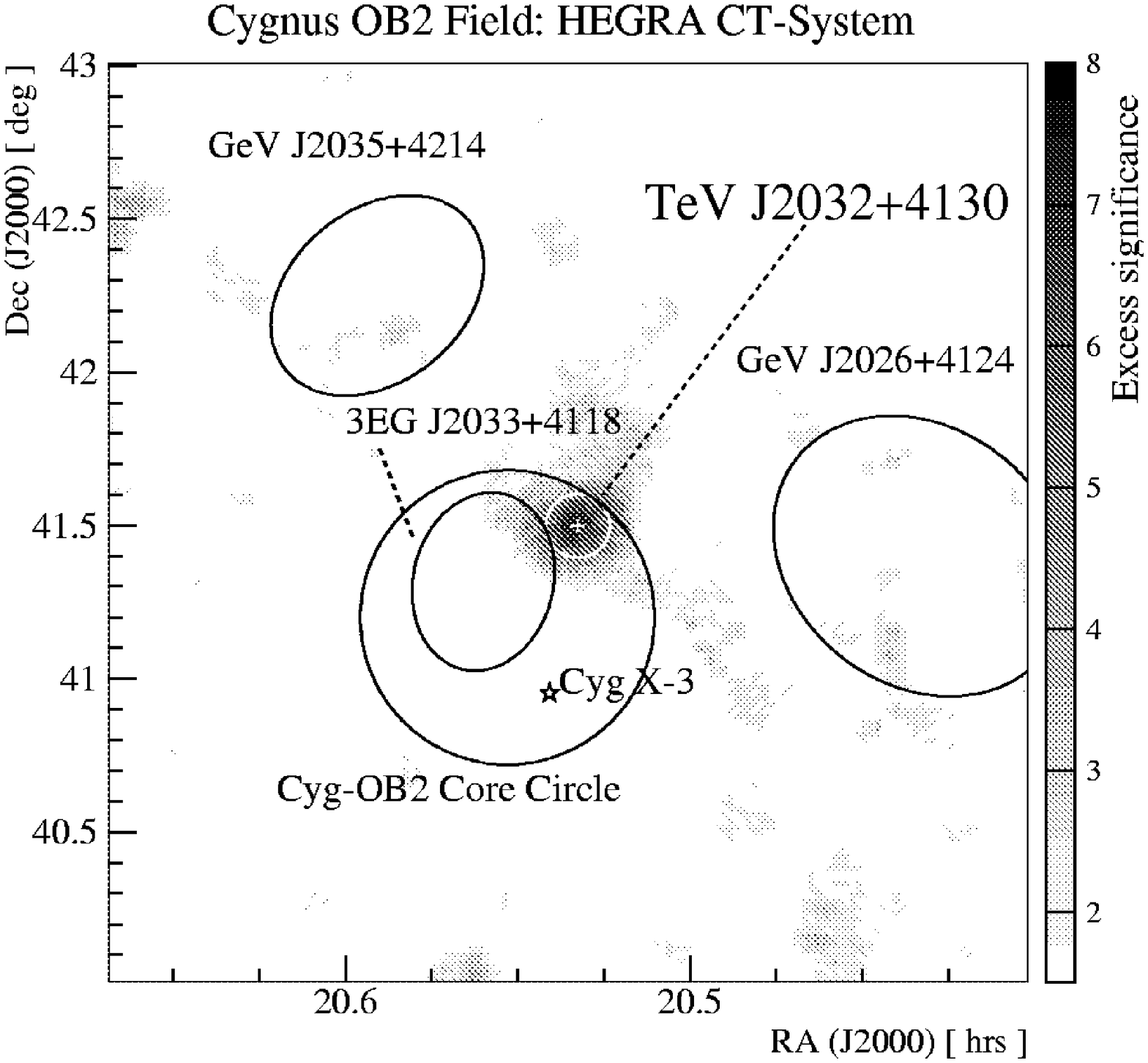}
 \end{center}
\end{minipage}\hspace{0.02\linewidth}
\begin{minipage}[t]{0.49\linewidth}
 \begin{center}
 \includegraphics[width=0.99\linewidth]{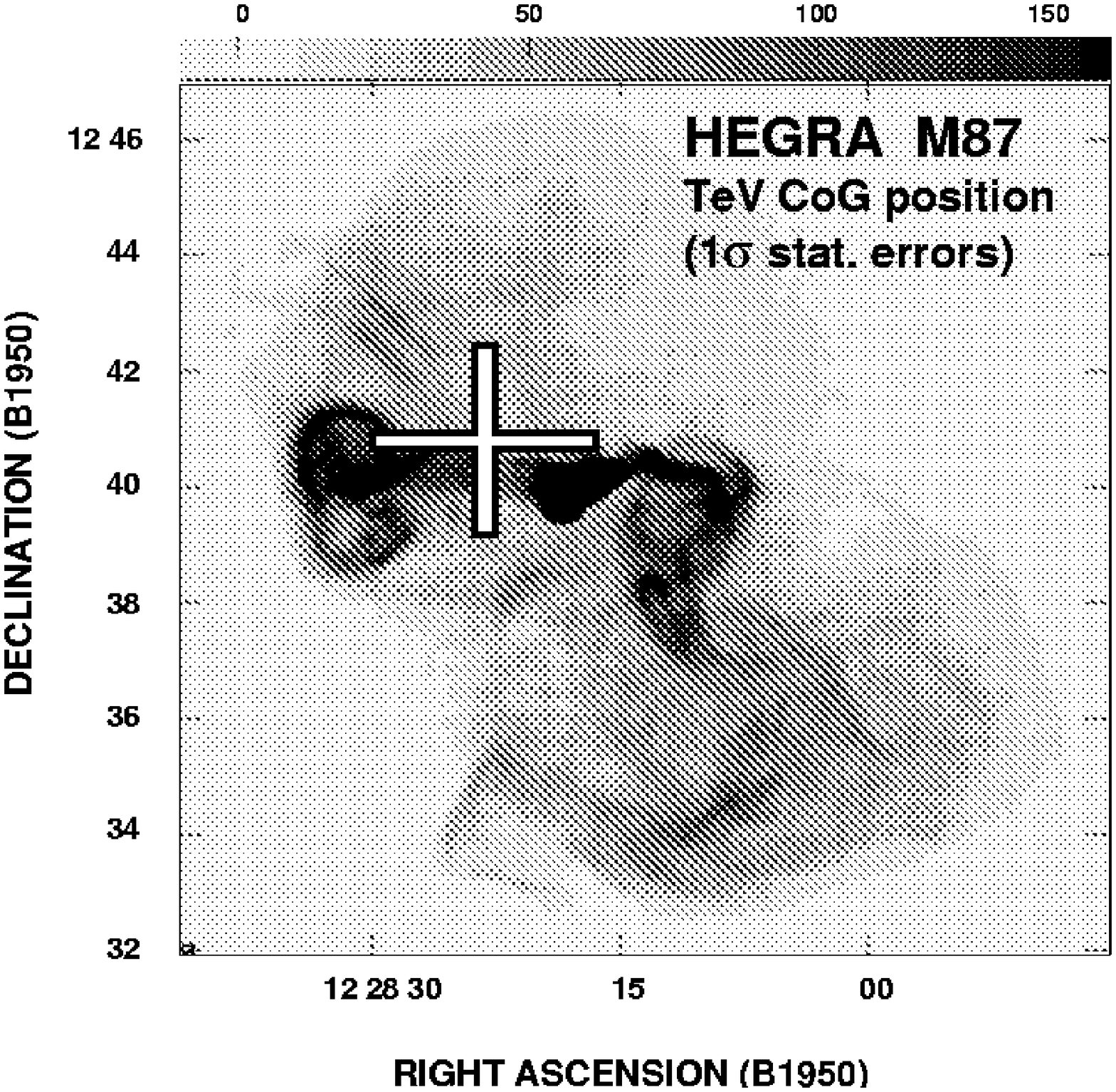}
 \end{center}
\end{minipage}
 \vspace{-0.5pc}
 \caption{\small {\bf \sc   Left:} Skymap of the Cygnus region  
   with the center of gravity and the 2$\sigma$ error circle for
   \mbox{\bf TeV J2032+4130} [6]. Also given are the 95\% error ellipses
   of EGRET sources, the core of Cygnus OB2 and the location of Cyg
   X-3. \newline {\sc Right:} Radio image of {\bf M87} at 90\,cm
   showing the structure of the M87 halo.  The center of gravity
   position of the TeV $\gamma$-ray excess from the HEGRA M87
   observations is marked by the cross (statistical 1\,$\sigma$
   errors) [11].  }
 \vspace{-1.2pc}
\end{figure}

\section{Extragalactic  sources}

Right after the completion of 4 telescopes ready for the stereoscopic
observation Mkn 501 showed strong flares in 1997 during the whole
observation period, allowing a measurement of an extended light curve
and energy spectra of unprecedented precision in this field up to 16
TeV [7].  Remarkably, the shape of the energy spectrum essentially did
not depend on the flux level in 1997. This is in contrast to Mkn 421
which exhibited also flaring states and showed examples for a
hardening of the spectrum with rising flux level in the HEGRA data
[8]. The Mkn 501 spectrum however became softer in the subsequent
years.  In the last two years two more blazars have now been well
established, namely 1ES1959+650 and H1426+428. 1ES1959+650, reported
by the Seven Telescop Array with 3.9$\sigma$ (1998), has been detected
by HEGRA end of 2001 with 5.2 $\sigma$ and showed subsequently strong
flares in May 2002 reaching a significance level above 20$\sigma$ [10] in
the HEGRA data (detected also by VERITAS and CAT). H1426+428 is the
most distant TeV $\gamma$-source up to now and has been detected by
all three northern instruments VERITAS, HEGRA and CAT (Table 2).

The precise HEGRA energy spectrum of Mkn 501 has been used by several
authors to infer the density of the extragalactic background light
(EBL) in the optical to infrared range due to the absorption process
$\gamma_{TeV} + \gamma_{EBL}$ $\rightarrow e^+e^- $. In fact, the
spectra of Mkn 501 (and Mkn 421) in the TeV range yield the best
limits for the EBL density in the mid-infrared range to date, which is
of great cosmological interest. With more precise data also from other
sources such as H1426+428, which is a factor four more distant (and
with a better understanding of the sources) even more stringent
results will be possible.  The present HEGRA data for H1426+428
already indicate a modulation of the measured energy spectrum [10] due
to the imprint of the absorption process mentioned above.
To understand the underlying acceleration processes in the jets of the
blazars several multi-wavelength campaigns have been performed e.\,g.
together with the X-ray satellite RXTE especially for Mkn 421 and
Mkn 501.  Correlated flux variations have been observed which indicate
that an inverse Compton origin is in general able to describe the TeV
data though other processes can not be excluded.

Analyzing the data from observations of further 52 objects from the
AGN class with a total observation time of 658 h yielded signals on
the 4 $\sigma$ level from the blazar 1ES2344+514 [11] (reported
earlier by Whipple) and very recently from the radio galaxy M87
(Fig.~1) [12] which still has to be confirmed.  M87 at the center of
the Virgo cluster is of specific interest since it is the first TeV
AGN being observed using the imaging technique and not belonging to the
blazar class, thus opening the field of TeV $\gamma $-astronomy to a
new class of sources. This very nearby object may also play an
important role for the acceleration of cosmic rays up to the highest
energies.  Further results will be presented at this conference.

\section{Conclusion}

The contributions of HEGRA to the TeV $\gamma$-astronomy on the way to
a mature field have been extremely important by pioneering new
experimental techniques and related analysis methods, by obtaining
high statistics and precise data for light curves and energy spectra
and by detecting new sources such as Cas A, TeV J2032+4130, 1ES
1959+650 and M87, thus promising a great future to the next generation
Cherenkov telescopes.

\begin{footnotesize}
\section*{References}
\re 1.      Mirzoyan, R.    et al.\  1994, NIM A  351, 513 
\re 2.      Daum, A.        et al.\  1997, Astropart. Phys., 8, 1
\re 3.      Aharonian, F.A. et al.\  2001, A\&Ap  370, 112  and P\"uhlhofer, G. these proc.             
\re 4.      Berezhko, E.G., P\"uhhofer, G. and V\"olk, H.J.\ 2003, A\&Ap 400,971  
\re 5.      Aharonian, F.A. et al.\  2001, A\&Ap  375, 1008                                             
\re 6.      Aharonian, F.A. et al.\  2002, A\&Ap  393, L37  and Rowell, G. et al.\ 2003,  these proc. 
\re 7.      Aharonian, F.A. et al.\  2001, A\&Ap  366, 62                            
\re 8.      Aharonian, F.A. et al.\  2002, A\&Ap  393, 89 
\re 9.      Aharonian, F.A. et al.\  2003, A\&Ap submitted and astro-ph/0305275 
\re 10.     Aharonian, F.A. et al.\  2003, A\&Ap  403, 523  and Horns, D. et al.\ 2003, these proc. 
\re 11.     Tluczykont, M.  et al.\  2003, these proceedings
\re 12.     Aharonian, F.A. et al.\  2003, A\&Ap  403, L1 and G\"otting, N. et al.\ 2003,  these proc.

\end{footnotesize}
\endofpaper

%% file: my_008827_1.tex
\chapter{Search for TeV Gamma-Rays from the Andromeda Galaxy
and for Supersymmetric Dark Matter in the Core of M31}

\author{%
%
%
W. Hofmann, D. Horns and H. Lampeitl, for the HEGRA collaboration \\
{\it Max-Planck-Institut f\"ur Kernphysik, D 69029 Heidelberg, P.O. Box 103989} \\
}

\section*{Abstract}

Using the HEGRA system of imaging atmospheric Cherenkov telescopes,
the Andromeda galaxy (M31) was surveyed for TeV gamma ray emission.
Given the large field of view of the HEGRA telescopes, three 
pointings were sufficient to cover all of M31, including also M32 and
NGC205. No indications for point sources of TeV gamma rays were
found. Upper limits are given at a level of a few percent of the
Crab flux. A specific search for monoenergetic gamma-ray lines from
annihilation of supersymmetric dark matter particles accumulating
near the center of M31 resulted in flux limits in the $10^{-13}$~cm$^{-2}$s$^{-1}$
range, well above the predicted MSSM flux levels except for models with
pronounced dark-matter spikes or strongly enhanced annihilation rates.

\section{Introduction}

M31, as the nearest large galaxy at a distance of 770\,kpc, 
most likely with a black
hole of $3~\cdot~10^7$~M$_\odot$ at its core
[7],
is an obvious target for TeV
observations. While conventional sources such as the Crab Nebula
would not be visible over the distance to M31, given the current
sensitivity of the instruments, the energy output of some of 
the X-ray sources detected in M31 is at a level which, if
continued to higher energy, might be detectable.
A particular mechanism for gamma-ray emission
is the annihilation of supersymmetric
dark matter particles accumulating at the core of M31;
rotation curves suggest that M31 contains a significant
amount of dark matter.

\section{The HEGRA telescope system and the M31 data set}

M31 was observed in August, September and November 2001, using
all five telescopes of the HEGRA system of Cherenkov
telescopes on the Canarian Island of La Palma. 
Each of the telescopes has a 8.5~m$^2$ mirror and is equipped
with a 271-pixel photomultiplier camera with a $4.3^\circ$
field of view. The system has an energy threshold around
500\,GeV and a $0.1^\circ$ angular resolution. 
The detection rate for gamma rays is rather uniform
within the central $2^\circ$ of the field of view,
and drops to 63\% of its peak value at $1.8^\circ$ from the 
optical axis.
In order to cover all of M31,
observation time was distributed between
three tracking positions, one centered on the core of
M31 at RA 0h 42' 44" DEC $41^\circ$~16'~9.12", one 
displaced by $0.56^\circ$ to the SW at
RA 0h 40' 30" DEC $40^\circ$~39'~0" and one 
displaced to the NE at RA 0h 44' 43" DEC
$41^\circ$~53'~0". After data cleaning, 20.1~h of
good observation time were selected, most of which were
taken at Zenith angles below $25^\circ$. For calibration and
reference, 9.7~h of Crab Nebula data taken in October and
November were selected.

In a first analysis step, the entire field was searched for
indications of gamma-ray point sources.
The data analysis and selection of gamma-ray candidates
follows to a large extent the procedures 
developed for earlier surveys [1,2].
For each grid point, gamma ray candidates with reconstructed directions
within $0.143^\circ$, consistent within the angular resolution, 
were counted.
Since no specific off-source data were taken,
the expected background for each grid point is determined from
the data itself, using appropriate background regions in 
equivalent locations of the field of view of the cameras.

No obvious excess of signal events is visible at
any grid point. The significance for each grid point is then
calculated based on [8].
The distribution of significances for all grid points
approximates a Gaussian distribution with unit width, as expected
in the absence of a genuine signal;
the point with the highest significance of 3.5 $\sigma$
is well compatible with the expected tail of the Gaussian
distribution, and should be interpreted as an upward
fluctuation.

We conclude that there is no statistically significant 
indication of a TeV gamma ray point source in M31, and
extract upper limits for the source flux.
The 99\% confidence level
on the number of excess events is calculated for each
grid point, and is converted into a flux using the 
measured gamma-ray rate from the Crab Nebula. 
The resulting flux limits are shown in 
Fig.~1 (left), and range from 3.3\% of the
Crab flux for the center of M31 to about 30\% at the periphery.
For two other objects in the survey range, M32 and NGC205,
limits of 4.4\% and 2.8\% of the Crab flux are derived,
respectively. More details can be found in [3].

\begin{figure}
\begin{center}
\mbox{
\includegraphics[height=15pc]{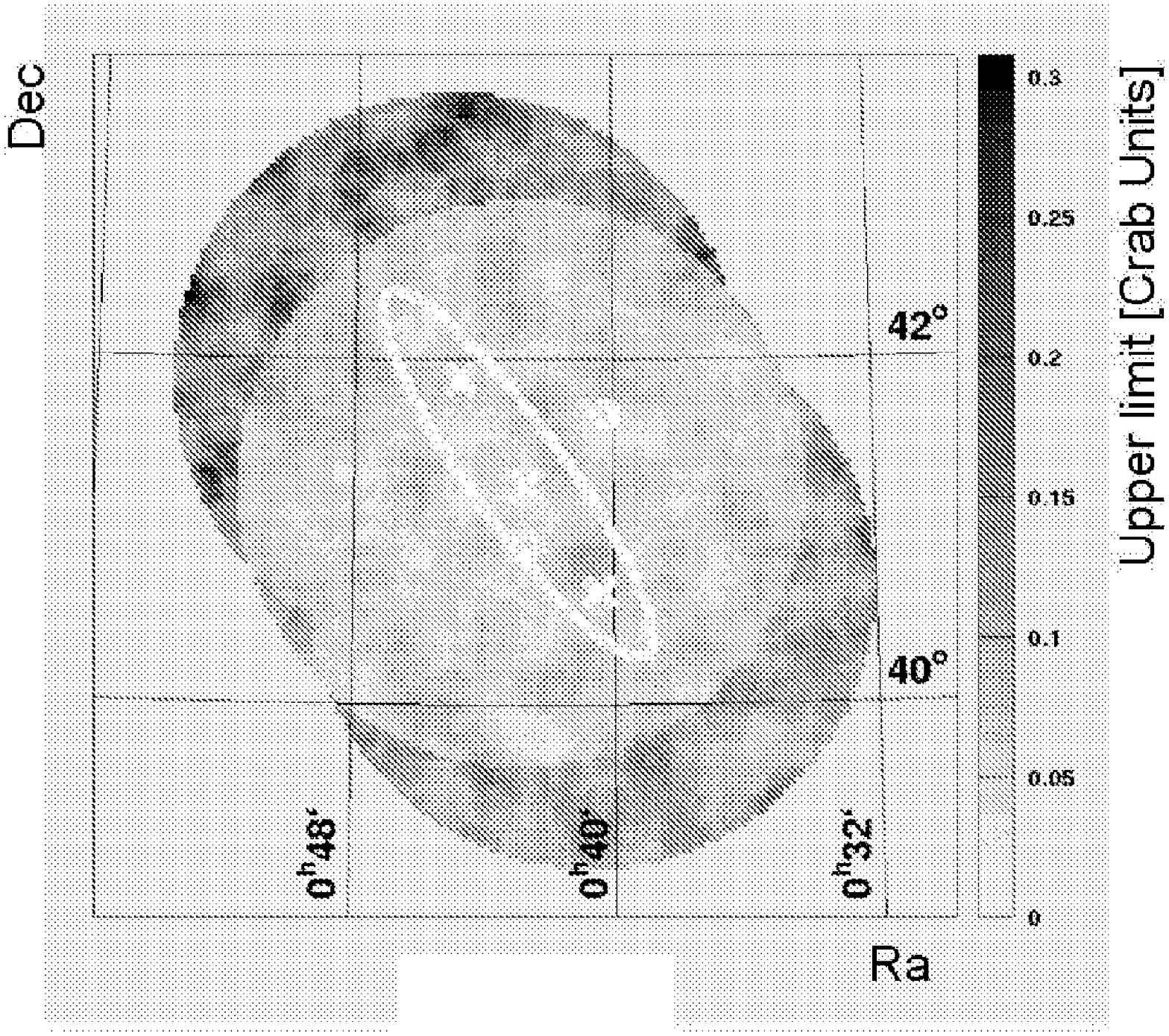}
\hspace{0.5cm}
\includegraphics[height=15pc]{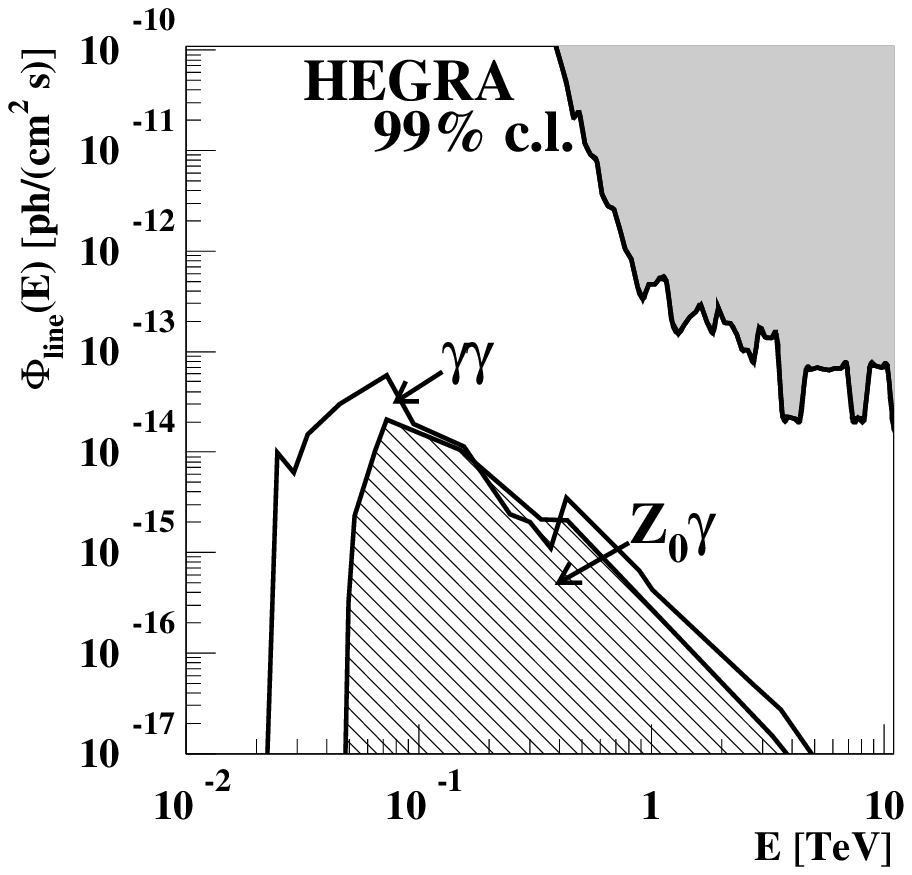}
}
\end{center}
\vspace{-0.5pc}
\caption{Left: Upper limits on the flux from TeV point 
sources in M31, expressed in units of the Crab flux.
For reference, the locations of the center of M31,
the 10~kpc dust ring and the two companion galaxies M32 and NGC205 
are indicated. Tracking positions are indicated by crosses.
Right: Exclusion region (grey shaded region) 
for the the gamma-ray line
flux from the center of M31. The two sets
of curves labeled $\gamma\gamma$ and $Z^0\gamma$ indicate the
upper range of model predictions for the minimal supersymmetric
models calculated in [4]. }
\label{fig_limits}
\end{figure}

\section{Search for supersymmetric dark matter in M31}

Beyond the general survey,
the good energy resolution of the HEGRA telescope system 
allows a dedicated search for supersymmetric dark matter in M31, 
looking for the line emission from the self-annihilation of the 
lightest stable particle (LSP). We have searched for line emission in the energy spectrum 
from the central part of M31 ($r<1.4$~kpc), and compare the results
with predictions of  the 
flux of gamma-rays from the annihilation of
neutralinos ($\chi_0$) in the framework of the minimal supersymmetric
standard model (MSSM) using 
a spatial distribution of the radial density of neutralinos in M31
in concordance with the measured rotation curve 
[6].

 In order to reduce the
background from isotropic cosmic ray events, in the
event selection a tight cut on the direction has been 
chosen
for  the central part of M31. Only events from within a cone of radius $0.105^\circ$
are accepted. At a distance of 770~kpc this
corresponds to the inner $1.4$~kpc of M31. 
The sensitivity for line emission depends on the energy resolution of the detector. 
For this analysis, an improved energy reconstruction algorithm has been applied. The relative energy resolution ($\Delta E/E$) reaches 
10\,\% for a wide energy band from the threshold of 500~GeV up to 10~TeV. 
The search bin in energy is $12~\%$ wide,
concentrating on the central section of a monoenergetic line in order to achieve best
signal to noise ratios.    
The energy bins are equally spaced on a logarithmic scale such that the bin centers
are separated by 0.025 on a decadic logarithmic scale, making the bins correlated. 

The background expectation of the measurement 
is determined using seven independent OFF-regions with similar acceptance to the
ON-region. For the calculation of upper limits on the number of excess events $N_{\gamma}^{99\,\%}$,
the 99\,\% c.l. upper limits were caculated 
and an upper limit on the
rate from a $\gamma$-ray line was derived.
The corresponding flux limits 
were calculated using collection areas $A_{eff}(E,\theta)$ 
derived from Monte Carlo simulations applying the same
reconstruction methods and event selection as for the data analysis. The resulting 
exclusion region is indicated in Fig.~1 (right) as the grey shaded region.

A prediction for the line flux emission was derived using scans of 
the 7-parameter space of MSSM performed by [4]. 
The expected flux 
is given by Eq.~(6) of [6] or the equivalent 
Eq.~(13) of [4] (scaled to the distance to M31):
$$
I_\gamma = (3.18 \cdot 10^{-13} \mbox{ photons cm$^{-2}$ s$^{-1}$})
\left( {\langle \sigma v \rangle N_\gamma \over 10^{-25} 
\mbox{ cm$^3$ s$^{-1}$}} \right)
\left( {500 \mbox{ GeV} \over m_\chi} \right)^2 \Sigma_{19}
$$
Apart from trivial factors, it
depends on the velocity-weighted annihilation
cross section $\langle \sigma v\rangle$ and on the line-of-sight
and solid-angle integrated squared LSP mass density $\rho$
$$
\Sigma = \int \int \rho^2~\mbox{d}s~\mbox{d}\Omega  
$$
written as $\Sigma_{19}$ in units of
$10^{19}$~GeV$^2$~cm$^{-5}$.
The
envelope of the allowed $\langle \sigma v\rangle$ vs. $m_{\chi}$ 
space was used as given in
Figs. 1 and 2 of [4]. 
The unitless parameter $\Sigma_{19} \approx 1.7$
corresponds to $\Sigma_{19}=3$ of  [6],
scaled to the inner 1.4 kpc on the basis of the $\rho(r)$
distributions of Fig. 3(b) of [6]
and using a distance of 770 kpc.
This value of $\Sigma_{19}$ represents the upper range of the different models
for the dark matter halo of M31 discussed in this reference.
The predicted line flux is 
given in Fig.~2 (right). 
We have indicated the two  individual final states with line emission ($\chi^0\chi^0\rightarrow
\gamma\gamma$ and $\chi^0\chi^0\rightarrow Z^0\gamma$) in the figure. The model prediction given here 
is based upon smoothly 
distributed dark matter. Within this model, no signal from M31 with the current sensitivity of 
Cherenkov detectors is to be expected. Very favorable conditions as for example
clumpiness of dark matter would lead to a considerable increase of the annihilation rate
[5,9]. It is conceivable 
that the flux level of some models would become detectable by tuning the distribution of the
dark matter halo and by invoking mechanisms to allow for larger cross sections. However, 
in the absence of a signal, no exotic speculations are required. 

\section*{Acknowledgement}

The support of HEGRA by the Spanish CICYT and 
the German Ministry for Education and Research BMBF is gratefully
acknowledged.

\vspace{\baselineskip}

\re
1.\ Aharonian, F.A., Akhperjanian, A.G., Barrio, J.A., et al. 2001,
A\&A 375, 1008
\re
2.\ Aharonian, F.A., Akhperjanian, A.G.,Beilicke, M., et al. 2002,
A\&A 395, 803
\re
3.\ Aharonian, F.A., Akhperjanian, A.G., Beilicke, M., et al. 2003,
A\&A 400, 153
\re
4.\ Bergstr\"om, L., Ullio, P., \& Buckley, J.H., 1998, Astropart. Phys. 9, 137
\re
5.\ Bergstr\"om, L, Edsj\"o, J, Gondolo, P, \& Ullio, P., 1999, Phys. Rev. D59, 043506
\re
6.\ Falvard, A., Giraud, E., Jacholkowska, A., et al. 2002,
astro-ph/0210184 
\re
7.\ Kormendy, J., \& Bender, R., 1999, ApJ 522, 772
\re
8.\ Li, T., \& Ma, Y., 1983, ApJ 272, 317
\re
9.\ Taylor, J.E., \& Silk, J. 2002, submitted to MNRAS, astro-ph/0207299

\endofpaper

%% file: my_008277-1.tex
\chapter{
Studies of the Crab Nebula based upon 400~hours of Observations with the HEGRA
System of Cherenkov Telescopes}

\author{%
%
%
D. Horns (\textit{dieter.horns@mpi-hd.mpg.de}), for the HEGRA collaboration \\
{\it MPI f. Kernphysik, Postfach 10\,39\,80, D-69117 Heidelberg, Germany}
}

\section*{Abstract}
\vspace*{-0.2cm}
      The Crab nebula has been observed extensively with the HEGRA system
of imaging air Cherenkov telescopes. Roughly 400 hours of prime observation
time at zenith angles between 5 and 65 degrees have been dedicated to the standard 
candle in TeV astronomy.  Based upon the data set gathered during the
5 years of operation of the HEGRA telescopes, the energy spectrum has been
reconstructed. The energy spectrum extends from 500 GeV beyond 20 TeV and
allows to constrain the position of a possible high-energy cut-off.
\section{Introduction}
\vspace*{-0.2cm}
The Crab nebula
has been a prime target of observations since the beginning of operation of the instrument primarily
 for  calibration and monitoring of the
instrument's $\gamma$-efficiency in the initial stage of operation and during the routine operation. The
observational results include the
energy spectrum [3], upper limits on the angular size of the emitting region [4], 
and constraints on the fraction of pulsed radiation [5]. After the end of the live-time of the instrument (it has
been dismantled in October~2002), an analysis of the entire dataset taken from the Crab nebula with increased statistics and improved
reconstruction technique with respect to previous publications [2] (the most recent published
spectrum from the Crab nebula is based upon $\approx 150$~hrs of observations) is justified and first preliminary results on
the energy spectrum are
presented here.

\section{Observations and Results}
\vspace*{-0.2cm}
     The observations had been carried out between September 1997 and September
 2002.  During it's life-time as a 4 (later
5) telescope system, the setup of the individual telescopes (including 
cameras and electronics) has not been modified.
 However, the degradation of the mirrors and the ageing of the photomultipliers
 have been compensated by an increase of the high voltage
on an annual basis.  The performance of the instrument has been closely monitored by 
various checks on the acceptance, absolute calibration with muon-rings, 
cut efficiencies, pointing calibration, and careful scrutinizing of the photomultiplier 
characteristics [7]. The benefit of the thorough calibration and 
understanding of the instrument's performance is the reliable data-taking that in turn
allows to combine data taken over many years with a minimum of systematics introduced
by changes of the detector.

 The simulations of the detector have been performed using two independent
Monte-Carlo type air-shower and detector simulations [8].  Both simulations agree within 10~\% on the relevant parameters (cut efficiency,
rate of cosmic ray events, angular resolution, energy resolution).  All
detector simulations have been carried out for individual periods 
(one moon cycle)
taking the degrading photomultiplier and optical gain into account to model the
response of the detector and particularly its change with time. Additionally, whenever required, the simulations have
been done with different telescope setups (3, 4, and 5-telescope setups) in case of failures of individual telescopes.
\begin{table}
\caption{\label{table:observations}  Summary of observations, split up into 4 bins of altitude.}
\begin{tabular}{lcccccc}
\hline
\hline
Season & alt$=65-85^\circ$ & $50-65^\circ$ & $40-50^\circ$ & $15-40^\circ$ & $\sum$ & $\sum$ \\
\hline
year   & $[$ksec$]$& $[$ksec$]$& $[$ksec$]$& $[$ksec$]$& $[$ksec$]$ & $[$hrs$]$ \\
\hline
1997/98  &  162.04 & 96.83&   54.48 &  49.67&  363.02&  100.84\\
1998/99  &  217.84 &  60.82&   92.15 & 115.72 & 486.52 & 135.15\\
1999/00  &  108.89 &   9.10&   0.66 &   0.15 &  118.80&   33.00\\
2000/01  &   129.09 &   39.59 &    17.02 & 2.96 &  188.66&   52.40 \\
2001/02  & 125.36 & 52.69&  32.90&  13.62& 224.58& 62.38  \\
\hline
\end{tabular}
\end{table}
 The observations are taken in wobble-mode, where the 
telescopes point with an offset of $0.5^\circ$ to the direction of the source, 
alternating the sign of the displacement from run to run.  A summary of the observational time 
is given in Table~\ref{table:observations}
 The listed observational times are selected for good weather conditions
and instrument performance. Overall, less than 10~\% of the runs have been rejected by requiring the event rate
to deviate by less than 25~\% from the expected value (for a given observational period and zenith angle).

 The analysis technique is based upon the method first introduced for HEGRA data in [2] with the
modification of applying a slightly tighter cut 
on the main $\gamma$/hadron separation quantity \textit{mean scaled width}$<1.1$. The \textit{mean scaled width} 
is related to the width of the images scaled to the expectation for a $\gamma$-ray induced event.

For individual events at a given
zenith angle, an energy estimate is calculated with a relative resolution of $\Delta E/E<12$~\% [6]. Using 
Monte-Carlo calculated tables of collection areas as a function of reconstructed energy for 5 discrete zenith angles
a collection area for each event is calculated using interpolation in zenith angle and energy. Finally, the sum
of the inverse of the collection areas is calculated for a given energy bin and using the dead-time corrected on-time, 
a differential flux is calculated.  The background is calculated by averaging over 5 regions in the camera with
the same distance to the camera center and diameter as the source region.

\begin{figure}[ht!]
\begin{center}
 \includegraphics[width=0.9\linewidth]{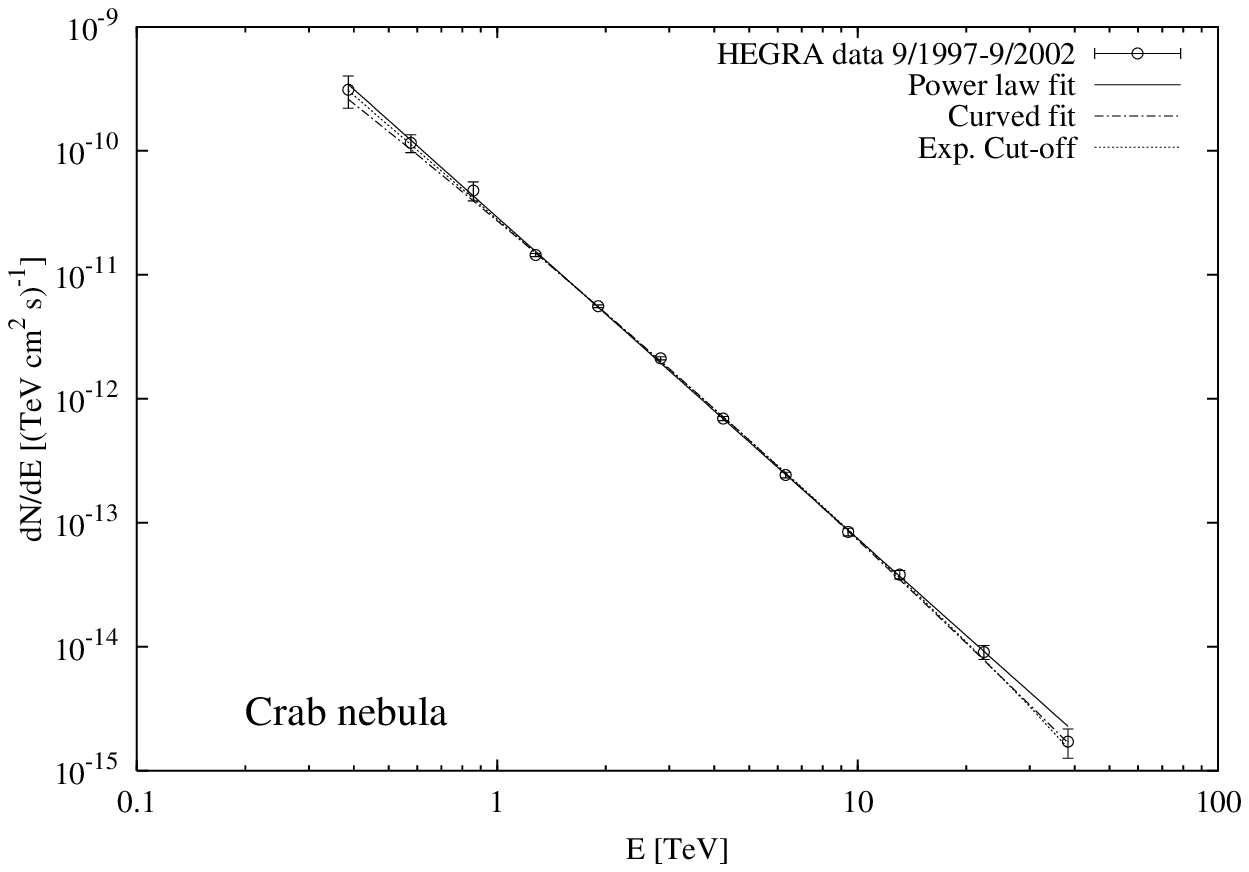}
\end{center}
 \caption{The differential photon energy spectrum for the Crab nebula. The preliminary result
\label{fig:Crab}
is shown with the statistical errors only. The systematic errors are dominant below 1~TeV and 
above 20~TeV and are not included. Additionally, the measurement is subject to an overall
uncertainty of the absolute energy scale estimated to be less than 15~\%. }
\end{figure}

 We have omitted to include the observations below an altitude of $40^\circ$. Further results on large zenith
angle observations will be presented at the conference.  The preliminary result is presented
in Fig.~\ref{fig:Crab} The deep exposure is sufficient to collect $\approx 10\,000$ photons used in the
analysis covering 2 decades in energy. The signal in the  highest energy bin centered on 39~TeV  is beyond 
$5~\sigma$. Inclusion of  large zenith angle observations will overcome the difficulties of saturation and boost
the statistics of photons up to even higher energies.

 The data is well described by a pure power-law fit $dN_\gamma/dE=(2.91\pm0.08)\cdot10^{-11}
(E/\mathrm{TeV})^{-2.59\pm0.02}$~ph/(cm$^2$ s TeV), $\chi^2_{red}(d.o.f.)=1.6(11)$.  
 A functional form with a curvature term describes the data slightly better: 
$dN_\gamma/dE=(2.72\pm0.08)\cdot 10^{-11}\cdot (E/\mathrm{TeV})^{-2.43\pm0.05-(0.14\pm0.04)\cdot\log_{10}(E/TeV)}$, 
$\chi^2_{red}(d.o.f.)=0.77(10)$. A similar $\chi^2_{red}(d.o.f)=1(7)$ is reached
for a power-law with an exponential cut-off: $dN_\gamma/dE=(2.73\pm0.08)\cdot10^{-11}\cdot
(E/\mathrm{TeV})^{-2.43\pm0.05}\cdot \exp(-E/(53\pm38)\mathrm{TeV})$. The very slight
curvature changes the photon-index from 2.43 at 1~TeV to 2.65 at 40~TeV. 

%
%
%
\section{Discussion}
\vspace*{-0.2cm}
 The observations presented here cover for the first time a wide range in energies (two decades) with
a single instrument. The advantage of such an observation is clearly that relative calibration uncertainties between
different measurements are absent. However, there are systematic uncertainties that need to be addressed
before conclusions on the physics are derived. 
The statistical errors of the measurement are very small ($<5$~\% for energies between 1 and 4~TeV) 
and therefore the systematic uncertainties are dominating. 

 The strong energy dependence of the response function of the telescopes in the threshold region is 
introducing uncertainties in the region below 1~TeV. Different reconstruction methods and event selections have
been carried out and the variation of the response at lower energies has been studied. The conservative 
estimate of the systematic uncertainty  below 1~TeV  is a $50$~\% uncertainty region.  This estimate is largely based
upon the remaining variations in the energy spectrum below 1~TeV in different zenith angle bands and over the years
of observation. 

 For the high energy end of the spectrum, saturation of the electronics used for digitization  of the pulses is crucial.
The dynamical range of the FADC electronics is limited to one order of magnitude before the individual time
slices of the FADC saturate. However, using the overall pulse shape makes a larger dynamical range accessible. 
This introduces possible systematic effects by either over- or under-compensating. From the measured image amplitudes
of cosmic ray induced events we expect these effects to be small ($<10$~\%). In a previous analysis [2]
of large zenith angle observations where the image amplitude even for events with energies beyond 10~TeV are
not saturating, no strong systematic effect was visible. 

 Further studies on the pulsed emission from the Crab pulsar and morphology of the $\gamma$-ray emitting
region at different energies are on the way.
\vspace*{-0.2cm}
\section{References}
\footnotesize
\vspace*{-0.2cm}
\re
1.\ Aharonian, F.A., et al. \ 1999, A\&A 346, 913
\re
2.\ Aharonian, F.A., et al. \ 1999, A\&A 349, 11
\re
3.\ Aharonian, F.A., et al. \ 2000, ApJ 539, 317
\re
4.\ Aharonian, F.A., et al. \ 2000, A\&A 361, 1073
\re
5.\ Aharonian, F.A., et al. \ 1999, A\&A 346, 913
\re
6.\ Hofmann, W. et al. \ 2000, Astrop.Physics 12, 207
\re
7.\ P\"uhlhofer, G. et al. \ 2003, submitted to Astrop.Physics
\re
8.\ Konopelko, A. et al. \ 1999, Astrop.Physics 10, 275, Bernl\"ohr, K. internal report, Horns, D. PhD 
thesis 2000
\endofpaper

%% file: my_008277-2.tex
\chapter{
High Energy Emission from H1426+428 and Absorption on the 
Extragalactic Background Light }

\author{%
%
%
D. Horns, F.A. Aharonian,  L. Costamante, for the HEGRA Coll. \\
{\it MPI f. Kernphysik, Postfach 10\,39\,80, D-69117 Heidelberg, Germany}
}
\vspace*{-0.25cm}
\section*{Abstract}
\vspace*{-0.25cm}
 The Blazar H1426+428 ($z=0.129$) has been
observed with the HEGRA stereoscopic system of imaging air Cherenkov telescopes 
for 270~hrs in the years 1999, 2000, and
2002. The object is detected with a significance of 7.5~$\sigma$.
The energy spectrum measured by HEGRA extends up
to $\approx 8$~TeV with indications for a change in the spectral slope
at energies above 2 TeV. The integral flux above 1~TeV in 1999 and 2000
amounts to $\approx8$~\% of the flux of the
Crab Nebula. In 2002, the flux drops to a level of $3$~\% of the Crab
Nebula.   The extragalactic background light (EBL) at
wavelengths between 1 and 10 microns causes substantial ($\tau=1\ldots 9$)
absorption of photons in the measured energy spectrum.
The intrinsic spectra after correction for absorption are discussed.
 Nearly simultaneous observations with the X-ray observatory RXTE
in 2000 and 2002 indicate flux and spectral variability in the X-ray band
between 2 and 20~keV.
\vspace*{-0.25cm}
\section{Introduction} 
\vspace*{-0.25cm}
The ongoing exploration of Blazar type objects over a broad region of the electromagnetic spectrum
has revealed surprising properties 
that are of relevance for the
study of injection, acceleration, and radiative cooling of particles in relativistic jets. However, a deep
understanding of these objects based upon the observations at TeV energies suffers from 
the ambiguity of the observed spectrum with respect to 
absorption effects on the extragalactic background light (EBL). Currently,
the object H1426+428 at a red shift of $z=0.129$ is the most distant emitter of
photons up to energies of $\approx 8$~TeV detected. 
 Unfortunately, the observational uncertainty on the measured energy density of the EBL 
is quite
large (up to a factor of 10 at $\lambda\approx 1~\mu$m, see also Fig.~1b).  Given the consequently large uncertainty
on $\tau(E)$, a reasonably accurate (better to within a factor of 10) 
correction for the effect of absorption is not possible. Here, another  approach is followed: 
Using widely different descriptions of the EBL, absorption corrections are applied to the
observed energy spectrum to infer the intrinsic source spectrum. The intrinsic source spectrum is then
checked for consistency with 
our current understanding of Blazar physics.
\vspace*{-0.25cm}
\section{Observations and Results} 
\vspace*{-0.25cm}
\subsection{X-ray observations} 
\vspace*{-0.15cm}
The instruments on-board the Rossi X-ray Timing
Explorer (RXTE) have been used to measure the X-ray flux from H1426+428 during
the years 2000, 2001, and 2002. 
The overall flux level is quite different between the
observations, varying by a factor of 4
between 2000 and 2001 (see Table~\ref{table_data} for a summary of the
pointed observations).  During all three pointings a hard spectrum described
by a power law with $dN/dE\propto E^{-\gamma}$ with $\gamma<2$ is observed. 
This indicates that the object is showing only  weak spectral variations, 
remaining in a hard state for different flux levels. 
Interpreting the corresponding spectral energy distribution
in the framework of a combined synchrotron and Compton dominated emission
model, this would imply that the position of the synchrotron peak is at or beyond
10~keV, and possibly similar to previous observations [3] located in the hard
X-ray band.  In 2000, the spectrum seems to soften above 8~keV.  In 2002 the spectrum seems to harden beyond 15~keV. 
A pure power-law does not satisfactorily describe the data in these cases.

The all-sky monitor (ASM) on-board the RXTE satellite offers a continuous coverage
of the object. Averaging the daily measurements  of the flux for the HEGRA observational nights 
during the 1999, 2000 data set
results in an ASM  rate of $0.33\pm0.08$~counts/sec whereas in 2002 the ASM count rate drops by a factor of 1.7
to $0.19\pm 0.04$~counts/sec. In the context of a leptonic emission model, a correlation of the X-ray flux
and the TeV flux is expected and is confirmed by the HEGRA data (see next section).
\begin{table}[t]
 \caption{Table of X-ray observations. The Galactic column
density has been kept fixed at $n_H=1.38\cdot10^{20}$~cm$^{-2}$.\label{table_data} The fit range for
a power law $dN/dE=N_0 (E/\mathrm{keV})^{-\gamma}$ starts at 4~keV.}
\begin{center}
\begin{tabular}{lc|ccc}
\hline
\small
Pointing 	& \small Exposure 		& \small $N_0$ & \small $\gamma$ & \small  $\chi^2/d.o.f.(d.o.f.)$ \\
\small $[$year$]$		& \small $[$ksec$]$		& \small $[$keV$^{-1}$s$^{-1}$cm$^{-2}]$ & & \\
\hline
\small 2000 &\small  12.7     &\small  $(9.1\pm9,7)\cdot10^{-3}$ &  \small  $1.96\pm0.04$  & \small $2.82(7)$  \\
\small 2001 &\small 4        &\small  $(30.1\pm1.6)\cdot10^{-3}$ &  \small  $1.81\pm0.03$  &\small  $0.75(23)$ \\
\small 2002 &\small  110      &\small  $(9.5\pm0.3)\cdot10^{-3}$ &   \small  $1.82\pm0.02$  &\small  $3.76(27)$ \\
\hline
\end{tabular}
\end{center}
\end{table}
\vspace*{-0.25cm}
\subsection{TeV observations} 
\begin{figure}[t]
 \mbox{
 \parbox{0.5\linewidth}{
    \includegraphics[width=\linewidth]{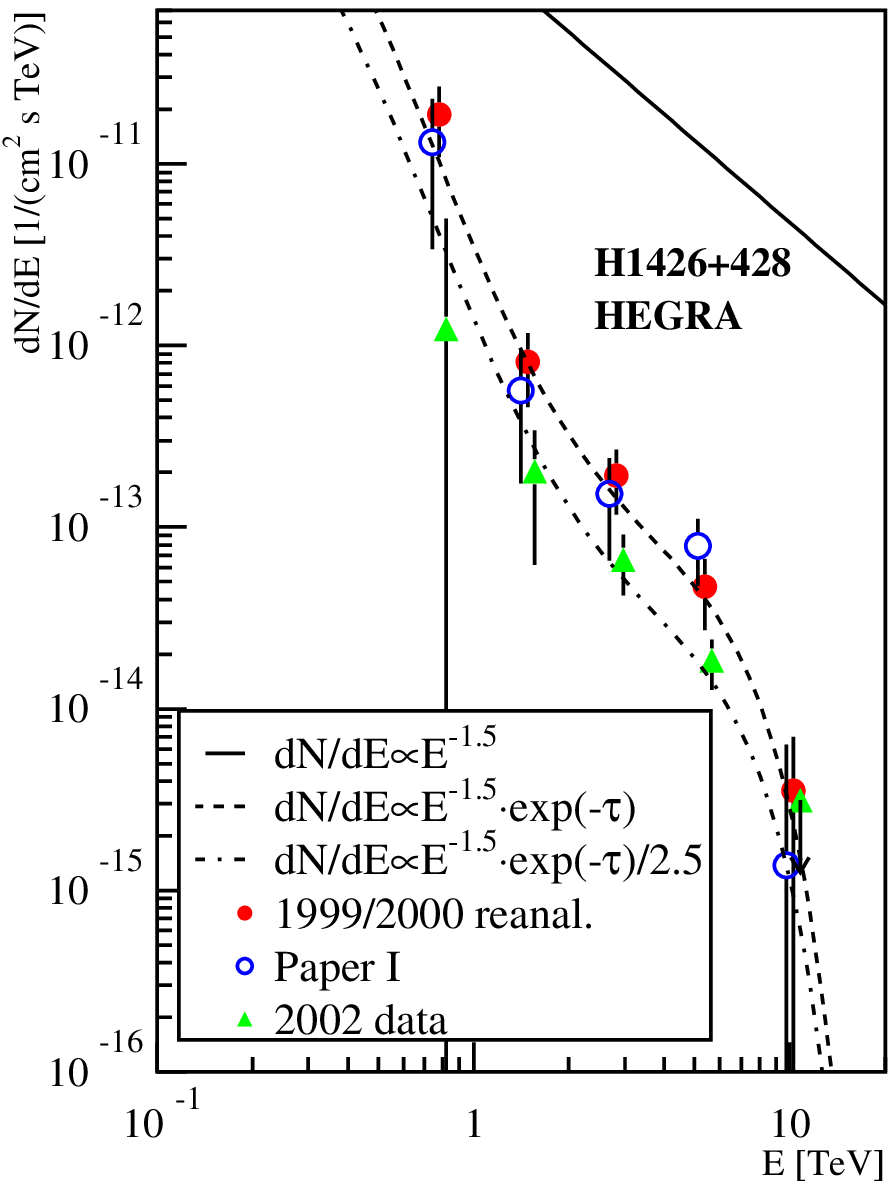}
 }
 \parbox{0.5\linewidth}{
    \includegraphics[width=\linewidth]{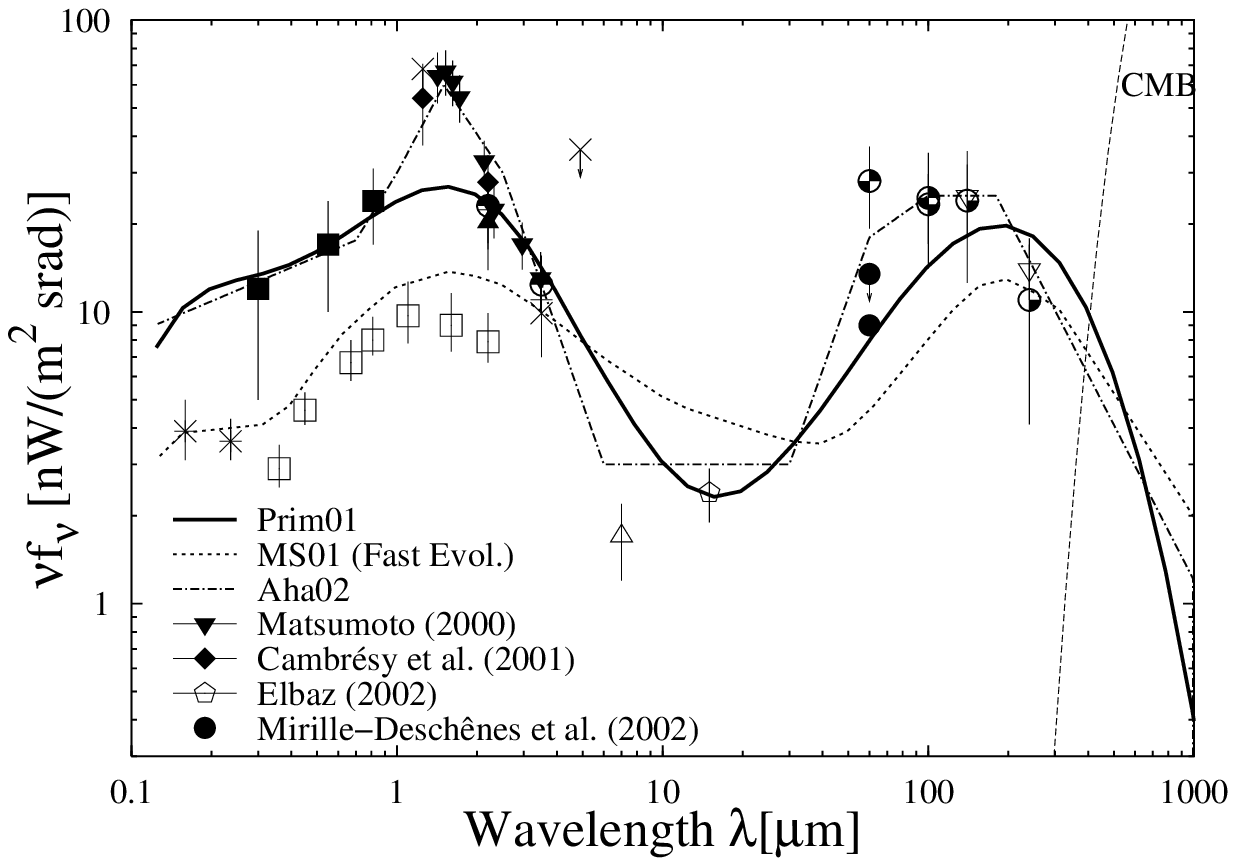}
  \caption{a) The differential energy spectrum of H1426+428 as measured by HEGRA in the two different
\label{fig:spec1}
flux states seen in 1999, 2000, and 2002. For comparison, the previously published results
are included. 
 b) A selection of measurements and models for the EBL from UV to far infrared
(taken from [2]).
 }
}
}
\end{figure}

\vspace*{-0.2cm}
The observations carried out with the HEGRA system of 5 imaging air Cherenkov
telescopes detected the object in 1999/2000 during a 40~hrs observation at a
flux level of 8~\% of the Crab Nebula [1]. In order to verify the observed energy
spectrum with better statistics, an extended observational campaign has been
carried out in 2002, accumulating 217~hrs of good data [2]. However, the source
showed a lower flux of merely 3~\% of the Crab Nebula. The spectral analyses
presented here are based upon an improved analysis technique tailored for
weak sources. The event-selection
criteria have been adopted to weak sources by applying tighter cuts on the arrival direction
and on \textit{mean scaled width}  in order to increase the signal-to-noise ratio.
An improved
energy reconstruction algorithm has been used which benefits from the
stereoscopical reconstruction of the position of the shower maximum and achieves a
relative energy resolution ($\Delta E/E$) of 15~\% at the threshold and
of 10~\% at higher energies. The new method has successfully been used for data taken on the
Crab Nebula and on the original 40~hrs data set of H1426+428 taken in 1999 and
2000.  The resulting data points are shown in Fig.~\ref{fig:spec1} Generally,
 good agreement between the previously published energy spectrum and the reanalysed 
data of 1999/2000 is seen. The 2002 data confirm a hardening in the energy spectrum above 1~TeV. Combining
the excess above 2~TeV of all the data results in a significance of $6~\sigma$.
 \begin{figure}[ht]
    \includegraphics[width=\linewidth]{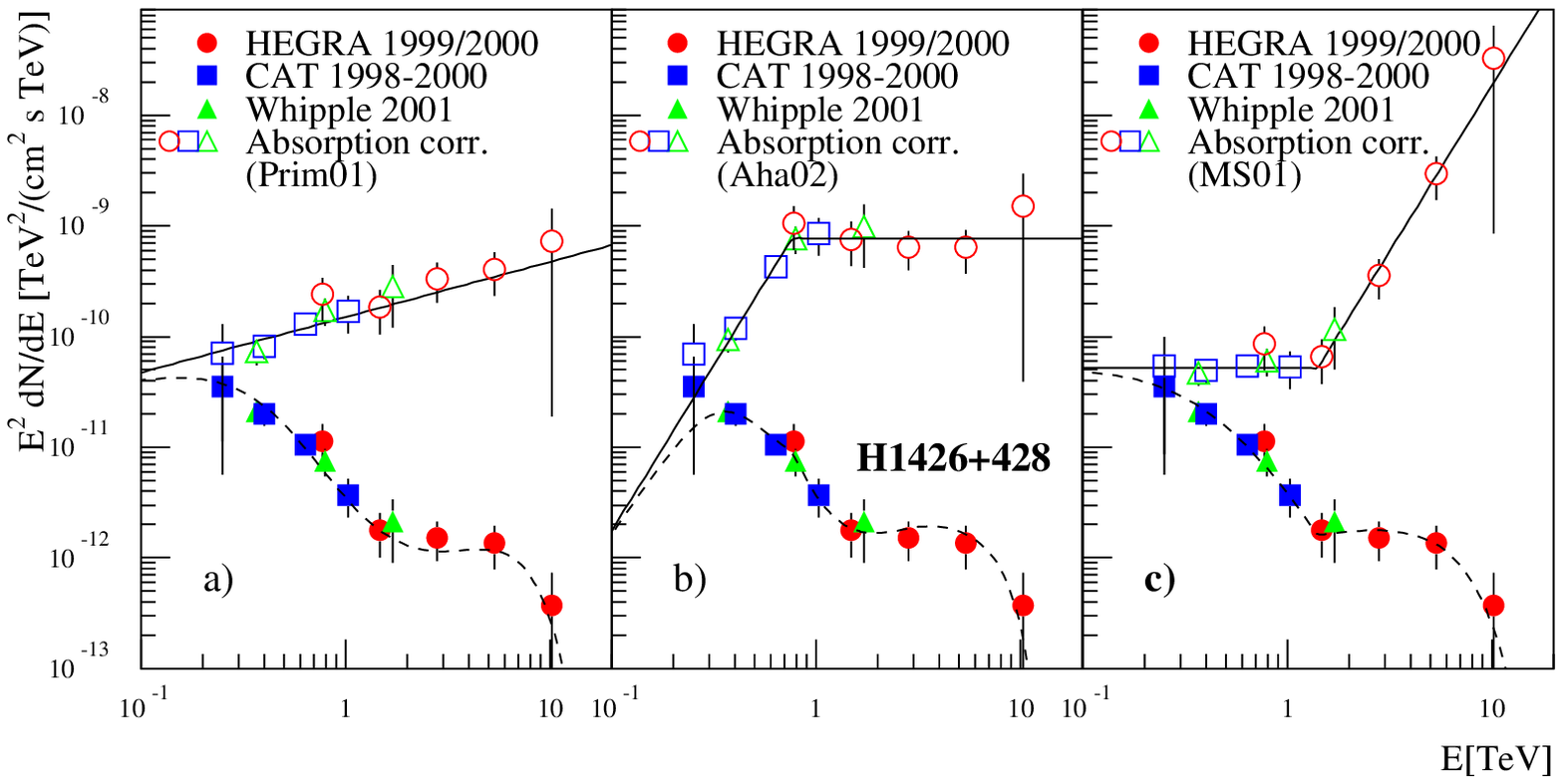}
  \vspace*{-0.7cm}
  \caption{The differential energy spectrum combined from the measurements
of the CAT, Whipple, and HEGRA (1999/2000) groups multiplied by $E^2$
(filled symbols). From
left to right (a-c): The observed and corrected spectra for three different
models of the extragalactic background light (Prim01, Aha02, MS01, see also Fig.~1b).
 The open symbols indicate the corrected spectrum. The solid lines are power
law and broken power-law fits to the open symbols. The dashed lines
are the same power-laws after multiplying the absorption term $\exp(-\tau(E))$.
 }
\end{figure}
\vspace*{-0.2cm}
\section{Conclusions}
\vspace*{-0.2cm}
 Based upon the fact that similar flux levels of H1426+428 
have been observed by the CAT, Whipple, and HEGRA group and 
that the ASM observations confirm a similar level for these observations at X-ray energies, 
it seems feasible to combine the different data sets to benefit from a wider energy band covered. 
Especially at energies below 1~TeV, a steep rise of the energy spectrum because
of absorption effects is expected and confirmed by the different groups [4]. The resulting energy
spectra are shown in Fig.~2(a-c) together with absorption corrected spectra using 
the SEDs of the EBL shown in Fig.~1b. Note, the models Aha02 and MS01 result in an unreasonable
rise of the source spectrum ($dN/dE\propto E$) either at the low energy part (Aha02, Fig.~2b) of the spectrum or the
high energy part of the spectrum (MS01, Fig.~2c). Consulting Fig.~1b it becomes clear that the Aha02 description
of the EBL data (especially the high flux between 1-2~$\mu$m) 
results in a steep rise of $\tau$ for 
$E>100$~GeV. The high value of the SED in the near-infrared as given by MS01 causes a quickly increasing
optical depth above $2$~TeV.

In summary, the TeV observations of extragalactic objects offer an indirect approach to constrain
the EBL. The HEGRA observations of H1426+428 with a clear detection of 6~$\sigma$ above 2~TeV are 
important to constrain the shape of the SED of the EBL in the near infra-red above 2~$\mu$m. Combined data from CAT, Whipple, and
HEGRA constrain the EBL at wavelengths below 2~$\mu$m.

\vspace*{-0.2cm}
\section{References}
\vspace*{-0.2cm}

\small 
\re
1.\ Aharonian, F.A., et al. \ 2002, A\&Ap 384, L23
\re
2.\ Aharonian, F.A., et al. \ 2003, A\&Ap in press
\re
3.\ Costamante, L., et al. \ 2001, A\&Ap 371, 512
\re
4.  Petry, D., Bond, I.H, Bradbury, S.M.~et al.\ 2002 ApJ, 580, 104, 
	Djannati-Ata{\" i}, A., Khelifi, B., Vorobiov, S.~et al.\ 2002, A\&Ap, 391, L25

\endofpaper

%% file: my_009254-1.tex
\chapter{
An new method to determine the arrival direction of individual air
showers with a single Air Cherenkov Telescope}

\author{%
%
%
Daniel Kranich,$^{1, 2}$ and Luisa Sabrina Stark,$^2$ for the HEGRA Collaboration\\
{\it (1) Max-Planck-Institut f\"ur Physik, M\"unchen, Germany\\
(2) Swiss Federal Institut of Technology Zurich, Zurich, Switzerland}\\
}

\section*{Abstract}

We present a new method to reconstruct the arrival direction of
individual air showers. The method is based in part on the arrival
direction reconstruction method of Lessard et al. [1] from the Whipple
collaboration, but yields a significant 30\% improvement in the
obtained angular resolution. The method was successfully tested on
Monte Carlo Simulations and real data of Mkn 421, Mkn 501 and 1ES1959
from the HEGRA Cherenkov Telescope CT1. Based on the same data an
angular resolution of $~0.1^\circ$ could be derived.

\section{Introduction}

The determination of the arrival direction of $\gamma$-ray photons is
essential for the investigation of extended sources and the search for
sources with unknown or inaccurate positions. In the case of an
observation with a few Cherenkov telescopes the images of at least 2
(sometimes 3) detectors are needed in order to determine the arrival
direction without ambiguity. However, even for a single telescope it
is possible to circumvent this problem. Since all $\gamma$-rays have a
similar orientation (limited by the source's extension) one can use
the images of several shower events to estimate the arrival direction
of an individual $\gamma$-ray photon.

\section{The Method}

The method to derive 2-dimensional source maps for a single telescope
is based on the procedure described in [1] and which can be
summarized as follows: The arrival direction for a given shower event
is estimated as the point on the major axis of the shower image
located at a distance DISP to the shower centroid. The DISP parameter
is a function of the elongation of the shower image and is defined as
$DISP = \xi \cdot \left( 1 - WIDTH/LENGTH \right)$. The scaling
parameter $\xi$ has to be determined from real observations or MC
data. The final source map is build up from the arrival directions of
all shower events. In the case of the CT1 data, the DISP parameter was
slightly modified to include the LEAKAGE parameter:

\begin{equation}
DISP = \xi \cdot \left( 1 - \frac{WIDTH}{LENGTH \cdot \left( 1 + \eta
\cdot LEAKAGE \right)} \right)
\label{eq_disp}
\end{equation}
The LEAKAGE parameter is defined as the ratio between the light
content in the camera edge pixels and the total light content of the
shower event. The inclusion of this parameter corrects for truncation
effects of the LENGTH parameter in the small CT1 camera (~$3^\circ$
diameter). The free variables in Equ. \ref{eq_disp}, \( \xi \) and
\( \eta \) were determined from Monte Carlo simulations as \( \xi =
1.3 \) and \( \eta = 6.6 \).

\begin{figure}[h]
  \begin{minipage}[c]{0.5\textwidth}
    \resizebox*{1.\textwidth}{0.33\textheight}{\includegraphics{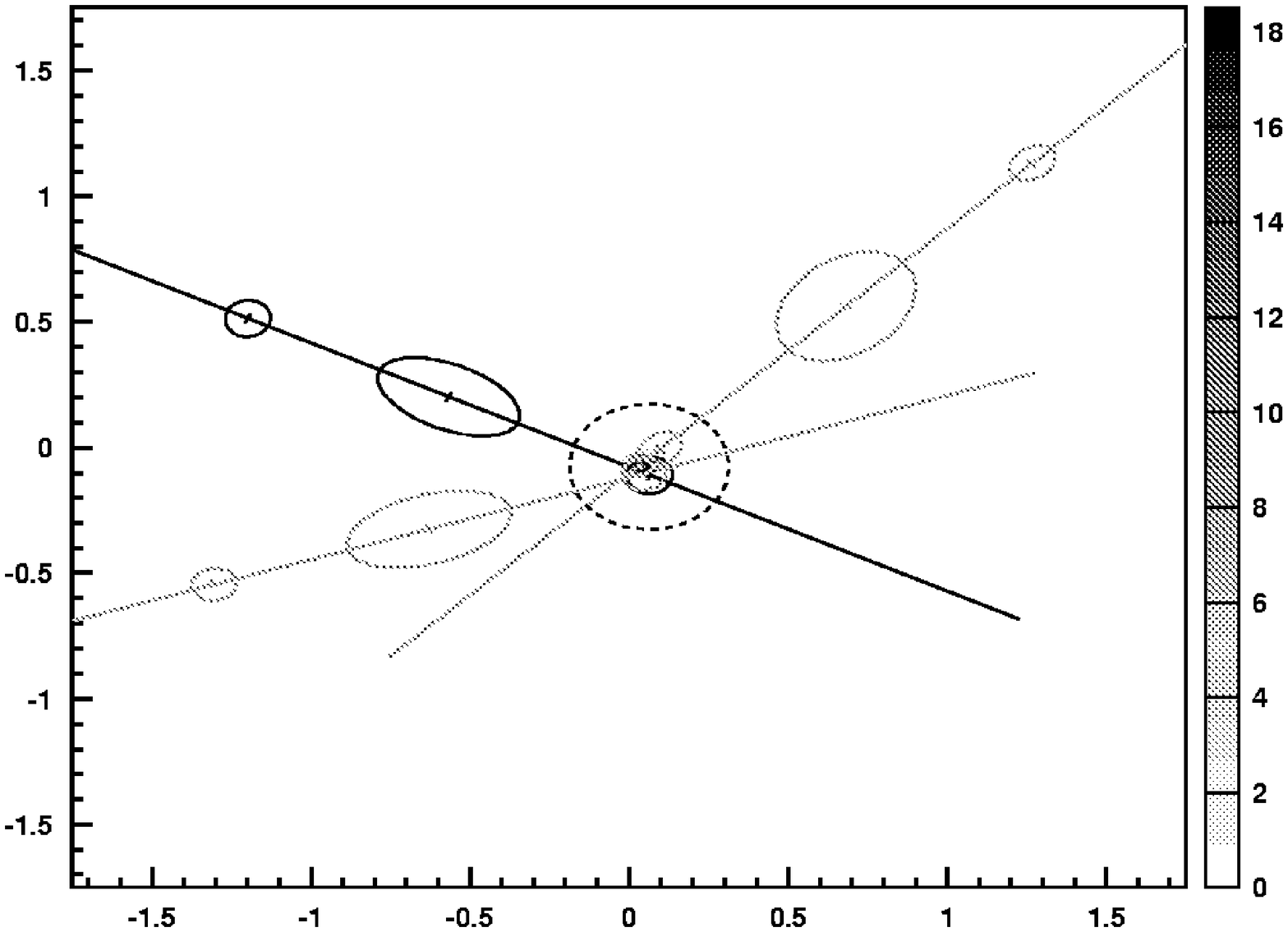}}
    \caption{Example of an event triple in the CT1 camera. The small
      ellipses denote the DISP based arrival direction and its
      uncertainty while the dashed circle gives the $1\sigma$ error of
      the intersection point. The 2 dimensional distribution denotes
      the final arrival distribution for the black shower event.}
    \label{fig_triple}
  \end{minipage}
  \begin{minipage}[c]{0.5\textwidth}
    \resizebox*{1.\textwidth}{0.38\textheight}{\includegraphics{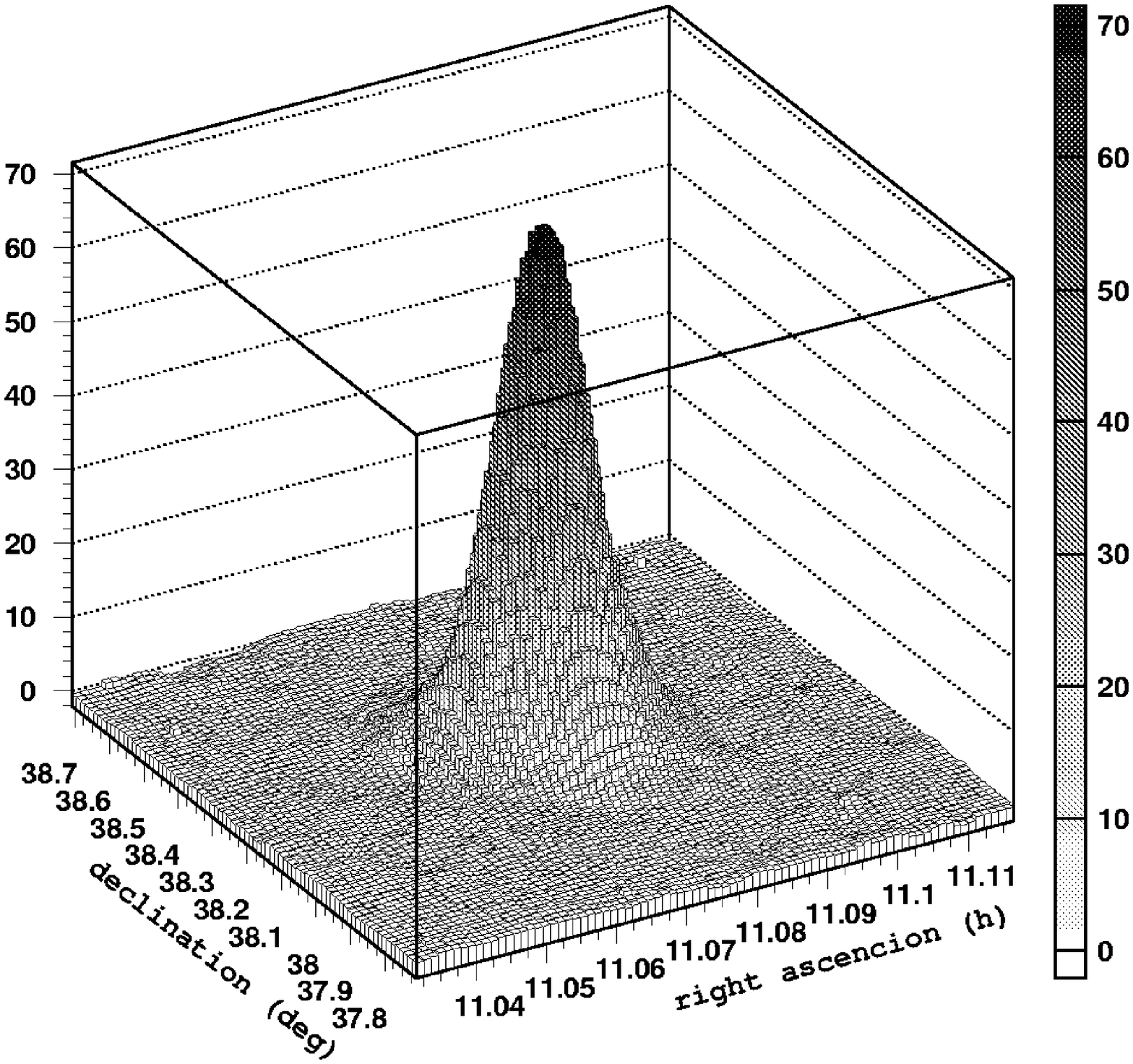}}
    \caption{The excess arrival distribution for the Mkn 421 CT1 data
      sample.}
    \label{fig:3don}
  \end{minipage}
\end{figure}

As a modification of this method we determine a set of possible
arrival directions for each shower event (in the following called
arrival distribution):\\
Taking into account the value and error of both the $DISP$ parameter
and the orientation of the major axis of each shower image one can
calculate the most probable intersection point for any (randomly
chosen) triple of shower images by means of a $\chi^2$  fit (see
Fig. \ref{fig_triple}). The arrival distribution for a given shower
event is then defined by the subset of all those intersection points,
where this event has been part of the triple\footnote{Due to the large
computational overhead of this method the number of intersection points
per shower event have been limited to 500.}. Poorly defined
intersection points with a $\chi^2 / dof. > 6 / 3$ were rejected. Each
derived arrival distribution was then normalized to unity. The final
ON source map is given by the superposition of arrival distributions
from all individual shower events and the final excess arrival
distribution is derived by subtracting the normalized OFF from the ON
source map.\\
The basic idea behind this approach is that all intersection points of
pure $\gamma$ shower triples should point towards the same sky region
whereas a more or less isotropic distribution is expected whenever
hadronic shower images are involved.

\section{Results and Conclusions}

\begin{figure}[t]
  \begin{center}
    \resizebox*{0.49\textwidth}{0.20\textheight}{\includegraphics{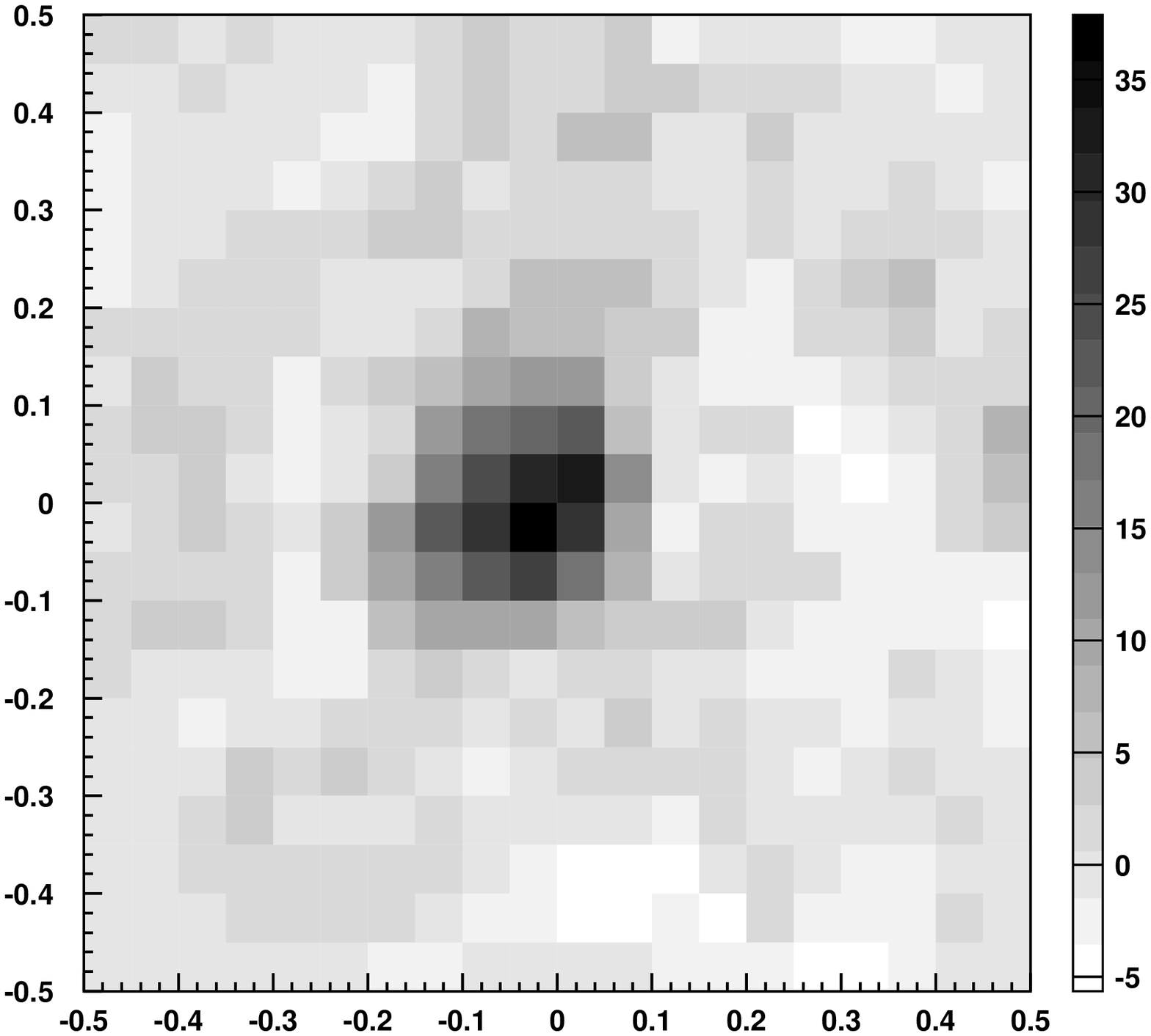}}
    \resizebox*{0.49\textwidth}{0.20\textheight}{\includegraphics{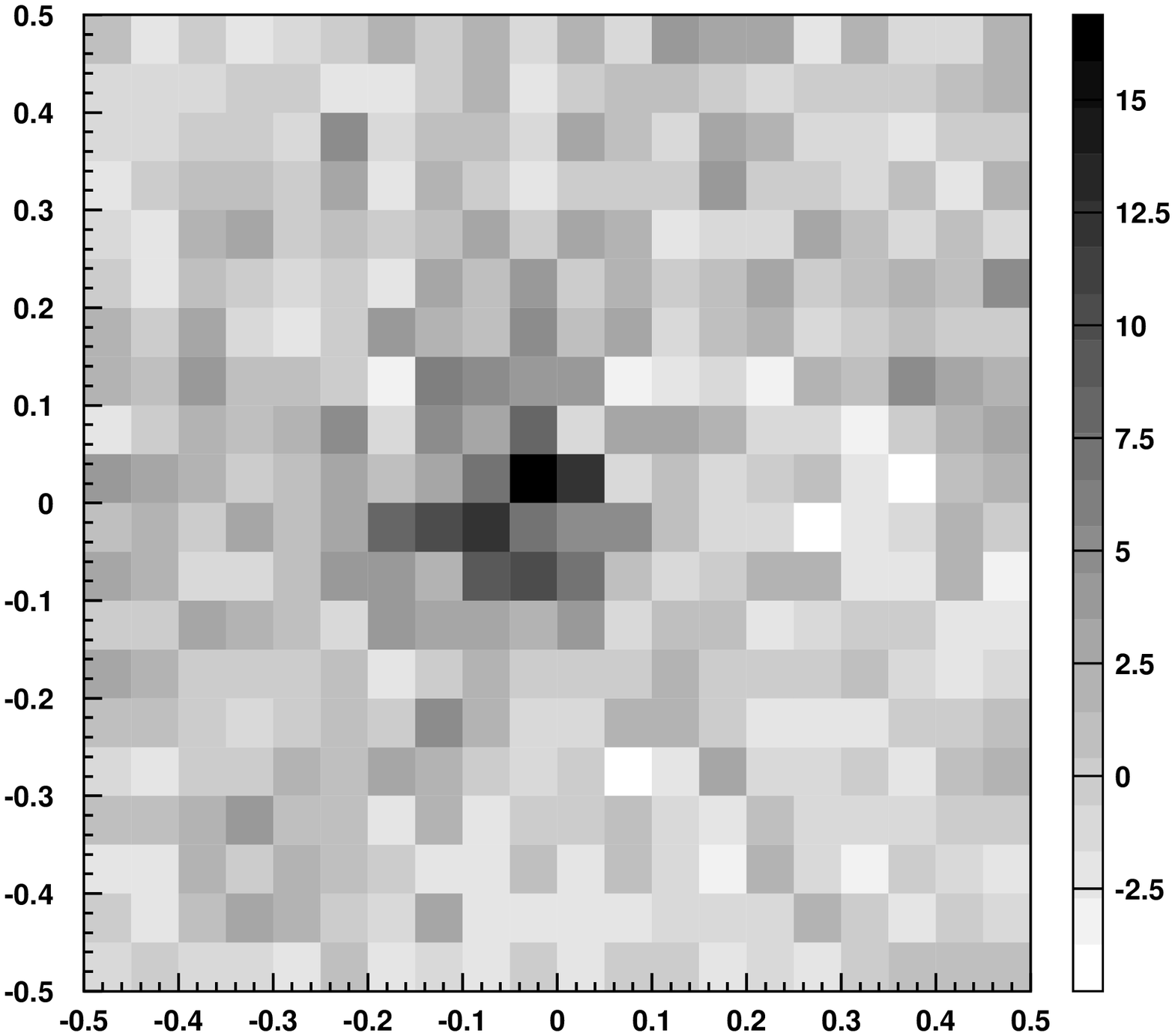}}
    \caption{The excess arrival distribution for the 1ES1959 data
      sample of 2002 as determined with the new (left) and original
      (right) method.}
    \label{fig:1es1959}
  \end{center} 
\end{figure}

The excess arrival distribution for the Mkn 421 CT1 data sample is
shown in Fig. \ref{fig:3don} A fit of a two dimensional Gaussian
onto the data has been used to derive estimates for the angular
resolution ($\sigma_{RA}$ and $\sigma_{DEC}$) and the reconstructed
source position ($RA_\textrm{rec}$ and $DEC_\textrm{rec}$). The
results from this fit on several CT1 data samples are shown in Table
\ref{results_tab} for both, the original method of Lessard et al. [1]
and the modified method as presented here. As can be seen from the
table, the new method yields a significant 30\% improvement in the
angular resolution for almost all sources. The reconstructed direction
of the objects coincides with the known position within the accuracy
of the shaft encoder steps ($0.02^\circ$). It is also evident
from Table \ref{results_tab}, that the angular resolution strongly
decreases for off-axis sources (i.e. where the telescope was pointing
aside the source). This effect seems not to be related to the method
itself since data which was artificially shifted towards an off-axis
position\footnote{Here the image parameters ALPHA and DIST were
recalculated relative to a shifted camera center while the WIDTH and
LENGTH parameters were kept.} didn't result in a decreased angular
resolution. One possible reason could be that for off-axis
observations the small CT1 camera (about $3^\circ$ diameter) leads to
a larger number of truncated images, which then lower the
effectiveness of both arrival direction estimation methods. This is
currently under investigation.\\
The decrease in angular resolution between the 1997 (Mkn 501) and the
2001/2002 (Mkn 421, 1ES1959) data is addressed to the CT1 mirror
upgrade at the end of 1997. The new CT1 mirror was increased from $5\
\mathrm{m}^2$ to $~10\ \mathrm{m}^2$ and shows some increased
aberration effects due to the larger mirror diameter.\\
Even though the derived angular resolution for 1ES1959 is similar for
both methods, the distribution of the excess arrival direction as
obtained with the new method is superior (see Fig. \ref{fig:1es1959})
and yields a much clearer excess peak.

In summary the presented method to determine the arrival direction of
individual photons for a single Cherenkov telescope gives a 30\%
improvement compared to the original method. The derived angular
resolution of $\sim 0.1^\circ$ is similar to the result one would
obtain from a system of Cherenkov telescopes [2]. This improvement
should also allow for an increased cut sensitivity, which, however,
has not been investigated yet.

\begin{table}
{\small
\centering
\caption{Results of the new (this paper) and original [1] arrival
  direction reconstruction method as derived from different CT1 data
  samples. $\sigma_{RA}$ and $\sigma_{DEC}$ denotes the angular
  resolution in RA and DEC direction, $RA_\textrm{rec}$ and
  $DEC_\textrm{rec}$ the corresponding reconstructed source
  position. $\Delta_{RA}$ and $\Delta_{DEC}$ denote the difference
  between the real and the reconstructed source position ($\Delta_{RA}
  := RA_\textrm{source} - RA_\textrm{rec}$ and $\Delta_{DEC} :=
  DEC_\textrm{source} - DEC_\textrm{rec}$). The term 'oa' denotes data
  where the telescope was not directed towards the source but at an
  angular distance of $0.3^\circ$.}
\label{results_tab}
\vspace{0.3cm}
\begin{minipage}{\linewidth}
\begin{center}
\begin{tabular}{l|cccc|cccc}
\hline\rule{0pt}{0.4cm}
{\bf data sample} & \multicolumn{4}{c}{\bf new method} & \multicolumn{4}{c}{\bf original method}\\
& $ \sigma_{RA} $ & $ \sigma_{DEC} $ & $ \Delta_{RA} $ & $ \Delta_{DEC} $
& $ \sigma_{RA} $ & $ \sigma_{DEC} $ & $ \Delta_{RA} $ & $ \Delta_{DEC} $\\[0.2cm]
\hline
Mkn 421, 2001 & 0.075 & 0.073 & 0.005 & 0.014  & 0.100 & 0.099 & 0.004 & 0.015\\
Mkn 501, 1997 & 0.058 & 0.056 & -0.023 & -0.035  & 0.078 & 0.077 & -0.023 & -0.035\\
Mkn 501, oa, 1997 & 0.082 & 0.068 & 0.017 & -0.010  & 0.109 & 0.101 & 0.008 & -0.030\\
1ES1959, 2002 & 0.077 & 0.085 & 0.035 & -0.001  & 0.082 & 0.088 & 0.052 & 0.003\\
MC & 0.052 & 0.052 & -0.010 & 0.022  & 0.073 & 0.072 & -0.010 & 0.023\\
MC oa & 0.081 & 0.081 & -0.040 & 0.019  & 0.119 & 0.123 & -0.039 & 0.022\\
\hline
\end{tabular}
\end{center}
\end{minipage}}
\end{table}

{\bf Acknowledgements: }
The support of the HEGRA experiment by the BMBF (Germany) and the
CYCIT (Spain) is acknowledged. We are grateful to the Instituto de
Astrofisica de Canarias for providing excellent working conditions on
La Palma.

\section{References}

\vspace{\baselineskip}

\re
1.\ Lessard R. W. et al.\ 2001, APh 15, 1
\re
2.\ Daum A. et al.\ 1997, APh 8, 1

\endofpaper

%% file: my_009018-2.tex
\chapter{Scans of the TeV Gamma-Ray Sky with the HEGRA System of Cherenkov Telescopes}

\author{%
%
%
Gerd P{\"u}hlhofer$^1$, for the HEGRA collaboration$^2$  \\
{\it (1) Max-Planck-Institut f{\"u}r Kernphysik, Heidelberg, Germany} \\
{\it (2) \tt http://www-hegra.desy.de/hegra}
}

\section*{Abstract}
{
Between 1997 and the end of the detector operation in fall 2002,
about 5500 hours of observational data were recorded with
the HEGRA system of Imaging Atmospheric Cherenkov Telescopes (IACT).
Besides dedicated scan observations of extended sky regions,
a considerable fraction
of the sky has been looked at as a side effect during the observations of selected source candidates.
Altogether, more than 3\% of the total sky has been observed with the HEGRA IACT system.
We report on a search for possible new TeV sources within this entire data set.
}

\section{Introduction}

A large fraction of the observation time of the HEGRA IACT system [3]
was dedicated to 
individual objects such as known supernova remnants and 
active galactic nuclei, which were known from other wavebands.
Given the system's large field of view (FoV) of 4.3$^{\circ}$
with a homogeneous $\gamma$-ray acceptance over more than 2$^{\circ}$ in diameter, 
scans of extended sky regions were also possible (e.g. [2]). 
Summing up scans as well as observations of individual objects,
more than 3\% of the entire sky has been observed with the HEGRA telescope system.

In the case of observations which were targeted at a point source,
a large fraction of the FoV is not affected by the possible emission from the target object,
and is usually only used to derive background estimates for the source candidate.
It is however obviously possible, in analogy to scanning observations, to regard any
position in the FoV as possible source candidate, and derive background estimates from 
other parts of the FoV.

\section{Analysis challenges}

\begin{figure}[t] 
\centering 
\includegraphics*[width=\textwidth]{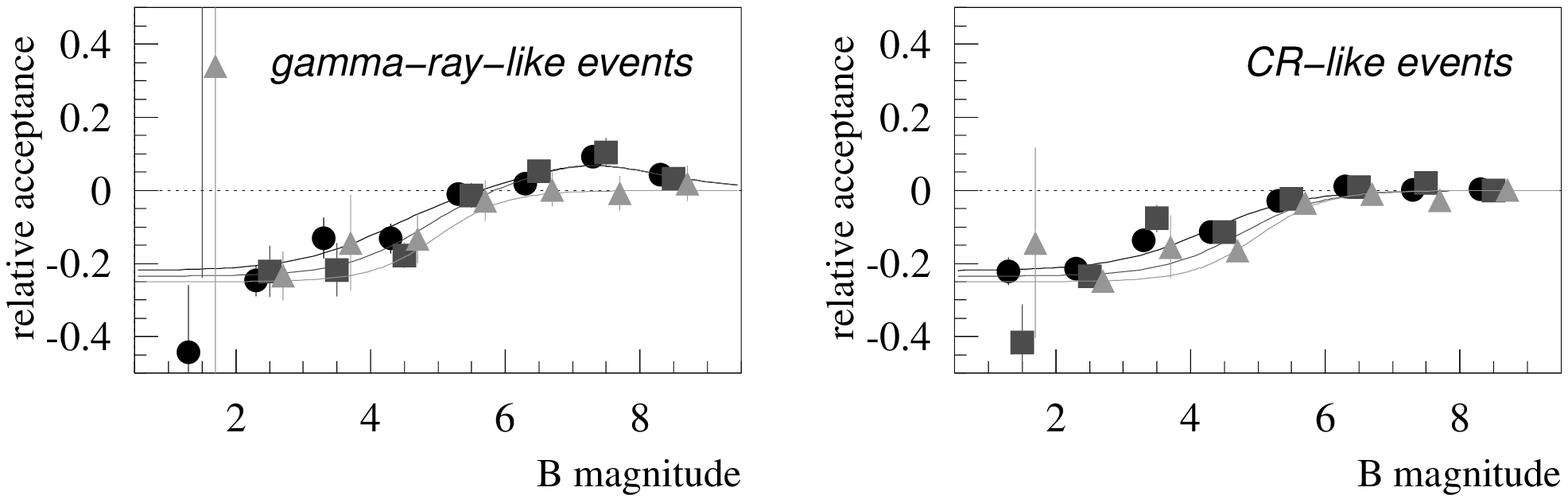}
\vspace*{-0.5cm}
\caption[]{Acceptance change in the field of view due to stars. The figures show the acceptance
relative to a ring segment around the star positions, as a function of the blue star magnitude.
The data points were derived on average for all stars in magnitude intervals. At the left panel,
$\gamma$-ray like events were used, on the right side hadron-like events. The different symbols
indicate different zenith angle bands. The lines are parametrisations which are used to account 
for the acceptance change.}
\end{figure}

The main analysis challenge is a reliable background estimate for an arbitrary position in the FoV. 
Gamma-ray event candidates are identified against the much larger hadronically induced background
solely by the shapes of the event images. The image cut typically rejects 92\% of the hadrons,
the remaining hadrons generate background in the $\gamma$-ray shape cut domain.
Depending on the total observation time and the angular extension of the search area, 
the syste\-matic accuracy of the determination of the absolute
background level at the source candidate location must be ~1\% for very deep observations. 

In HEGRA system data analysis, the standard approach to determine a background estimate 
uses only $\gamma$-ray event candidates after image cuts;
then different positions in the FoV are used as background control regions.
An alternative method (the so-called 'template model' [1])
uses hadron candidates which were reconstructed
to originate from the same direction as the source candidate. In this case, a normalisation
is needed to estimate the background level in the $\gamma$-ray regime from the event counts in the hadron domain;
this value 
must be derived from an average across the whole FoV.

Both background estimates rely on the homogeneity of the ($\gamma$-ray or ha\-dron) acceptance across the FoV.
While detector acceptance inhomogeneities
are typically of the order of 5\% or less,
they may reach 10-20\% in special cases such as large zenith angle observations or strong sky brightness variations
due to stars. Much effort went into the development of an acceptance correction
which accounts for as many as possible known systematic effects.
Figure 1 shows as an example the acceptance at the positions of stars in the FoV relative to the 
surrounding acceptance, 
as a function of the B-band of stellar magnitude. The values were derived from data, averaging over all
stars in the respective magnitude interval which have no other stars nearby.
The lines show empirical parametrisations which are used as part of the
overall acceptance correction. The functional dependence for the $\gamma$-ray and hadron regimes is quite similar,
supporting the idea that the 'template' background estimate may be better able to cope with 
inhomogeneities (of whatever type) at individual pointings.
Fortunately, the two different discussed background estimates are affected by
different uncertainties of the acceptance determination; 
since the results derived with both methods are generally in good agreement, 
we conclude that both estimates are reliable.

The 'template' background estimate is in principle applicable to the investigation of source candidates
which cover a large portion of the FoV, or to look for diffuse $\gamma$-ray emission.
However, the cited systematic problems in the FoV homogeneity prevent us currently from
searching for source candidates which cover a large portion of the FoV.

\section{Search for new gamma-ray sources}

\begin{figure}[t] 
\centering 
\includegraphics*[width=25pc]{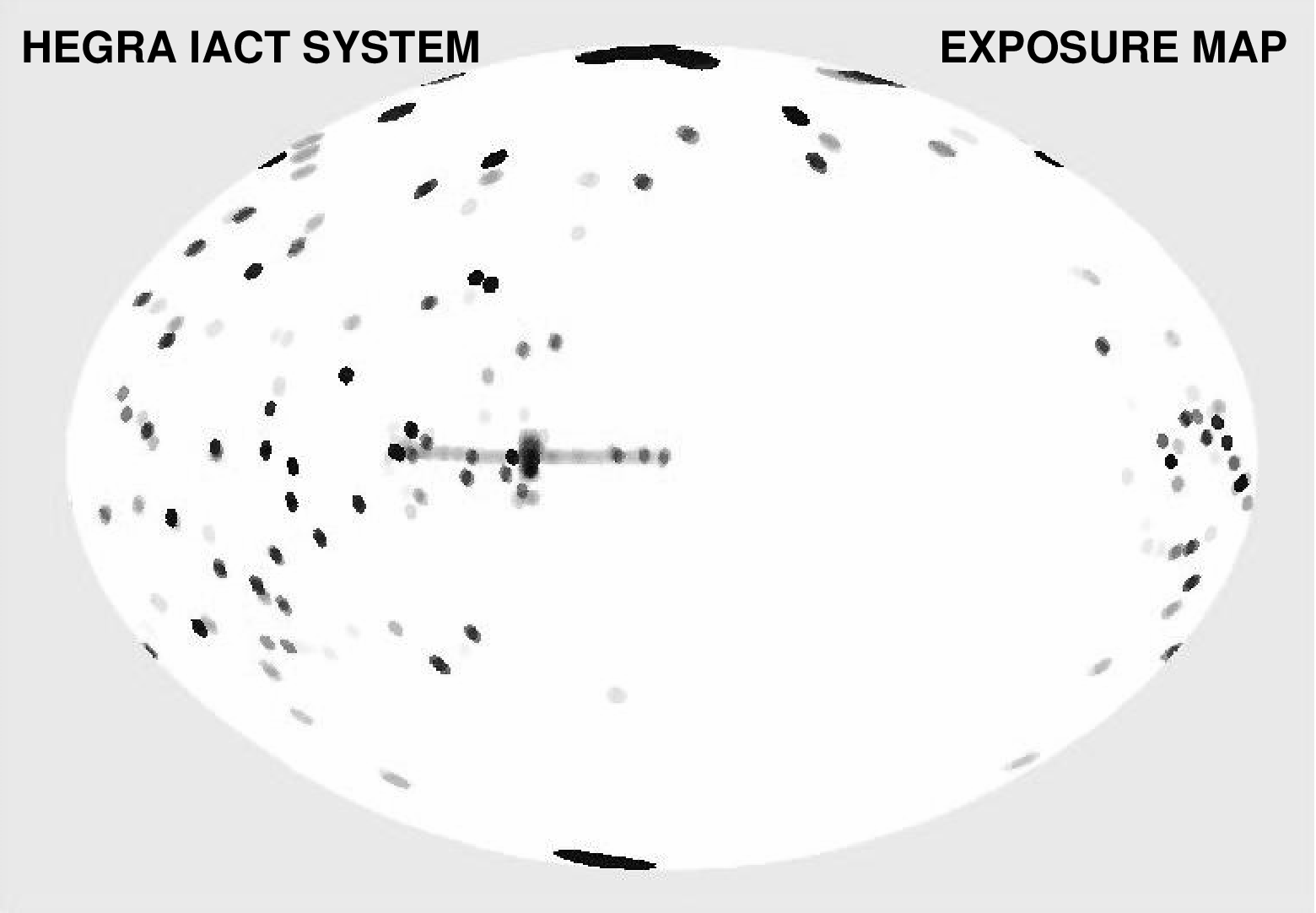}
\caption[]{Exposure sky map of the HEGRA IACT system for the years 1997 to 2001, in Galactic coordinates. 
           Darker colour denotes longer observations.}
\end{figure}

\begin{figure}[t] 
\centering 
\includegraphics*[width=33pc]{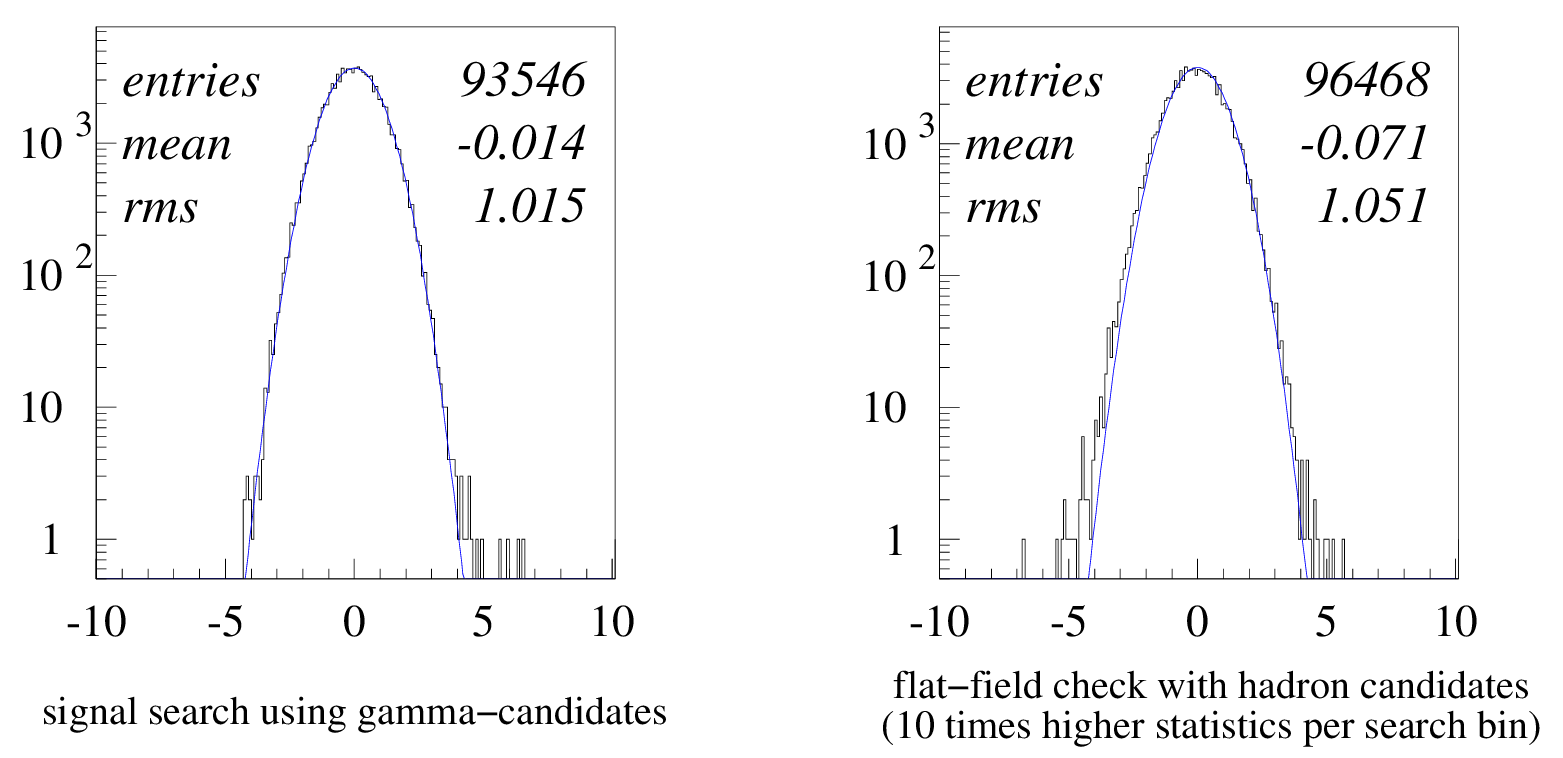}
\caption[]{Significance distributions for the $0.125^\circ$ grid. 
On the left panel, the result of the actual search for $\gamma$-ray sources is shown;
in this case, $\gamma$-ray-like events in a ring surrounding the search position were used
for the background estimate. The solid line indicates the expectation for a purely Gaussian distribution.
On the right side, hadron-like events were treated in the same way as the $\gamma$-candidates 
to perform a check of the flat-field. Since the event statistics is 10 times higher than in the
$\gamma$-ray regime, the nearly Gaussian behaviour of this distribution shows that the background 
estimate is well under control.
}
\end{figure}

In the current analysis,
two different grid search patterns were applied: a dense grid of $0.0625^\circ$ spacing with a tight angular
cut optimized for point sources, and a wider grid of $0.125^\circ$ with a wider angular cut for 
slightly extended sources.
In both cases, the significances for neighbouring grid points are correlated since the cuts are a bit larger than the grid spacing.
Strong known $\gamma$-ray sources were, a priori, excluded from the search grid.

Figure 3 shows significance distributions for the wider grid
from a preliminary 
analysis containing data up to 2001.
The plot on the left hand side represents the search for new $\gamma$-ray sources.
Here, $\gamma$-candidates in a ring surrounding the search position 
were used for background estimate.
An excellent agreement
with the Gaussian expectation for most of the scan positions is obtained, proving that the method works well.
A few scan points show a $\gamma$-ray signal at a level of $5-7\,\sigma$; these include objects
that were detected as TeV emitters already in earlier analyses which were dedicated to
these source candidates (e.g. Cas\,A, H\,1426+428). 
A few new $\gamma$-ray source candidates have been identified;
final results for the entire data set are under way and will be presented at the conference.
Strong unknown TeV emitters in the sky -- at least where HEGRA was pointed -- 
can however be excluded.

The analysis was also performed for hadron events where no localized
sources are expected.
The hadron domain has ten times more counts per search bin than the $\gamma$-ray regime, 
the significance distribution (right panel of Fig.~2) is hence a very sensitive tracer for
possible systematic errors in the flat-field procedure. 
The distribution is nearly Gaussian, which is a good additional check that
systematic acceptance variations are well under control.





\section{References}
\vspace{-0.7cm}
\vspace{\baselineskip}
\small
\re
1. Aharonian, F., Akhperjanian, A., Beilicke, M. et al., A\&A 393 (2002), L37 
\re
2. Aharonian, F., Akhperjanian, A., Beilicke, M. et al., A\&A 395 (2002), 803.
\re
3. Daum, A., Hermann, G., He{\ss}, M., et~al., Astropart. Phys 8 (1997), 1.


\endofpaper

%% file: my_009018-1.tex
\chapter{The Technical Performance of the HEGRA IACT System}

\author{%
%
%
Gerd P{\"u}hlhofer$^1$, for the HEGRA collaboration$^2$  \\
{\it (1) Max-Planck-Institut f{\"u}r Kernphysik, Heidelberg, Germany} \\
{\it (2) \tt http://www-hegra.desy.de/hegra}
}

\section*{Abstract}
{
Between the beginning of 1997 and fall 2002, the HEGRA collaboration
was operating a stereoscopic system of 4 (later 5) Imaging Atmospheric
Cherenkov Telescopes (IACT) on the Canary Island La Palma.
We review the calibration schemes which were developed for the system,
and report on the performance of the detector over this period.
The system had an energy threshold of 500\,GeV
under optimum detector conditions at zenith.
With the calibration schemes described here, a systematic accuracy
of 15 percent on the absolute energy scale has been achieved.
The continuous sensitivity monitoring provides a relative accuracy
of a few percent, and shows that the threshold did not exceed 600\,GeV
throughout the whole period of operation. The readout electronics
and the imaging quality of the dishes were well monitored and stable.
The absolute pointing had an accuracy of better than 25$^{\prime\prime}$.
}

\section{Introduction}

The HEGRA
stereoscopic Imaging Atmospheric Cherenkov Telescope system was located
in the Canary Islands,
at 2200\,m above sea level on the Roque de los Muchachos on La Palma (17$^{\circ}$52$^{\prime}$34$^{\prime\prime}$\ West,
28$^{\circ}$45$^{\prime}$34$^{\prime\prime}$\ North).
The system consisted of 5 identical telescopes (CT\,2 - CT\,6), which operated in coincidence for the
stereoscopic detection of air showers in the atmosphere.
Now, after the disassembly of the system,
we present an account of the long-term performance
and stability of the system.
A far more detailed description can be found in [2].

\section{Telescope pointing and point spread function}

\begin{figure}[t]
  \begin{center}
    \includegraphics[height=18pc]{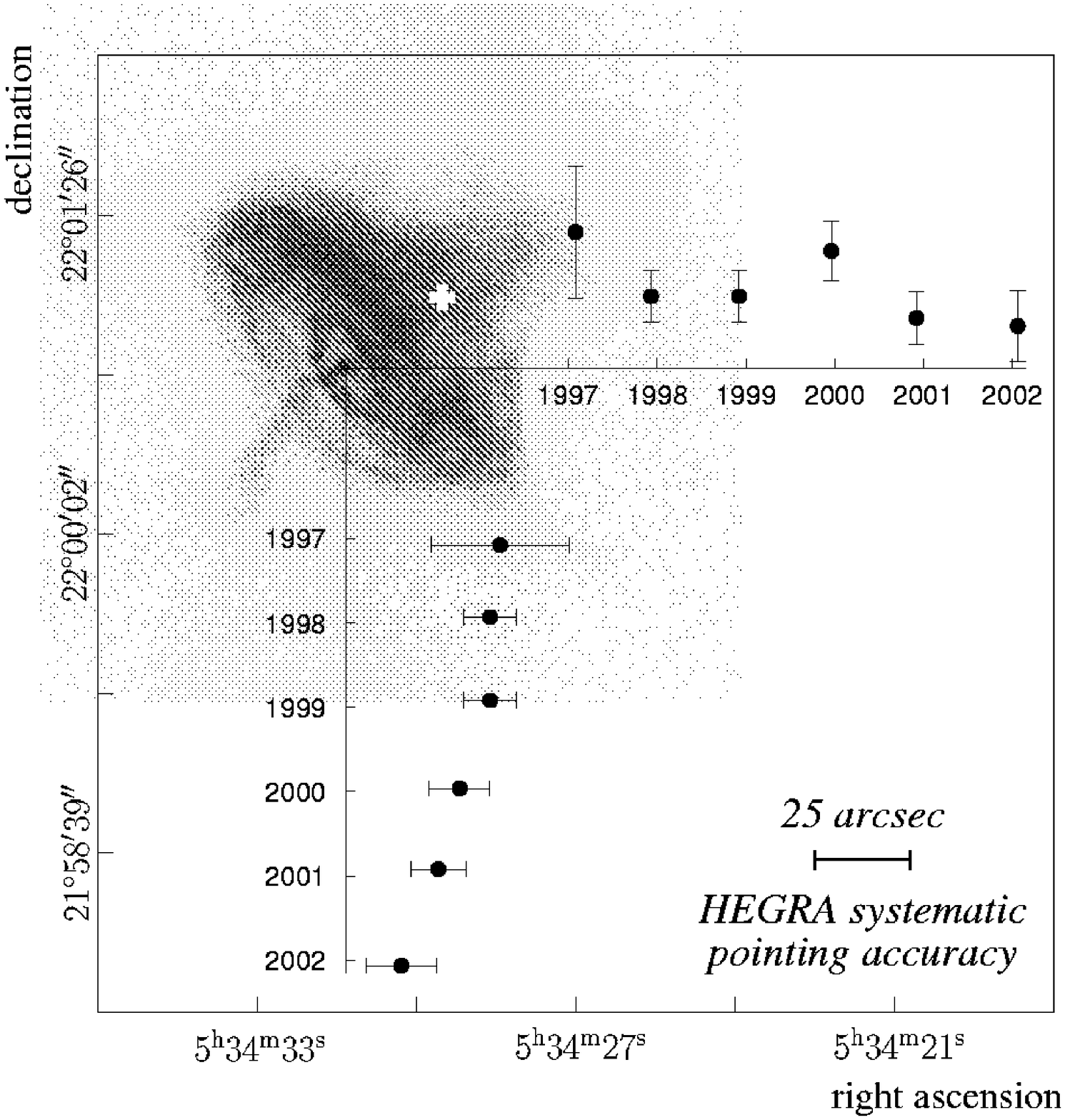}
  \end{center}
  \vspace{-0.5pc}
  \caption{Reconstructed center of TeV emission from the Crab Nebula, and the Chandra X-ray image for reference.
           The white cross denotes the result for the entire HEGRA data set, the insets show the deviation
	   from the expected position in yearly time intervals.}
\end{figure}

The 3.4\,m reflectors were azimuthally mounted and driven by stepper motors.
The telescopes' pointing was calibrated and monitored off\-line by
dedicated calibration runs, so called {\it point runs}, which used
stars as reference objects. Those calibration runs were typically performed every few months.
During data taking, the tracking algorithm relied solely
on the position of the telescopes' axes, which were measured by 14 bit optical shaft encoders.
The pointing was corrected off\-line using an analytical
model of the mechanical structure of the telescopes, the parameters of which 
were determined from the {\it point runs}.
The resulting pointing accuracy was 25$^{\prime\prime}$\ in right ascension and declination.

The temporal stability and accuracy of the pointing calibration
was verified for example by the examination of 
the center of gravity of the TeV emission of the Crab Nebula.
In Figure 1, the white cross denotes the reconstructed center of the
TeV emission for the whole HEGRA data sample; for reference, the
Chandra X-ray image was placed at the year 2000 coordinates of the pulsar.
Additionally, the HEGRA data were split up into yearly intervals, with
results as shown in the insets.
Within the systematic and statistical errors, the centers of the
X-ray and TeV $\gamma$-ray emission are in agreement over the full lifetime of the experiment.

Each telescope dish held 30 spherical glass mirrors with diameters of 60\,cm;
the total mirror area per telescope was $8.5\,\mathrm{m}^{2}$.
The quality of the optical point spread function ({\it psf})
was monitored using the same {\it point runs}. 
The influence of the {\it psf} on $\gamma$-ray induced shower images is best  
obtained from the regular observations of strong TeV $\gamma$-ray sources.
For two short periods only, the mirror alignment had been distorted by ice
such that the image shape
parameters needed to be tuned, until the mirror alignment was corrected.
For the rest of the system lifetime, the image parameters were stable.

\section{Camera electronics calibration, relative sensitivity monitoring}

Each telescope held a camera with 271 photomultipliers, viewing 4.3$^{\circ}$ of the sky.
The signals were digitized with 120\,MHz FADCs.
Less than 2\% of channels were excluded due to technical problems in 95\% of the runs.
The relative gain of the camera pixels
and the relative pixel timing was calibrated by {\it laser runs},
in which the camera was illuminated
uniformly by laser flashes. These runs were performed every night, 
to provide a continuous FADC gain flat-field.
In addition, the individual pixel HV's were adjusted every few months, based on the laser run results;
a trigger acceptance flat-field of 5\% or better was 
thereby 
achieved.

\begin{figure}[t]
  \begin{center}
    \includegraphics[width=30pc]{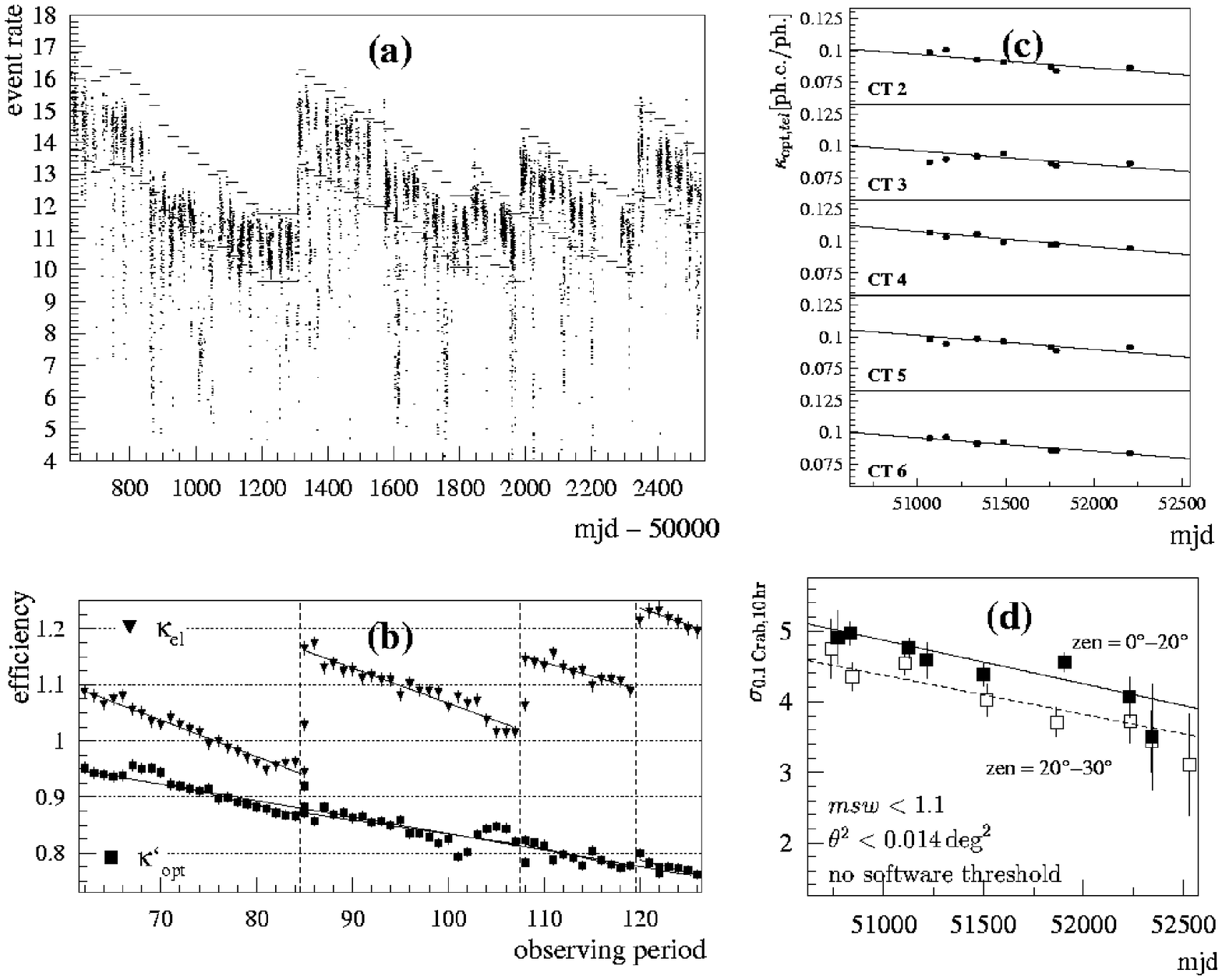}
  \end{center}
  \vspace{-0.5pc}
  \caption{HEGRA system sensitivity results. (a) cosmic ray event rate from runs below 
  30$^{\circ}$\ zenith angle (values rescaled to a 4-telescope setup); rates far below the expectation
  were taken under poor weather conditions.
  (b) electronic ($\kappa_{\mathrm{el}}$) and relative optical ($\kappa'_{\mathrm{opt}}$) efficiency,
  averaged over all telescopes. (c) absolute optical efficiency, as obtained from {\it muon runs}.
  (d) Gamma-ray sensitivity of the HEGRA system (4-telescope setup), derived from Crab observations.}
\end{figure}

Continuous monitoring of changes in the light sensitivity of the complete detectors
was provided by the event trigger rate which originates from the steady background flux of charged cosmic rays (CR)
(see Fig.\,2a). 
The numbers directly translate into a change of the energy threshold of the system. Assuming an
energy threshold of 500\,GeV under optimum conditions (see next section), the threshold
deduced from the CR rate never exceeded 600\,GeV.

The main contribution to the sensitivity decrease of the detectors came from the electronics chain,
presumably due to aging of the PM's last dynodes.
The absolute gain of the camera electronics $\kappa_{\mathrm{el}}$ was
monitored using the {\it laser runs}, values are given in Fig.\,2b.
To compensate for these losses, the global HVs of all cameras were increased three times
(indicated by the vertical dashed lines).
The remaining part of the sensitivity loss was caused by losses in the optical 
throughput, presumably mirror aging (cf. $\kappa'_{\mathrm{opt}}$ in Fig.\,2b, and also Fig.\,2c).

\section{Absolute energy threshold, gamma-ray sensitivity}

The absolute energy threshold $E_{\mathrm{thr}}$ of the system was obtained by shower and detector simulations [1]; 
for early detector conditions, a value of $E_{\mathrm{thr}}=500\,\mathrm{GeV} \pm 15\%\mathrm{(sys)}$ 
was estimated.
The standard method of comparing the CR event rate to expectations from simulations is limited by uncertainties
in the CR energy spectrum and composition. However, the detector sensitivity itself can be derived from
the investigation of muon images in the cameras, and also from special laser calibration setups
using a calibrated photodiode as reference. Figure 2c shows results from {\it muon runs}, which were
performed a few times per year; the values already include the correction by $\kappa_{\mathrm{el},tel}$.
The {\it muon run} results agree within 15\%
with the values used for the CR event rate simulations, confirming the estimated energy threshold.

Finally, in Fig.\,2d we show the sensitivity of the HEGRA system for weak $\gamma$-ray sources,
derived from Crab observations.
The values denote the expected significances for a ten hour observation,
in units of standard deviations $\sigma$, 
for a $\gamma$-ray source which has 10\% of the Crab's source strength.
The slight sensitivity decrease over the detector lifetime
is presumably caused by a deterioration in the background rejection power, 
which could not be compensated for by the
global HV increases [2].

To conclude, the telescope system was very well understood,
regarding both the absolute calibration and the slight performance changes
over the years.					 
We believe that with the HEGRA system, an important contribution has been made to the
effort of establishing ground based imaging Cherenkov telescopes as precision
detectors, well suited within the broad range of astronomical instruments.

\section{References}

\vspace{\baselineskip}

\re
1. Konopelko A., Hemberger M., Aharonian F., et~al. 1999, Astropart.\,Phys.,\,275
\re
2. P{\"u}hlhofer G., Bolz O., G{\"o}tting N., et~al. 2003, accepted in Astropart.\,Phys.

\endofpaper

%% file: my_009463-1.tex
\chapter{
The New Unidentified TeV Source in Cygnus (TeV~J2032+4130): 
HEGRA IACT-System Results}

\author{%
%
%
Gavin Rowell$^1$, Dieter Horns $^1$, for the HEGRA Collaboration$^2$ \\
{\it (1) MPI f\"ur Kernphysik, Postfach 103980, D-69029 Heidelberg, Germany\\
(2) see {\tt http://www-hegra.desy.de/hegra/}}
}

\section*{Abstract}
The first unidentified TeV source in Cygnus is confirmed by follow-up observations
carried out in 2002 with the HEGRA stereoscopic system of \v{C}erenkov Telescopes.
Using all $\sim$279 hrs of data, this new source {\bf TeV J2032+4130}, 
is steady over the four years of data taking, is extended with radius 6.2$^\prime$, 
and has a hard spectrum with photon index $-1.9$. Its location places it at the edge of the core 
of the extremely dense OB association, Cygnus~OB2. Its integral flux above energies $E>1$ TeV amounts
to $\sim$3\% of the Crab nebula flux.
No counterpart at radio, optical and X-ray energies
is as-yet seen, leaving TeV~J2032+4130 presently unidentifed. Summarised here are observational parameters of 
this source and brief astrophysical interpretation.

\section{Introduction \& Data Analysis}
Analysis of archival data ($\sim 121$ h) of the HEGRA system of Imaging 
Atmospheric \v{C}erenkov Telescopes (HEGRA IACT-System see for e.g. [13]) devoted to the 
Cygnus region revealed the presence of a new TeV source [1]. This serendipitous discovery
is now confirmed in
follow-up observations from 2002 ($\sim$158 h) by the same telescopes. Given the lack of a counterpart
at other energies, TeV~J2032+4130 may represent a new class of particle accelerator.

\begin{figure}[p]
{\begin{center}
 \includegraphics[width=0.7\linewidth]{skymap_icrc2003.epsi}
 \includegraphics[width=0.8\linewidth]{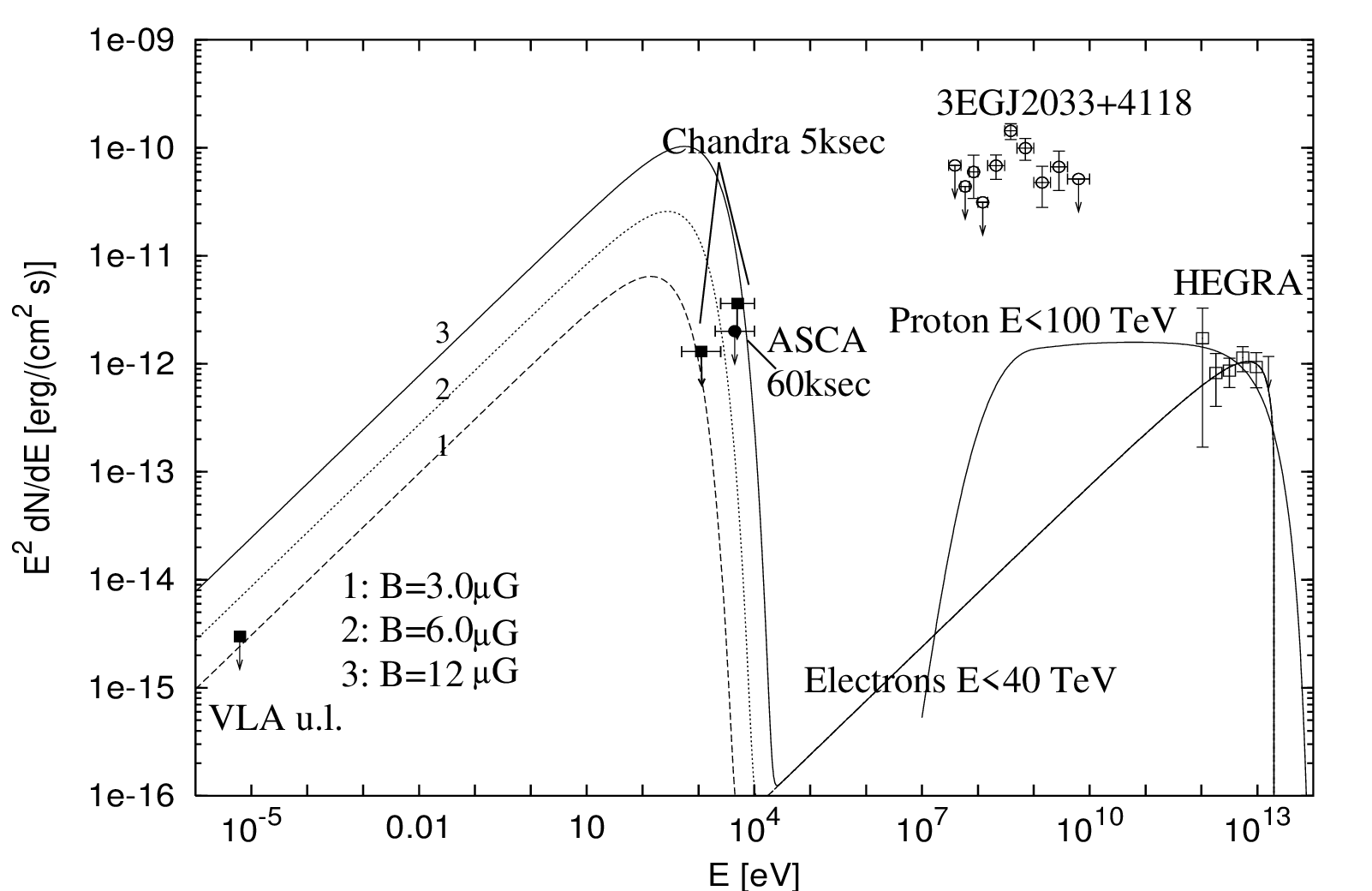}
 \end{center}
}
    \caption{{\bf Upper:}
     Skymap of event excess significance ($\sigma$) from all HEGRA CT-System data 
     (3.0$^\circ \times 3.0^\circ$ FoV) centred on TeV J2032+4130. 
     Some nearby objects are indicated (GeV sources with 95\% contours). The TeV source centre of 
     gravity (CoG) with 
     statistical errors, and error circle (65\% confidence) for the extension (std. dev. of a 2D
     Gaussian, $\sigma_{src}$) are indicated by the white cross and white circle 
     respectively. {\bf Lower:} Spectrum of TeV J2032+4130 (labelled HEGRA) compared with
     purely hadronic (Protons E$<100$ TeV) and leptonic (Electrons E$<40$ TeV) models. 
     Upper limits, contraining the synchrotron emission, are from the VLA and 
     Chandra [3] and ASCA [1]. 
     EGRET data points are from the 3rd EGRET catalogue [4].} 
     \label{fig:skymap} 
\end{figure}

For these analyses, cosmic-ray (CR) background events are rejected via a cut on the {\em mean-scaled-width} 
parameter $\bar{w}$ [2]. Event directions are reconstructed using the 
so-called `algorithm 3' [6], and a cut is made on the angular separation $\theta$ between the 
reconstructed event and assumed source direction.  So-called {\em tight cuts} are implemented: 
$\theta<0.12^\circ$, $\bar{w}<1.1$, and also demanding a minimum $n_{\rm tel}\geq3$ 
images for the $\theta$ and $\bar{w}$ calculation. 
The background is estimated using the {\em template} 
model [1,14], and consistent results are also obtained using an alternative
{\em displaced} background model which employs ring-segments 
within the FoV. For the centre of gravity (CoG) and source extension determination,
an additonal cut on the estimated error in reconstructed direction ($\epsilon\leq0.12^\circ$) is applied, reducing
systematic effects (e.g. [7]). The CoG and source extension are estimated by fitting a 2D
Gaussian convolved with the instrument's point spread function to  a histogram of 
$\gamma$-ray-like ($\bar{w}<1.1$) events
binned over a $1^\circ\times 1^\circ$ FoV. At the CoG, the excess significance now exceeds
7$\sigma$ from all 278.2 h of data, and the source extension is confirmed as non point-like.
The reconstruction of event energy follows the method of [8] and we use tight cuts plus a cut on
the reconstructed air-shower core distance of the event {\em core}$<$200 m. A pure power law explains well the 
energy spectrum, showing no indication for a cut-off. A lower limit to the cut-off 
energy $E_c\sim3.6$, 4.2 and 4.6 TeV is however estimated when fitting a power law+exponential cutoff 
term $\exp(-E/E_c)$ 
and fixing the power index at values $\gamma=$1.7, 1.9 and 2.2 respectively. The integral flux
for energies $E>1$ TeV is $5.9\,\, (\pm3.1_{\rm stat} \times 10^{-13}) ,\,\,{\rm ph\,\,cm^{-2}\,s^{-1}}$ 
or about 3\% of the Crab nebula flux. 
Results are summarised in Table~\ref{tab:numbers} and Fig~\ref{fig:skymap} (upper panel). 
\begin{table}[t]
\caption{Numerical summary for TeV J2032+4130 (preliminary). (a) Centre of Gravity (CoG) and 
      extension $\sigma_{\rm src}$
     (std. dev. of a 2D Gaussian); (b) Event summary. The values $s$ and $b$ are
     event numbers for the $\gamma$-ray-like and bakground (from the Template and Displaced models, see text) 
     respectively, 
     and $s-\alpha b$ is the excess using a normalisation $\alpha$. $S$ denotes the
     excess significance using Eq. 17 of [9]; (c) Events after spectral cuts; (d)
     Fitted power law.}
\label{tab:numbers}
\parbox{0.5\linewidth} {
    {\centering \scriptsize
     \begin{tabular}{l}
     \fbox{\bf \boldmath (a) CoG \& Extension ($\epsilon\leq0.12^\circ$)}\\[2mm]
     \end{tabular}
     \begin{tabular}{lll}
      RA $\alpha_{2000}$:   &  20$^{\rm hr}$ 31$^{\rm m}$ 57.0$^{\rm s}$ & $\pm 6.2^{\rm s}_{\rm stat}$ $\pm 13.7^{\rm s}_{\rm sys}$ \\
      Dec $\delta_{2000}$:  &  41$^\circ$   \,\,29$^\prime$  \,\,56.8$^{\prime\prime}$ & $\pm 1.1^\prime_{\rm stat}$ $\pm 1.0^\prime_{\rm sys}$ \\
      $\sigma_{src}$        &  6.2$^\prime$  & $\pm 1.2^\prime_{\rm stat}$ $\pm 0.9^\prime_{\rm sys}$\\
     \end{tabular}
     \begin{tabular}{l}\\
     \fbox{\bf \boldmath (b) Tight cuts: $\theta<0.12^\circ$, $\bar{w}<1.1$, $n_{\rm tel} \geq 3$} \\[2mm]
     \end{tabular}
     \begin{tabular}{lccccc} 
     Backgr.       &  $s$ & $b$   & $\alpha$ & $s-\alpha\,b$ & $S$ \\ \hline 
     Template       &  1245 & 5926  & 0.168    &  252      & {\bf +7.1} \\        
     Displaced      &  1245 & 15492 & 0.065    &  243      & {\bf +7.1} \\ \hline 
     \end{tabular} \\
   }
 } 
\parbox{0.5\linewidth} {
    {\centering \scriptsize
     \begin{tabular}{l} \\
     \fbox{\bf (c) \boldmath Spectral Cuts: Tight Cuts + core$\leq 200$m}\\[2mm]
     \end{tabular}
     \begin{tabular}{lccccc}
     Backgr.   &  $s$ & $b$   & $\alpha$ & $s-\alpha\,b$ & $S$   \\ \hline
     \multicolumn{6}{c}{----- Energy estimation method: See [8] -----}\\
     Displaced   &  974 & 5122  & 0.143   &   242        & {\bf +7.9} \\ \hline 
     \end{tabular}\\
     \begin{tabular}{c} \\
     \fbox{\bf (d) Fitted Spectrum: Pure Power-Law}\\[2mm]
     \end{tabular}
     \begin{tabular}{ccl}
      $dN/dE$ & = & $N\,\,(E/1\,{\rm TeV})^{-\gamma} \,\,\,\,{\rm ph\,\,cm^{-2}\,s^{-1}\,TeV^{-1}}$\\
      $N$     & = & $5.3\,\, (\pm2.2_{\rm stat} \pm1.3_{\rm sys})\, \times 10^{-13}$   \\
      $\gamma$& = & $1.9\,\, (\pm0.3_{\rm stat} \pm0.3_{\rm sys})$ \\
     \end{tabular}\\
    }
 }
\end{table}

\section{Modelling TeV J2032+4130}
Possible origins of TeV~J2032+4130 have been discussed in literature [1,4,12].
One interpretation involves association with the stellar
winds of member stars in Cygnus OB2, individually or collectively, which provide conditions conducive to
strong and stable shock formation for particle acceleration.
Certainly the existence of TeV emission suggests particles accelerated to multi-TeV energies. 
We have therefore matched the spectral energy distribution of TeV J2032+4130 with coarse leptonic and
hadronic models (Fig.~\ref{fig:skymap} lower panel). Another scenario involves particle acceleration at a
termination shock, which are expected at the boundary where a relativistic jet meets the interstellar
medium. TeV J2032+4130 actually aligns well within the northern error cone of the bi-lobal jet of the 
famous microquasar Cygnus X-3 [10,11].

For simplicity we assume the TeV emission arises from either
a pure sample of non-thermal hadronic or leptonic parent particles. Under the hadronic scenario 
the $\pi^\circ$-decay prediction explains well the TeV flux when using a parent proton power law 
spectrum of index $-$2.0 with a sharp limit up to energies 100 TeV.
The neighbouring EGRET source 3EG J2033+4118 (likely not related to TeV J2032+4130) provides no 
constraint on this
model. Associated synchrotron X-ray emission would also be expected from tertiary electrons 
($\pi^\pm \ldots \rightarrow \mu^\pm  \ldots \rightarrow e^\pm \ldots$). We have not yet modelled 
this component  which essentially represents an absolute lower limit on any synchrotron emission visible. 
Assuming a pure leptonic scenario, TeV data are matched well by an 
inverse-Compton spectrum (up-scattering the cosmic microwave background) arising from
an uncooled electron spectrum with power law index $\sim\, -2.0$ and hard cutoff at 40 TeV. This allows us  
to predict the synchrotron emission as a function of local magnetic field, constrained by the available upper
limits at radio and X-ray energies. The most conservative synchrotron
prediction arises from the $B_0=3.0 \mu$G choice, which is realistically the lowest such field expected
in the Galactic disk. But in fact, much higher fields ($B_0>10\mu$G) are generally expected in such regions 
containing young/massive stars with high mass losses and colliding winds (e.g. [5]). 
Deep observations by XMM and Chandra will provide strong contraints on the leptonic component.

\section*{References}

{\footnotesize

\noindent 1. Aharonian F.A, et~al. 2002 A\&A 393, ~L37\\ 
2. Aharonian F.A,  et~al. 2000 ApJ 539, 317\\ 
3. Butt Y. et al. 2003 ApJ {\em submitted} ({\tt astro-ph/0302342})\\
4. EGRET Catalogue (3rd) {\tt http://heasarc.gsfc.nasa.gov/}\\
5. Eichler D., Usov V. 1993 ApJ 402,~271\\
6. Hofmann W., et.al 1999 Astropart. Phys. 10, ~275\\
7. Hofmann W., et.al 2000 A\&A 361, 1073\\   
8. Hofmann W., et.al 2000 Astropart. Phys. 12, 207 \\  
9. Li T.P., Ma Y.Q. 1983 ApJ 272, ~317\\
10. Mart\'{\i} et al. 2000 ApJ 545, 939\\
11. Mart\'{\i} et al. 2001 A\&A 375, 476\\
12. Mukherjee R. et al. 2003 ApJ {\em in press}, ({\tt astro-ph/0302130})\\
13. P\"uhlhofer G. et. al 2003 Astropart. Phys. {\em submitted} \\
14. Rowell G.P. 2003 A\&A(Supp) {\em submitted}\\

\endofpaper

%% file: my_009463-2.tex
\chapter{
TeV Observations of Selected GeV Sources with the HEGRA IACT-System}

\author{%
%
%
Gavin Rowell$^1$, for the HEGRA Collaboration$^2$ \\
{\it (1) MPI f\"ur Kernphysik, Postfach 103980, D-69029 Heidelberg, Germany\\
(2) see {\tt http://www-hegra.desy.de/hegra/}}
}

\section*{Abstract}
Results from a search for TeV $\gamma$-ray emission from the vicinity of five X-ray-studied GeV
sources is presented. A number of possible X-ray counterparts
have been suggested based on ASCA and ROSAT observations (e.g. the SNR CTA 1, and X-ray 
binary LS~I+61$^\circ$303). 
Our search has yielded no convincing evidence 
for TeV emission at these GeV source positions and also a number possible counterparts. 
Preliminary upper limits in the range $\sim$1\% to 10\% of the Crab flux have been estimated above various
energy thresholds 0.7 to 1.3 TeV.

\section{Introduction \& Data Analysis}
A number of EGRET sources visible above 1 GeV [8,10] 
were earlier studied in X-rays with ASCA [12],
revealing possible counterparts. The large field of view (FoV) of the HEGRA IACT-System 
(Imaging Atmospheric \v{C}erenkov Telescope System, see [11]) allows good
coverage of many GeV sources even if they were not the original targets of observation.
We have selected five of these GeV sources for further analysis:\\
\noindent {\bf GeV J0008+7304}: This source could be associated with the nearby ($\sim0.2^\circ$ distant) 
supernova remnant CTA~1 (radio shell 100$^\prime$ diameter) and may contain a radio quiet 
pulsar [3]. This pulsar may power 
the extended, centre-filled non-thermal X-ray emission from CTA~1. Our observations were centred on the ROSAT point source 
RX~J0007.0+7302 [15]. Previous TeV observations by the CAT [7] and Whipple [5] collaborations
give upper limits for CTA~1.\\  
\noindent {\bf GeV J0241+6102}: The GeV contours are quite consistent with the unusual X-ray binary 
LS~I+61$^\circ$303 [4]. An association with the COS B
source 2CG135+01 in the past has led to various $\gamma$-ray production models for LS~I+61$^\circ$303. 
Our observations were centred on 2CG135+01.\\
\noindent {\bf GeV J1907+0557}: This GeV source is located $\sim2^\circ$ from the microquasar SS-433.
ASCA studies reveal a weak source possibly comprised of two pointlike components within the GeV 95\%
contour. Our observations were taken from the HEGRA SS-433 archive.\\
\noindent {\bf GeV J2026+4124}: ASCA studies reveal a single point source within the 95\% contour. This
 GeV source is $\sim1^\circ$ from Cyg X-3.\\
\noindent {\bf GeV J2035+4214}: Two pointlike (possibly non-thermal, labelled Src1 \& Src2) and one 
marginally extended 
(thermal, Src3) ASCA sources are seen within the GeV 68\% contour. HEGRA observations of these latter two 
GeV sources come from the extensive archive on Cyg X-3 and TeV J2032+4130 [14].
  
In these analyses we employ the {\em mean-scaled-width} $\bar{w}<1.1$ [2]
cut as a means to reject the cosmic-ray background, and also a cut in the angular separation $\theta$ 
between reconstructed (employing the so-called `algorithm 3' [6]) and assumed 
arrival directions for each event. A point-source search utilises a $\theta_{\rm cut}<0.12^\circ$ cut. Extended 
sources with radius $\sigma_{\rm src}$ utilise a cut 
$\theta_{\rm cut} < \sqrt{\sigma^2_{\rm src} + 0.12^2}^\circ$. 
A minimum of three telescope images per event $n_{\rm tel}\geq3$ are demanded in
calculating $\bar{w}$ and $\theta$. The cosmic-ray background is estimated using the so-called {\em template} 
model [1,13]. Quite consistent results are also obtained using
independent (e.g. so-called {\em displaced}) background models. 

\begin{table}[h]
 \begin{center}
 \footnotesize 
 \caption{Numerical results for various GeV sources and other positions of interest. The values
          $s$ and $b$ are respectively the source counts and background estimate from the template
          model.}
 \begin{tabular}{lcccccccc} \hline 
  Source & Obs. time & $^a E_{\rm th}$ & $\theta_{\rm cut}$ & $s$ & $b$ & $\alpha^b$ & $S^c$ & $^d \phi^{99\%}$\\
         &   [h]     &  [TeV]       &    [deg]           &     &     &             & [$\sigma$]   \\ \hline
 \multicolumn{8}{c}{----- GeV J0008+7304 -----}\\ 
  GeVJ0008+7304        & 26.0 & 1.3 & 0.120 & 72   & 615   & 0.114 & \bf +0.2   & \bf 0.74 \\ 
  RXJ0007.0+7302       & 26.0 & 1.3 & 0.120 & 70   & 598   & 0.114 & \bf +0.2   & \bf 0.73 \\ 
  CTA 1$^e$            & 26.0 & 1.3 & 0.195 & 181  & 1618  & 0.114 & \bf $-$0.2 & \bf 1.09 \\ 
 \multicolumn{8}{c}{----- GeV J0241+6102 -----}\\ 
  GeVJ0241+6102        & 28.8 & 0.8 & 0.120 & 55   & 338   & 0.130 & \bf +1.5   & \bf 1.74 \\ 
  LS~I+61$^\circ$303   & 28.8 & 0.8 & 0.120 & 59   & 380   & 0.130 & \bf +1.3   & \bf 1.50 \\ 
 \multicolumn{8}{c}{----- GeV J1907+0557 -----}\\ 
  GeVJ1907+0557        & 114.1 & 0.7 & 0.120 & 183  & 1019 & 0.164 & \bf +1.1   & \bf 0.66 \\ 
  AXJ1907.4+0549       & 114.1 & 0.7 & 0.120 & 187  & 1101 & 0.164 & \bf +0.4   & \bf 0.62 \\ 
 \multicolumn{8}{c}{----- GeV J2026+4124  -----}\\ 
  GeVJ2026+4124        & 275.0 & 0.7 & 0.120 & 541  & 3429 & 0.168 & \bf $-$1.3 & \bf 0.29 \\ 
  AXJ2027.6+4116       & 275.0 & 0.7 & 0.120 & 771  & 4497 & 0.168 & \bf +0.6   & \bf 0.42 \\ 
 \multicolumn{8}{c}{----- GeV J2035+4214  -----}\\ 
  GeVJ2035+4214        & 275.0 & 0.7 & 0.120 & 784  & 4457 & 0.168 & \bf +1.2   & \bf 0.45 \\ 
  AXJ2036.0+4218(Src1) & 275.0 & 0.7 & 0.120 & 620  & 4005 & 0.168 & \bf $-$1.8 & \bf 0.22 \\ 
  AXJ2035.4+4222(Src2) & 275.0 & 0.7 & 0.120 & 663  & 4083 & 0.168 & \bf $-$0.7 & \bf 0.27 \\ 
  AXJ2035.9+4229(Src3) & 275.0 & 0.7 & 0.120 & 584  & 3626 & 0.168 & \bf $-$0.9 & \bf 0.28 \\ \hline 
  \multicolumn{8}{l}{\scriptsize a. Estimated $\gamma$-ray threshold energy for mean zenith $\bar{z}$: 
             $E_{\rm}=0.5\cos(\bar{z})^{-2.5}$.}\\
  \multicolumn{8}{l}{\scriptsize b. Normalisation factor for the background $b$.}\\
  \multicolumn{8}{l}{\scriptsize c. Statistical significance using Eq. 17 of [9].}\\
  \multicolumn{8}{l}{\scriptsize d. $\phi^{99\%}_{ph} = $99\% upper limit $E>E_{\rm th}$ [$\times 10^{-12}$ ph 
      cm$^{-2}$s$^{-1}$]}\\
  \multicolumn{8}{l}{\scriptsize e. For CTA~1, a radius $\sigma_{\rm src}=0.153^\circ$ is used (the inner region for ROSAT spectral analysis [15]).}\\
 \end{tabular}
 \end{center}
 \label{tab:numbers}
\end{table}
\begin{figure}
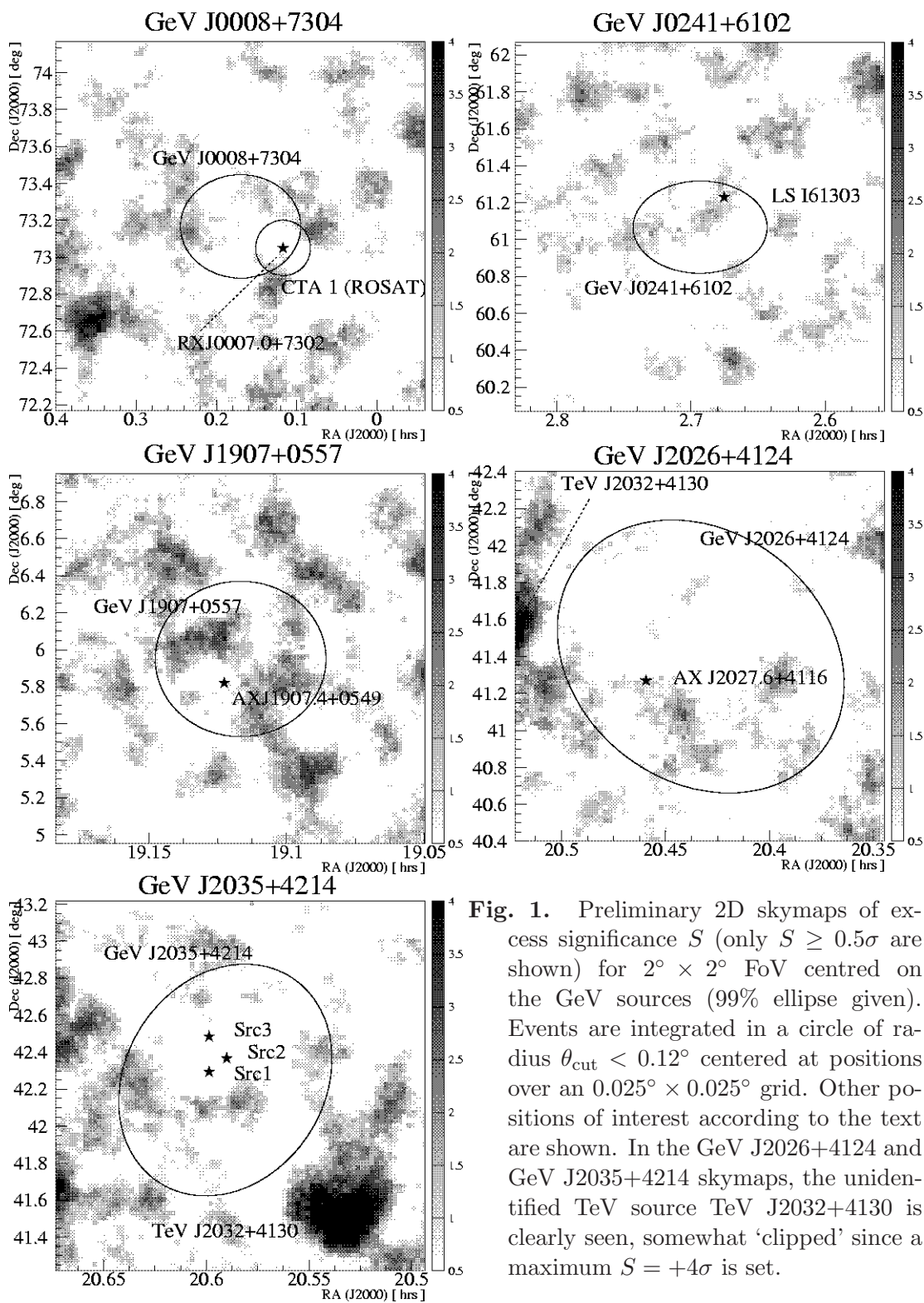

 \parbox{0.5\linewidth}{\includegraphics[width=\linewidth]{skymap_gevj0007.epsi}}
 \parbox{0.5\linewidth}{\includegraphics[width=\linewidth]{skymap_gevj0241.epsi}}
 \parbox{0.5\linewidth}{\includegraphics[width=\linewidth]{skymap_gevj1907.epsi}}
 \parbox{0.5\linewidth}{\includegraphics[width=\linewidth]{skymap_gevj2026.epsi}}
 \parbox{0.5\linewidth}{\includegraphics[width=\linewidth]{skymap_gevj2035.epsi}}
\parbox{0.5\linewidth}{
  \caption{Preliminary 2D skymaps of excess significance $S$ (only $S\geq0.5\sigma$ are shown) for
      2$^\circ \times 2^\circ$ FoV centred on the GeV sources (99\% ellipse given). 
      Events are integrated in a circle of radius $\theta_{\rm cut}<0.12^\circ$ centered at positions 
      over an $0.025^\circ\times 0.025^\circ$ grid.
      Other positions of interest according to the text are shown. 
      In the GeV~J2026+4124 and GeV~J2035+4214 skymaps, the unidentified
      TeV source TeV~J2032+4130 is clearly seen, somewhat `clipped' since a 
      maximum $S=+4\sigma$ is set.}}
 \label{fig:skymaps}
\end{figure}

\section{Results \& Discussion}
As {\em a-priori} targets, we have chosen the ASCA sources identified by [12], 
some pre-identified objects (e.g. CTA~1, LS~I+61$^\circ$303 etc.), and also a point source centred
on each GeV source (SIMBAD coordinates). Numerical results are presented in Table~\ref{tab:numbers} and
2D skymaps of excess significances are given in Fig~\ref{fig:skymaps}
For all of our {\em a-priori}-chosen positions, no convincing evidence for TeV emission was
found. Upper limits at the 99\% confidence level were estimated for each position
using the expected Crab event rate from Monte Carlo simulations scaled 
according to an estimate of the $\gamma$-ray acceptance in the FoV. 
These preliminary upper limits are in the range 1\% to 10\% of the Crab flux at various energy thresholds 
($E_{\rm th} \sim$ 0.7 to 1.3 TeV, based on the mean zenith angle of events). 
Further discussion and comparison with models will be presented at the conference.

\section*{References}

{\normalsize
 \noindent1. Aharonian F.A, et~al. 2002 A\&A 393, ~L37 \\
 2. Aharonian F.A,  et~al. 2000 ApJ 539, 317\\ 
 3. Brazier K.T.S. et~al. 1998, MNRAS 295, ~819\\
 4. Frail D.A., Hjellming R.M. 1991 AJ 101, 2126\\
 5. Hall T.A. et al. 2001 Proc 27th ICRC (Hamburg) OG167\\ 
 6. Hofmann W., et.al 1999 Astropart. Phys. 10, ~275\\
 7. Khelifi B. et al. 2001 Proc 27th ICRC (Hamburg) \\
 8. Lamb R.C., Macomb D.J. 1997, ApJ 488, ~872\\
 9. Li T.P., Ma Y.Q. 1983 ApJ 272, ~317\\
 10. Macomb D.J., Lamb R.C. 1999 Proc. 25th ICRC (Salt Lake City) OG 2.4.21\\ 
 11. P\"uhlhofer G. et. al 2003 Astropart. Phys. {\em submitted} \\
 12. Roberts S.E., Romani R.W., Kawai N. 2001, ApJS 133, ~451\\
 13. Rowell G.P. 2003 A\&A(Supp) {\em submitted}\\
 14. Rowell G.P., Horns D. et al. 2003 {\em these proceedings}\\
 15. Slane P. et~al. 1997, ApJ 485, ~221\\

}

\endofpaper

%% file: tluczykont_agn.tex
\chapter{Observations of 54 Active Galactic Nuclei with the HEGRA Cherenkov Telescopes}

\author{M.~Tluczykont$^1$, N.~G\"otting$^1$ and G.~Heinzelmann$^1$ for the HEGRA Collaboration$^2$
{\it (1) Institut f\"ur Experimentalphysik, Luruper Chaussee 149, 
22761 Universit\"at Hamburg}\\
{\it (2)} {\tt http://www-hegra.desy.de/hegra}
}
\normalsize
\section*{Abstract}
In total 54 Active Galactic Nuclei (AGN)
apart from Mrk-421 and Mrk-501
have been observed between 1996 and 2002 with the HEGRA Cherenkov Telescopes 
in the TeV energy regime.  Among the observed 54 AGN 
are the meanwhile well established BL Lac type objects 
\mbox{H\,1426+428} and \mbox{1ES\,1959+650}.
The BL Lac object \mbox{1ES\,2344+514} and the radio galaxy M\,87 
show evidence for a signal on a 4\,$\sigma$ level. 
Fluxes resp. upper limits are given for each of the 54 AGN.
\section{Introduction}
In the commonly adopted view AGN are powered by 
a central super massive black hole with
$\sim 10^9$\,${\mathrm{M_\odot}}$ surrounded 
by an accretion disk. 
Two relativistic plasma outflows (jets) perpendicular to
the accretion disk are pointing 
in opposite directions away from the center black hole $[1]$. 
Most detections of extragalactic 
TeV ${\mathrm \gamma}$--rays
refer to objects of the BL Lac type, 
i.e. AGN having their jet
pointing close to the observers line of sight. 
In this work the results of an analysis of the observations of 
\mbox{54 AGN} with the HEGRA telescope system 
are presented.
The HEGRA collaboration was operating a stereoscopic system of 
imaging air Cherenkov telescopes (IACT-System)
$[2]$ and one stand alone telescope $[3]$ (not used for 
the present analysis) on the Canary island of 
La Palma (28.75$^\circ$\,N, 17.90$^\circ$\,W) at an altitude of 2200 m a.s.l. 
The stereoscopic reconstruction technique first introduced by
HEGRA allows the complete reconstruction of the shower geometry and
results in improved angular and energy resolutions as well as
a very good $\gamma$-hadron separation (hadron rejection up to a factor of 100) 
using the so called mean scaled width (mscw) parameter.
The data set used for this work amounts to a total exposure time 
before rejection of runs with poor quality
of $\approx$~1150~hours (without Mkn--421 and Mkn-501) 
corresponding to more than one year of continuous 
observations in moonless nights with the HEGRA IACT-System. 
After quality cuts, 1017~hours remain in the analysis.
The applied event cuts are optimized on data of the 
Crab Nebula which is used as a calibration source in TeV astronomy.
The significance of an excess is calculated following
$[4]$.
Observed Crab-$\gamma$--rates are used to compute integral flux 
upper limits $[5]$.
\section{Results}
A distribution of the significances for 
steady state emission (DC) of all 
analyzed objects is shown in Fig.\,1. 
The objects 
\mbox{1ES\,1959+650}, \mbox{H\,1426+428}, \mbox{1ES\,2344+514} and M\,87
show significances deviating from the expectation of
a Gaussian distribution in case of pure background
fluctuations. 
\begin{figure}[t]
  \begin{center}
  \includegraphics[width=\textwidth]{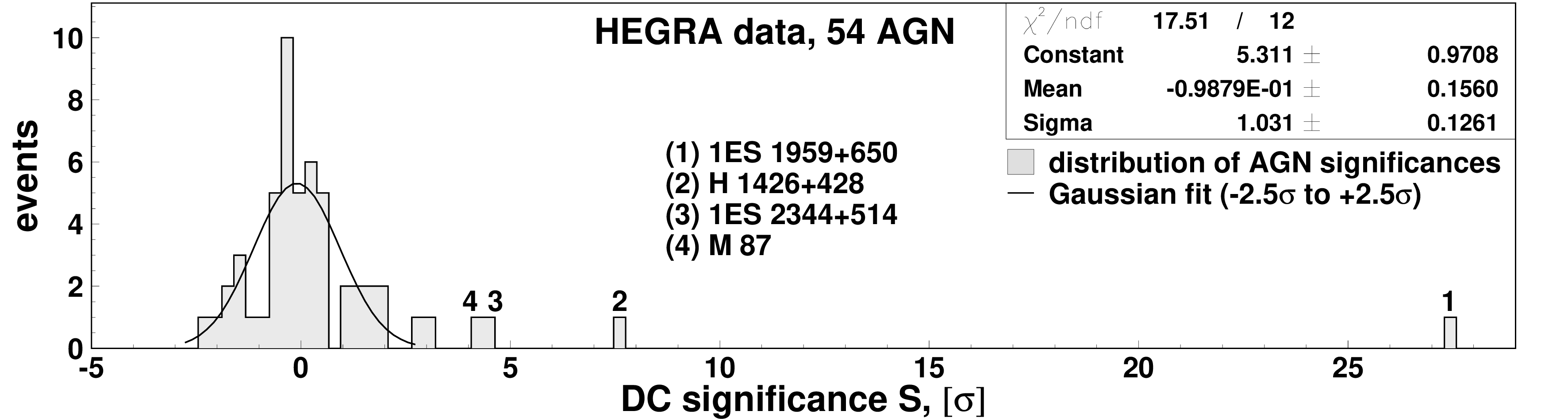}
  \end{center}
  \small
  \caption{Distribution of significances for all AGN analyzed in this work. 
           The objects \mbox{1ES\,1959+650} (1) and \mbox{H\,1426+428} (2)
	   show a clear deviation from the background expectation,
	   represented by a Gaussian fit from -2.5\,$\sigma$ to 2.5\,$\sigma$. 
           Further two objects, \mbox{1ES\,2344+514} (4.4\,$\sigma$, (3))
           and M\,87 (4.1\,$\sigma$, (4)) also show 
           a deviation from the background 
           expectation and thus for emission of TeV $\gamma$-radiation.} 
  \label{distribution}
\end{figure}
In Tab.\,1
a list of all objects analyzed in this work ordered by ascending 
redshift is shown along with their observation time and 
the results of this analysis.
\begin{table}[Ht]
 \begin{center}
 \scriptsize
 \caption{Results of all objects of the HEGRA AGN data sample. 
          The object names are given as well as 
          observation time, number of ON- and OFF-events,
          energy threshold, upper limits in Crab units and in
	  flux units and fluxes in Crab units.}
\begin{tabular}{|lcrrrcccc|}\hline
Object			
& $z$
& time	
& $\mathrm{N_{on}}$ 
& $\mathrm{N_{off}}$  
& E$_\mathrm{thr}$   	
& $\mathrm{F_{\tiny UL}^{99\%}(>E_{thr})}$ 		
& $\mathrm{\Phi_{\tiny UL}^{99\%}(>E_{thr})}$    
& $\mathrm{F(>E_{thr})}$\\
                        
&
& $[$h$]$
& \#  
&  \#   		
& $[$TeV$]$	
& $[Crab]$     	
& $[\mathrm{10^{-12} cm^{-2} s^{-1}}]$ 
& $[Crab]$    \\\hline

1ES 0647+250	&---- 	&   4.1  & 12  & 10   & 0.74	&  0.13 	&  3.65  &   \\
MG 0509+0541 	&---- 	&  15.8  & 60  & 50   & 0.79	&  0.11 	&  2.61  &   \\
M87		&.004 	&  70.0  & 241 & 184  & 0.76	&     		&        & 0.033  \\    
NGC 315		&.016 	&  14.6  & 31  & 33   & 0.75	&  0.05 	&  1.29  &   \\
NGC 1275	&.018 	&  87.6  & 231 & 236  & 0.75	&  0.03 	&  0.84  &   \\
H 1722+119	&.018 	&   5.1  & 21  & 14   & 0.76	&  0.21 	&  5.51  &   \\
PKS 2201+04	&.028 	&  17.8  & 59  & 46   & 0.79	&  0.08 	&  1.89  &   \\
V Zw 331	&.029 	&   4.1  & 9   & 9    & 0.75	&  0.09 	&  2.44  &   \\
NGC1054		&.032 	&  57.9  & 134 & 155  & 0.75	&  0.02 	&  0.46  &   \\
3C 120		&.033 	&  25.4  & 64  & 70   & 0.78	&  0.05 	&  1.14  &   \\
NGC 4151	&.033 	&   7.0  & 16  & 18   & 0.74	&  0.07 	&  1.99  &   \\
UGC01651	&.037 	&  14.3  & 44  & 36   & 0.74	&  0.07 	&  1.79  &   \\
UGC03927	&.041 	&   6.3  & 7   & 16   & 0.86	&  0.09 	&  1.93  &   \\
1ES 2344+514	&.044 	&  72.5  & 235 & 171  & 0.80	&   		&        & 0.033  \\
Mkn0180		&.046 	&   9.8  & 29  & 32   & 1.38	&  0.12 	&  1.24  &   \\
1ES 1959+650	&.047 	& 163.7  & 1202& 454  & 1.17	&   		&        & 0.053-2.200  \\
3C 371.0	&.050 	&   5.4  & 16  & 18   & 1.40	&  0.19 	&  1.88  &   \\
B2 0402+37	&.054 	&   6.7  & 9   & 17   & 0.74	&  0.05 	&  1.32  &   \\
I Zw 187	&.055 	&  16.0  & 44  & 32   & 0.78	&  0.09 	&  2.21  &   \\
Cyg-A (3C 405.0)&.057 	&  59.0  & 159 & 161  & 0.77	&  0.03 	&  0.83  &   \\
1ES 2321+419	&.059 	&  22.3  & 53  & 66   & 0.76	&  0.03 	&  0.86  &   \\
3C 192.0	&.060 	&   2.9  & 8   & 7    & 0.78	&  0.20 	&  5.01  &   \\
4C+31.04	&.060 	&   3.0  & 8   & 9    & 0.74	&  0.14 	&  4.01  &   \\
BL Lacertae	&.069 	&  26.7  & 94  & 67   & 0.87	&  0.28 	&  5.97  &   \\
1ES 1741+196	&.083 	&  10.2  & 28  & 26   & 0.78	&  0.07 	&  1.88  &   \\
4C+01.13	&.084 	&   7.7  & 30  & 31   & 0.81	&  0.10 	&  2.45  &   \\
PKS 2155-304 	&.116 	&   1.8  & 4   & 4    & 4.62	&  0.27 	&  0.39  &   \\
1ES 1118+424	&.124 	&   2.0  & 5   & 4    & 0.79	&  0.24 	&  5.88  &   \\
1ES 0145+13.8	&.125 	&   3.2  & 2   & 1    & 0.75	&  0.06 	&  1.73  &   \\
EXO0706.1+5913	&.125 	&  33.7  & 81  & 85   & 0.86	&  0.06 	&  1.32  &   \\
H 1426+428	&.129 	& 258.5  & 796 & 624  & 0.77	&   		&        & 0.024-0.060  \\  
3C197.1		&.130 	&  15.0  & 22  & 24   & 0.79	&  0.05 	&  1.17  &   \\
1ES 1212+078	&.130 	&   2.4  & 6   & 8    & 0.78	&  0.17 	&  4.26  &   \\
1ES 0806+524 	&.138 	&   1.0  & 2   & 2    & 0.86	&  0.29 	&  6.19  &   \\
1ES 0229+200	&.139 	&   3.0  & 11  & 8    & 0.78	&  0.17 	&  4.25  &   \\
RBS 0958	&.139 	&   3.8  & 18  & 9    & 0.75	&  0.28 	&  7.57  &   \\
1ES 1255+244	&.140 	&   5.9  & 14  & 14   & 0.79	&  0.12 	&  2.88  &   \\
MS1019.0+5139	&.141 	&  17.5  & 44  & 43   & 0.78	&  0.07 	&  1.78  &   \\
1ES 0323+022	&.147 	&  14.3  & 24  & 32   & 0.81	&  0.04 	&  1.00  &   \\
OQ 530		&.152 	&   9.4  & 32  & 30   & 0.89	&  0.10 	&  2.05  &   \\
3C 273.0	&.158 	&  12.2  & 44  & 46   & 0.91	&  0.09 	&  1.82  &   \\
1ES 1440+122	&.162 	&  13.1  & 35  & 41   & 0.77	&  0.08 	&  1.95  &   \\
PKS 0829+046	&.180 	&  18.0  & 59  & 55   & 0.81	&  0.06 	&  1.35  &   \\
PG 1218+304	&.182 	&   3.9  & 7   & 8    & 0.75	&  0.12 	&  3.22  &   \\
1ES 0347-121	&.185 	&   1.9  & 13  & 7    & 1.34	&  0.56 	&  5.90  &   \\
1ES 0927+500	&.186 	&  13.3  & 30  & 29   & 0.79	&  0.06 	&  1.44  &   \\
PKS 2254+074	&.190 	&  16.3  & 44  & 48   & 0.77	&  0.05 	&  1.27  &   \\
MS0317.0+1834	&.190 	&   2.7  & 5   & 6    & 0.74	&  0.12 	&  3.30  &   \\
1ES 1011+496	&.200 	&   2.0  & 1   & 3    & 0.81	&  0.11 	&  2.58  &   \\
1ES 0120+340	&.272 	&  18.9  & 36  & 44   & 0.75	&  0.04 	&  1.03  &   \\
2E 0414+0057	&.287 	&   4.5  & 18  & 15   & 0.81	&  0.13 	&  3.07  &   \\
S5 0716+714 	&.300 	&   1.7  & 6   & 4    & 1.46	&  0.38 	&  3.54  &   \\
3C 066A		&.444 	&   1.3  & 3   & 3    & 0.75	&  0.17 	&  4.70  &   \\
PKS 0219-164    &.698         &   1.7  & 5   & 10   & 1.67	&  0.27 	&  2.04  &   \\\hline
\end{tabular}
\normalsize
 \label{table1}
 \end{center}
\end{table}
The results of the observations of the BL Lac objects
\mbox{1ES\,1959+650} and \mbox{H\,1426+428} as well as
a tentative detection of a signature of absorption by pair production
on the diffuse extragalactic background radiation in the TeV
spectrum of \mbox{H\,1426+428} were reported elsewhere $[6,7]$.
The results of a dedicated analysis of the M87 data 
are presented separately at this conference
and have been published recently $[8,9]$.
The first TeV detection of \mbox{1ES\,2344+514}
was reported by the Whipple group in 1998 with 6\,$\sigma$
in one night only $[10]$.
The results of the HEGRA observations of \mbox{1ES\,2344+514} 
of the year 1997 and 1998 were first reported by 
$[11]$ with a DC significance of 3.3\,$\sigma$. 
The analysis presented here includes the complete dataset (1997 -- 2002) and
results in an excess of $\mathrm{63\pm 14}$ $\gamma$--like excess events
(N$_{\mathrm{on}}$ = 235, $\big<$N$_{\mathrm{off}}$$\big>$ = 171) 
on a significance 
level of \mbox{4.4\,$\sigma$} and no evidence for burst-like behaviour. 
However, the excess is accumulated almost exclusively
in the observation periods of \mbox{August to November 1998}
and September 2002 ($61\pm12$ excess events). 
In Fig.\,2. the distribution of the squared angular distance of
the reconstructed shower events to the object position as well as
a plot of the excess rate vs. observation periods are shown.
\begin{figure}[t]
  \includegraphics[width=7.3cm]{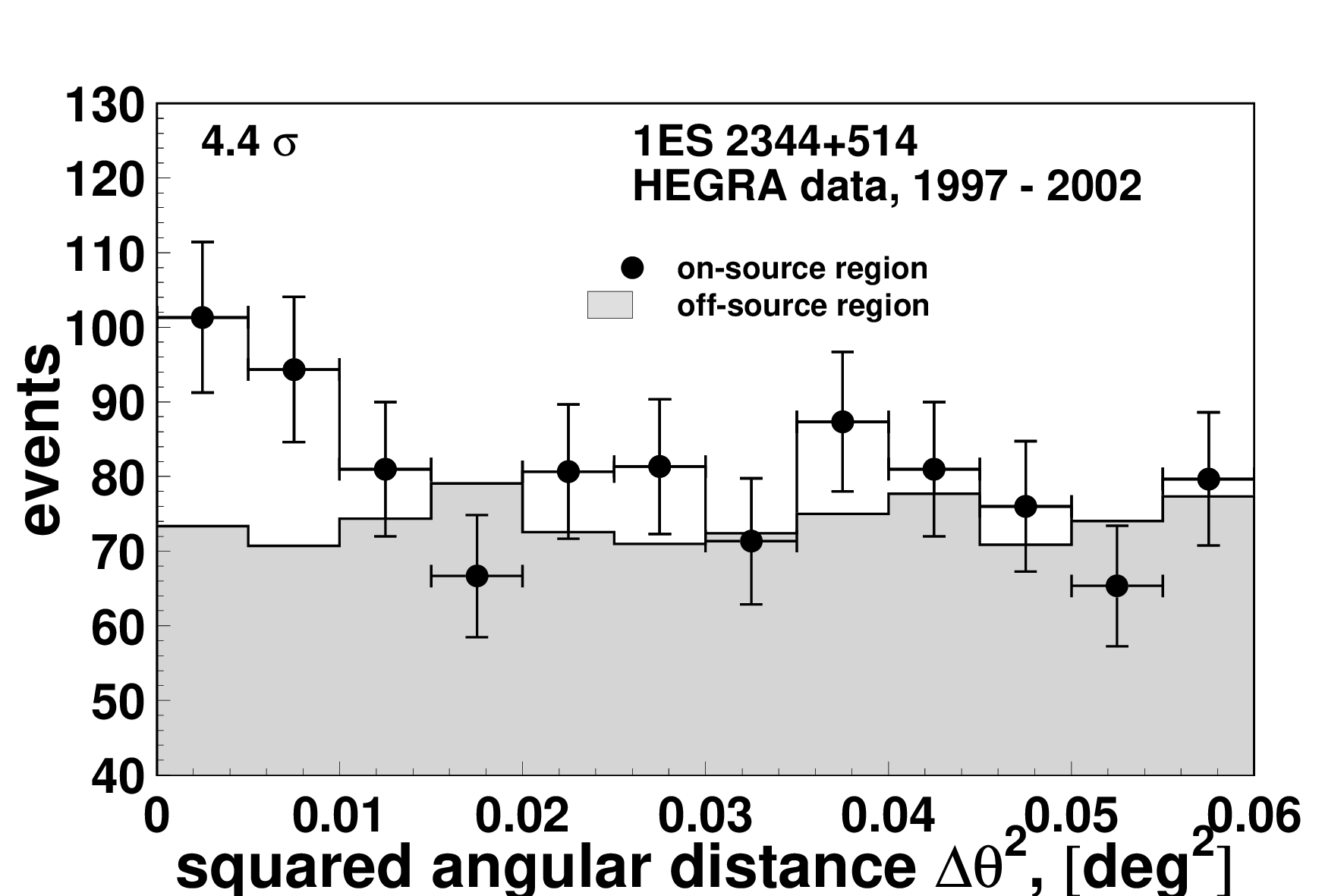}
  \includegraphics[width=7.3cm]{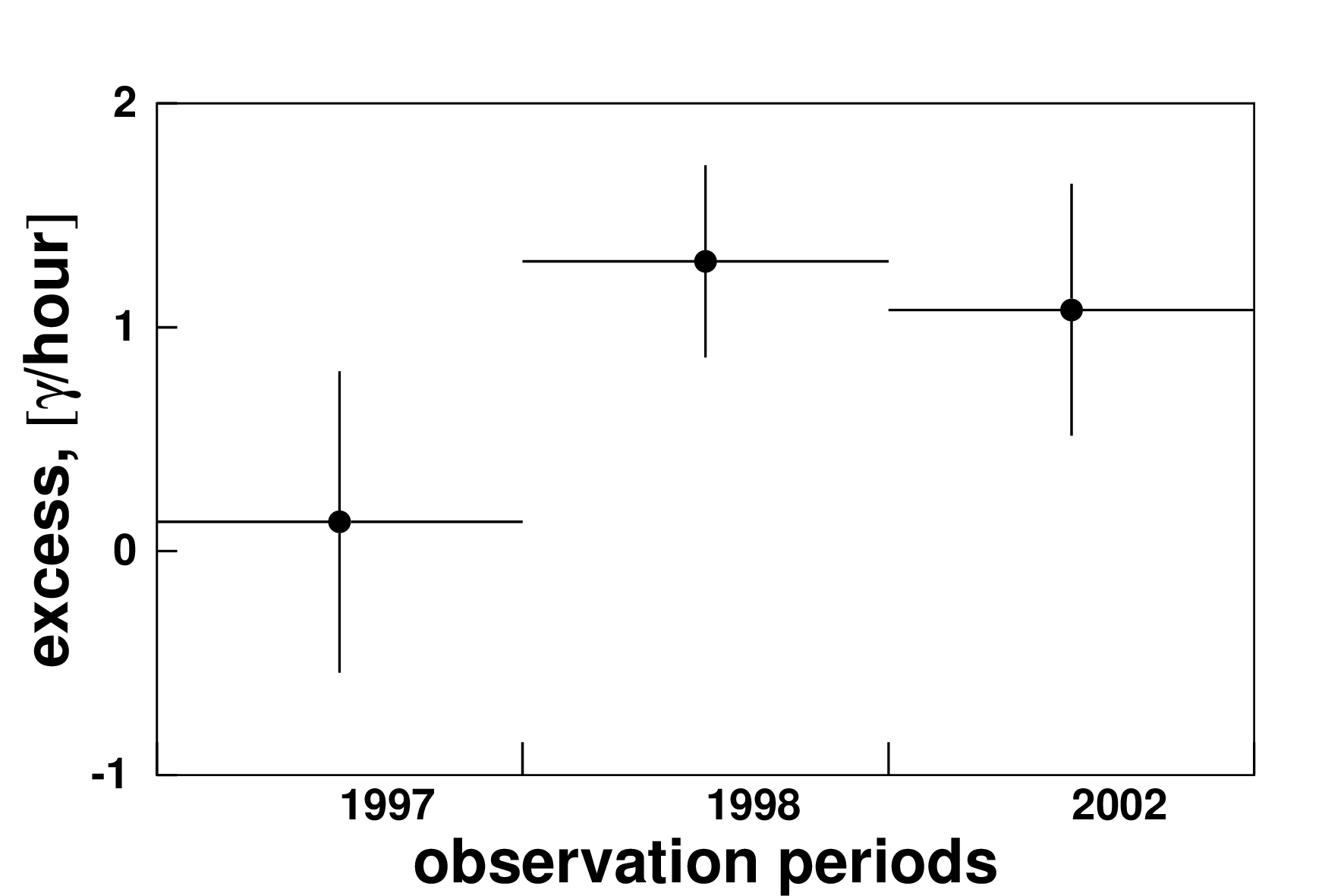}
  \caption{Left: Distribution of reconstructed squared angular distances of the
           \mbox{1ES\,2344+514} data. The distribution of the on--source 
           events is represented by the data points. The background is shown 
           as a shaded histogram. Right: The excess rate in $\gamma$/hour for
           the three observation campaigns on 1ES\,2344+514. The excess is
           accumulated almost exclusively in 1998 and 2002.}
\end{figure}
The observed excess results in a flux of
$\Phi$(E$>$0.80 TeV) = $\mathrm{(0.8 \pm 0.3) \cdot
            10^{-12}\,cm^{-2}\,s^{-1}}$
corresponding to 
3.3\% of the Crab-Nebula flux.
\section{Summary}
The two objects \mbox{1ES\,1959+650} and \mbox{H\,1426+428} 
can be considered as well established
sources of TeV ${\mathrm \gamma}$-radiation. 
The evidence for \mbox{1ES\,2344+514} being a TeV source 
is strong taking the detection by the Whipple collaboration into account. 
The detection
of TeV~${\mathrm \gamma}$--rays from M87, if confirmed, 
would be the first detection of photons
in the TeV energy regime of an AGN not 
classified as a BL Lac object.

  \paragraph{Acknowledgements}
  \small
  The support of the German Federal Ministry for Education an Research 
  BMBF and
  of the Spanish Research Council CICYT is gratefully acknowledged. 
  We thank the Instituto de Astrof\'{\i}sica de Canarias (IAC)
  for the use of the HEGRA site at the Observatorio del Roque de los 
  Muchachos (ORM) and for supplying excellent working conditions on La
  Palma.

\vspace{\baselineskip}

\re
1.\ Rees, M.\,J.\ 1984, Ann. Rev. Astron. \& Astroph. 22, 471
\re
2.\ Daum, A. et al.\ 1997, Astropart. Phys. 8, 1
\re
3.\ Mirzoyan, R. et al.\ 1994, Nucl. Inst. Meth. A 351, 513
\re
4.\ Li, T. P. and Ma, Y. Q.\ 1983, ApJ 272, 317
\re
5.\ Helene, O.\ 1983, Nucl. Inst. Meth. 212, 319
\re
6.\ Aharonian, F.\,A. et al.\ 2003 submitted to A\&Ap, astro-ph/0305275
\re
7.\ Aharonian, F.\,A. et al.\ 2003, A\&Ap 403, 523
\re
8.\ G\"otting, N.\ 2003, These Proceedings
\re
9.\ Aharonian, F.\,A. et al.\ 2003, A\&Ap 403, L1
\re
10.\ Catanese, M. et al.\ 1998, ApJ 501, 616
\re
11.\ Konopelko, A.\ 1999, Proc. 26th ICRC 3, 426

\endofpaper

%% file: my_009271-1.tex
\chapter{
Study of the VHE Gamma Ray Emission from the AGN 1ES1959+650 with the HEGRA Cherenkov Telescope CT1}

\author{%
%
%
Nadia Tonello$^1$ and Daniel Kranich$^1$, for the HEGRA Collaboration\\
{\it (1) Max-Planck-Institute for Physics (Werner-Heisenberg-Institute), Foehringer Ring 6, Munich 80805, Germany} \\
}
\vspace{-0.5pc}

\section*{Abstract}

The BL Lac object 1ES1959+650 has been observed by the HEGRA Collaboration in the years 2000, 2001 and 2002. Here we report on the results obtained in 2002 with the standalone CT1 telescope.
While the source has only shown weak TeV activity until mid-May 2002, significant TeV emission has been observed later in 2002. During 2002 the source has been monitored for about 200 hours, of which 30 hours data were taken under moonlight conditions. The signal from the 'dark' time  2002 data set showed a significance of 11$\sigma$. Preliminary results on the light curve and the spectrum of the 2002 'dark' time data up to 45$^{\circ}$ zenith angle are presented.

\section{Introduction}

The galaxy 1ES1959+650 has a strong emitting jet from the nucleus, oriented along the line of sight. 
It was predicted by [8] to be a TeV $\gamma$-ray emitter candidate  from scaling its X-ray observed spectrum, and from studies based on the synchrotron-self Compton model [2]. 
It was detected for the first time as a TeV $\gamma$-ray emitter by the Telescope Array experiment [7], that reported a signal with a significance of 3.9$\sigma$ after 56.7 hours of observation in 1998. 

In 2000 and 2001 the HEGRA collaboration observed 1ES1959+650 with the CT System for 94 hours. The source was at low state and an averaged signal of 5.4$\sigma$ of significance level was recorded [3]. 
In May 2002 the VERITAS Collaboration reported a strong TeV $\gamma$-ray signal from 1ES1959+650 [11]. The detection had a significance of 13$\sigma$ from observation in 2 consecutive nights. 
The followed observations by the HEGRA Collaboration confirmed 1ES1959+650 as a highly variable VHE $\gamma$-ray source [1,4]. The results of observations with the HEGRA CT System are reported elsewhere on this conference.

\section{Observations with the HEGRA CT1 Telescope and Data Analysis}

The HEGRA Cherenkov Telescopes  are installed at the Roque de los Muchachos Observatory on the Canary Island of La Palma (28.75$^{\circ}$ N, 17.89$^{\circ}$ W, 2200 a.s.l.). The telescope CT1 [6] has a mirror area of 10.3 m$^2$ and a camera of 127 pixels of 0.25$^{\circ}$  diameter each. The camera has a FOV of 3$^{\circ}$ diameter. In zenith position, the telescope has an energy threshold  of 750 GeV and a typical source discovery sensitivity of 3.5 $\sigma$ $\times\sqrt{t}$, with $t$ in hours, for a Crab-like source. 
Table 1 summarizes the CT1 observation times of different background light conditions of 1ES1959+650. The observation angle ranged between 36.4$^{\circ}$ and 54.2$^{\circ}$ (with rather few data above 45$^{\circ}$). A total of 1.6 million triggers have been recorded in 2002. 

\begin{table}[t]
 \caption{Observation times of 1ES1959+650 with HEGRA CT1}
\begin{center}
\begin{tabular}{ll|rrr}
\hline
     & {\small \bf Observation conditions}   &{\small \bf 2000} &{\small \bf 2001} &{\small \bf 2002}\\
\hline
{\small nom }  & {\small good 'dark' night conditions, nominal HV } & {\small 58 h} & {\small 86 h }&  {\small156 h}\\
{\small hv00 } & {\small weak moonlight, no HV reduction   } & {\small   - } & {\small 24 h }& {\small 8.1 h}\\
{\small hv04 } & {\small moonlight, 4\% HV reduction       } & {\small   - } & {\small 12 h }& {\small 16.1 h}\\
{\small hv08}  & {\small moonlight, 8\% HV reduction      }  & {\small   - } & {\small  8 h }& {\small 6.1 h}\\ 
{\small hv12}  & {\small moonlight, 12\% HV reduction     }  & {\small   - } & {\small  7 h} & {\small 2.9 h}\\
\hline
\end{tabular}
\end{center}
\end{table}

Quality cuts are applied to the raw data in order to reject accidental noise triggers and nights with bad atmospheric conditions. The image parameters of the recorded events were calculated. The adopted $\gamma$/hadron separation method is described in [5]: the so-called dynamical super-cuts in the parameters DIST, WIDTH and LENGTH are applied, in order to take into account variations of the image parameters with the zenith angle, the energy of the primary particle and the impact parameter of the shower (estimated mainly from the image parameter DIST). 
The ALPHA parameter distribution after the dynamical super-cuts for 'dark' nights data, taken up to a zenith angle of 45$^{\circ}$, is shown in Figure 1. The number of recorded $\gamma$ events has been estimated by fitting this distribution with a Gaussian function in the signal region 0$^{\circ}<\mathrm{ALPHA}<13^{\circ}$ and a smooth background (BG) has been estimated by fitting the same data sample with a polynomial of second order (without linear term) in the range   $20^{\circ}<\mathrm{ALPHA}<80^{\circ}$, where no signal is expected. The BG was then extrapolated down to ALPHA = 0$^\circ$. In Figure 1b the BG parametrization is verified for a pure BG sample.  

\begin{figure}[t]
\begin{minipage}[l]{0.53\linewidth}
\centering
\includegraphics[,height=14pc, width=\linewidth ]{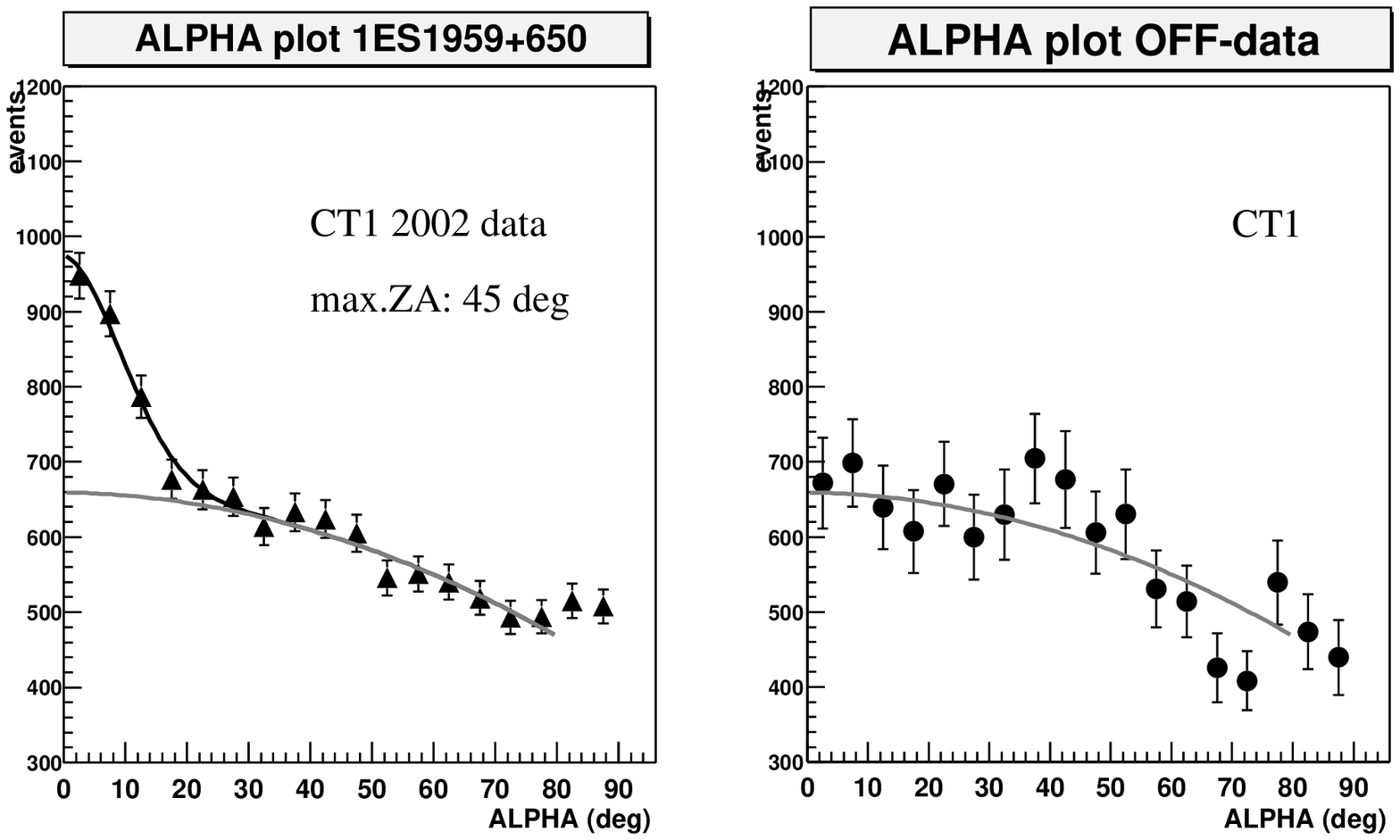}
\caption{Left: Distribution of the image parameter ALPHA for the 2002 data; maximum zenith angle: 45$^\circ$. 
Right: ALPHA distribution of a normalized sample of off-data, plotted here for comparison.}
\end{minipage}\hfill
\begin{minipage}[r]{0.45\linewidth}
 \begin{center}
    \includegraphics[height=12pc, width=\linewidth]{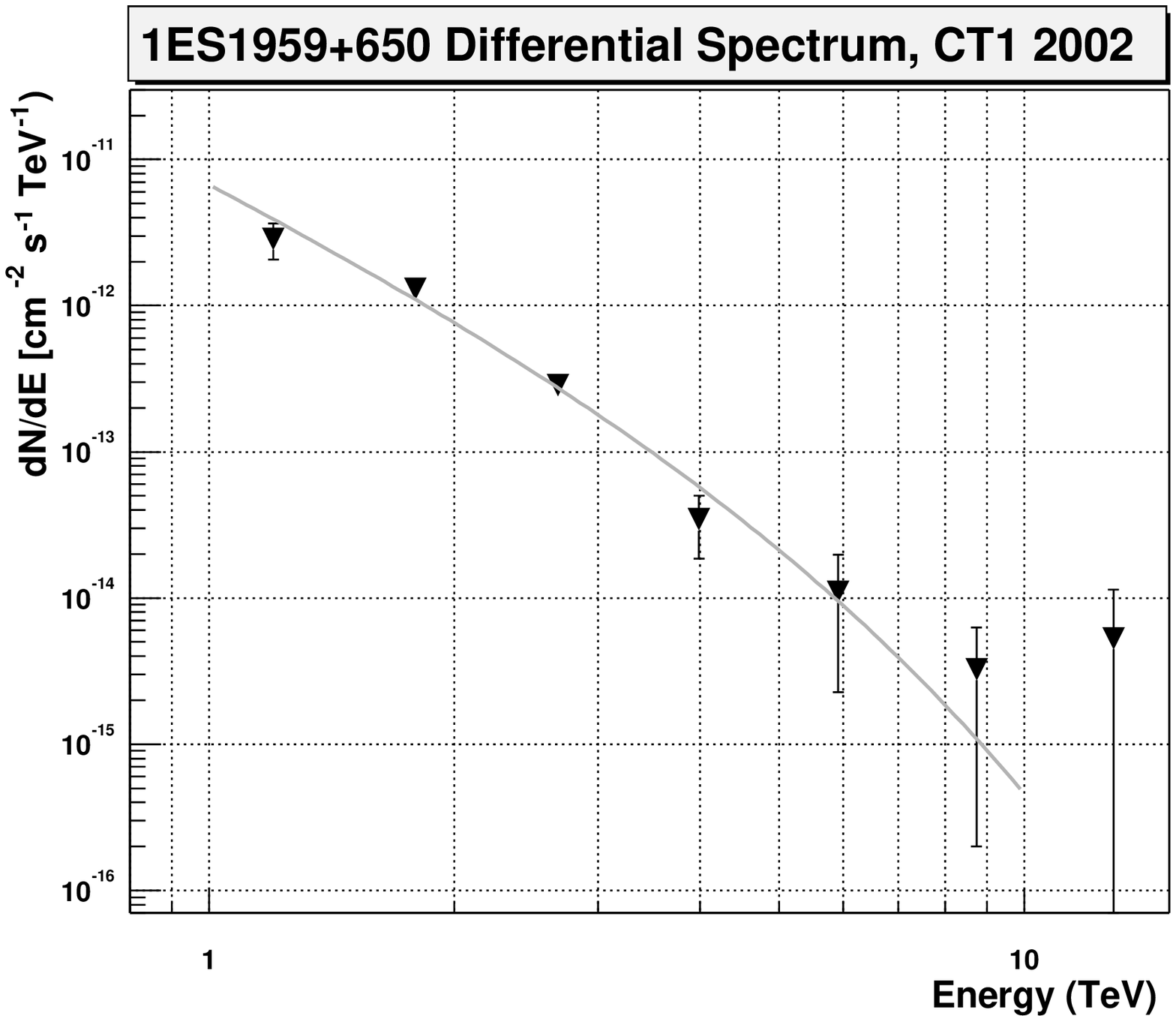}
 \end{center}
  \caption{Differential spectrum of the AGN 1ES1959+650 (HEGRA CT1 2002 data set, no moonlight, max.ZA 45$^\circ$). The line that fits the data is a power law with a fixed cutoff energy of 2.4 TeV.}
\end{minipage}
\end{figure}

\section{Results}

The CT1 data taken in 2000 and 2001 (not shown here) reveal that the source was in a low state. The averaged estimated $\gamma$-ray flux above 1TeV was between 5\% and 6\% of the Crab flux at the same energy range [9,10].

The ALPHA distribution analysis for the data taken during the year 2002 for $\sim$ 150 h of observation (Figure 1) shows a $\gamma$-ray signal from 1ES1959+650 with a significance of 11.1$\sigma$, with 664 $\pm$ 40 excess and 1672 $\pm$ 37 BG events for ALPHA $<$ 13$^\circ$. The error takes the BG extrapolation uncertainty into account.

The TeV light curve of the year 2002 is shown in Figure 3 together with the 2-12 keV activity monitored by RXTE ASM [12]. Two periods of stronger activity were recorded in that year: one in May (MJD 52414), when the Crab flux was reached,  and a flare in July (MJD 52460 and 52470). The first comparisons between the X-ray fluxes and the TeV fluxes show no evident correlation between the two considered energy ranges.
The differential spectrum of the source is shown in Figure 2. It can be fitted with a power law function, with  a spectral index $\alpha$ = 3.49 $\pm$ 0.19 ($\chi^{2}$/d.o.f. = 9.1 /6). On the other hand, the Extragalactic Background Light should already affect the spectrum above a few TeV. We tried also a power law approach modified by an exponential energy cutoff. Using a cutoff of E$_c$ = 2.4 TeV [3] resulted in a somewhat better fit with $\alpha$ = 2.54 $\pm$ 0.21, $\chi^{2}$/d.o.f. = 7.1 /6. The data quality is too poor above a few TeV to draw decisive conclusions.

\begin{figure}[t]
  \begin{center}
    \includegraphics[height=17pc,width=0.9\textwidth]{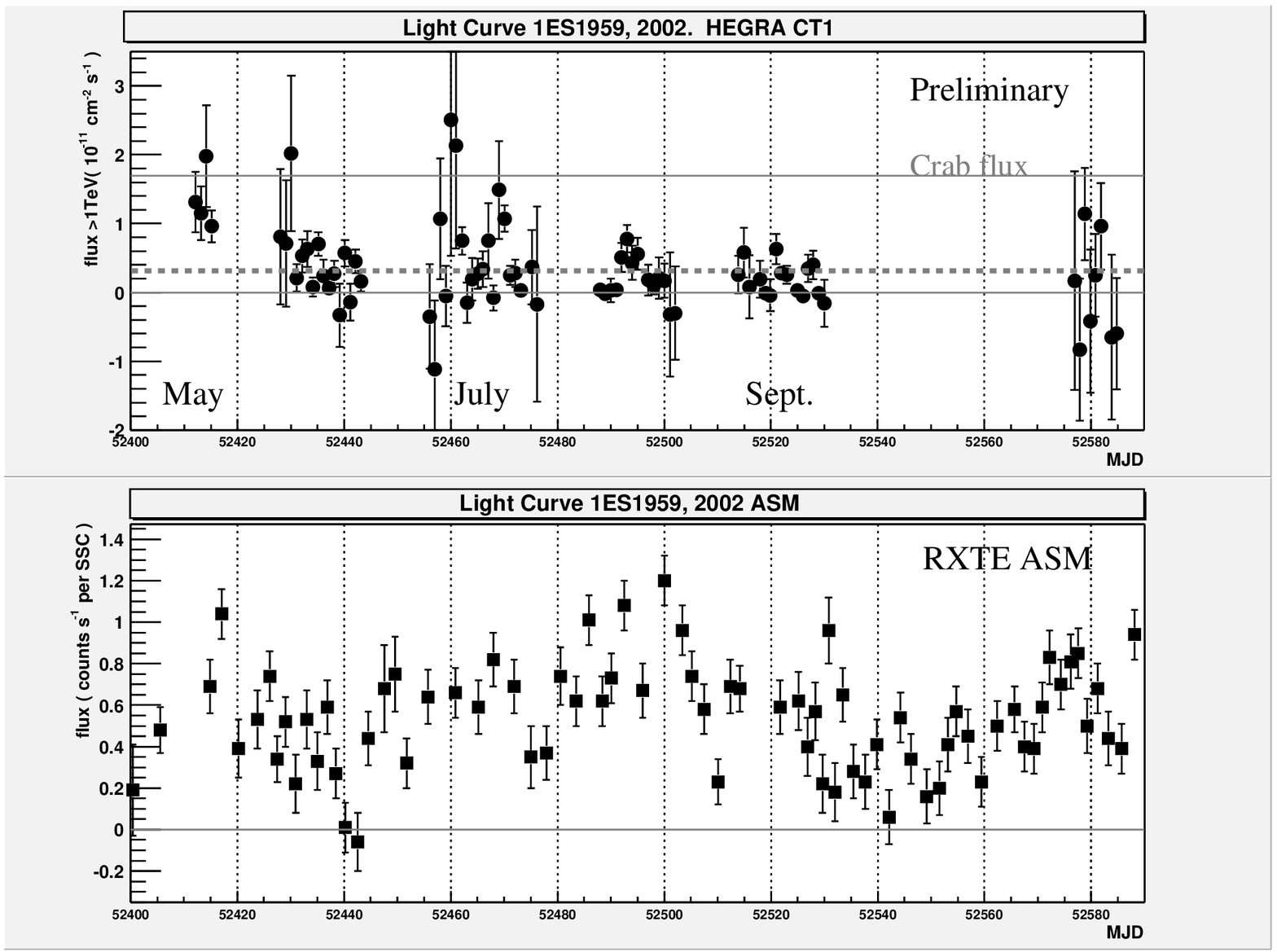}
  \end{center}
  \vspace{-1.5pc}
  \caption{The light curve of the data taken with HEGRA CT1 in 2002 (above), in absence of moonlight. Below: data taken with RXTE ASM in the same period. }
\end{figure}

\section{Discussion and Conclusions}

The AGN 1ES1959+650 has been observed with the HEGRA CT1 telescope. In May and July 2002, a short period of high emissions in the TeV range was observed, followed by a gradual decrease. The analysis of the 150 'dark' hours of observations with good conditions show a signal with a significance of 11.1$\sigma$. 
The averaged spectrum can be described by an unbroken power law, but a cutoff around 2.4 TeV is equally likely.
A simple attempt to study X-ray-TeV correlation showed no significant correlation. In particular, we note that the X-ray flux did not vary significantly around the TeV flare at MJD 52460-52470. It is expected that next generation low threshold IACTs might allow detailed clarifying studies.

\vspace{\baselineskip}
{\footnotesize {\bf Acknowledgments}. 
We would like to thank the IAC for the excellent working conditions on the La Palma Observatorio Roque de los Muchachos. We acknowledge the rapid availability of the RXTE ASM data. This work was supported by the German Ministry of Education and Research, BMBF, the Deutsche Forschungsgemeinschaft, DFG, and the Spanish Research Foundation, CICYT.}


\vspace{\baselineskip}
\re
{\bf References}
\re
1.\ Aharonian F. et al.\ 2003, submitted to A\&A Letters, astro-ph$/$0305275 
\re
2.\ Costamante L., Ghisellini G.\ 2002, A\&A 384, 56
\re
3.\ Horns D.\ et al.\ 2002, The Universe Viewed in Gamma-Rays, Kashiwa, Japan
\re
4.\ Konopelko A. et al.\ April 2002, APS/HEAD Meeting, Albuquerque
\re
5.\ Kranich D.\ 2001a, PhD Thesis, TUM, Munich, Germany  
\re
6.\ Mirzoyan R.\ et al.\ 1994, NIM A 351, 513
\re
7.\ Nishiyama T.\ et al.\ 1998, Proc. 26th ICRC , OG.2.1.21
\re
8.\ Stecker F.W., de Jager O.C., Salamon M.H.\ 1996, ApJ 473, L75-L78
\re
9.\ Tonello N., Kranich D.\ March 2002 , DPG Tagung, Aachen, Germany 
\re
10.\ Tonello N.\, PhD Thesis in preparation, TUM, Munich, Germany 
\re
11.\ Weekes T. et al.\ 17 May 2002, IAU, Circular No.7903  
\re
12.\ http:$//$heasarc.gsfc.nasa.gov$/$cgi-bin$/$xte$\_$weather$/$
\endofpaper

%% file: my_009109-1.tex
\chapter{Observation of VHE Gamma Rays from the Remnant of SN 1006 with HEGRA CT1}

\author{Vincenzo Vitale,$^1$, for the HEGRA Collaboration \\
{\it (1) Max Planck Institute for Physics,
F\"ohringer Ring 6, Munich, Germany\\
}}

\section{Abstract}

The HEGRA collaboration observed the shell-type remnant SN1006 with the telescope CT1 for 219 h.
For the observations we pointed the telescope towards the SNR center.
An excess is observed from NE part of the rim  with a  statistical significance of $\approx$ 5 $\sigma$.
Only large zenith angle (ZA)  observations ( $>$70$^{\circ}$)  are possible from the telescope site.
The energy threshold is $\approx$ 18 TeV.
We discuss the analysis and the problems related to large ZA observations.

\section{Introduction}

Supernova remnants (SNRs) are considered as the most 
likely sources of galactic cosmic ray.
The shell type remnant G327.6+14.6 is originated from the SN 1006.
This remnant is nearly circular (angular diameter $\approx$ 0.5$^{\circ}$,
radius of 9.5 pc, distance 2.18 $\pm$ 0.08 Kpc )[1],
limb-brightened and shows  the NE and SW regions significantly brighter 
than the rest of the shell (bipolar or barrel-shaped)[1].
The shell expands with velocity 2890 $\pm$ 100 km/s [2] into a low density ISM.
ASCA and ROSAT detected non-thermal X rays from the NE (the brighter one) and SW regions [5],[6]; 
while Cangaroo [7] detected TeV $\gamma$ rays only from the NE part of the  rim.
Non-thermal X rays were explained as synchrotron emission from 10-100 TeV electrons .
The possible origins of the TeV radiation are :
1) the above electron population, which can produce TeV $\gamma$ rays via inverse Compton 
scattering on the CMBR  and local IR photons ;
2) a possible contribution from accelerated hadrons ($\pi_{0} 
\Rightarrow \gamma \gamma $)[6].

\section{Observations and Analysis}

We observed SN 1006 for 219 hours from 1999 to 2001 with the HEGRA telescope CT1. 
CT1 is part of the HEGRA cosmic ray detector complex on La Palma (28.75$^{\circ}$ N, 17.9$^{\circ}$ W, 2225 m asl). 
SN 1006 (Ra 15:02:48.8, Dec -41:54:42) culminates at less than 20$^{\circ}$ elevation when observed from La Palma. 
Therefore only large zenith angle (ZA) observations are possible. 
We collected in total 346000 so-called ON-source events restricting observations to a narrow zenith range 
between 71-73$^{\circ}$ ZA as well as a small set of OFF-source data at the same ZA. 
We recorded also a large background sample of muons passing through the camera. 
Such muons leave narrow images, which might mimic $\gamma$ events.
The data reconstruction proceeded in the following way: 
At first we applied the usual filter cuts rejecting accidental noise triggers or
event samples recorded during non-optimal atmospheric conditions [8,9,10]. 
As a next step the image parameters were reconstructed and various cuts applied to select $\gamma$ events. 
Because of the large ZA observations we had to modify considerably our standard $\gamma$/h separation procedure. 
The image dimensions shrink with the ZA (roughly  $\propto \frac{1}{cos(ZA)}$ ). 
The shape of small images is determined with less precision than in the zenith position because 
the pixel diameter is constant. 
Therefore the  $\gamma$/h separation power decreases. 
The change of the image parameters could be either evaluated by Monte Carlo simulations or by studying 
real hadron events and assuming that  $\gamma$ image parameters would scale in the ratio. 
We opted for the second procedure and studied a large sample of hadronic images covering the entire
ZA range between 0$^{\circ}$ and more than 70$^{\circ}$ [10]. 
Up to 45$^{\circ}$ we could verify the scaling of the $\gamma$ image parameters with events from a large
sample of Crab Nebula and Mkn 421 data.  
We assumed that the scaling is also applicable to data at 72$^{\circ}$ ZA. (
An error in the extrapolation should not fake a possible source but would mainly result in a less than 
optimal selection of $\gamma$s). 
The scaled image cuts in Width and Length were applied to the data. 
Muons are more likely to generate background at large ZA observations because muons are more likely to 
pass simultaneously the glass envelope of many PMs. 
We studied the muon rate with closed camera and found a raw rate of 5.1$\times$10$^{-3}$ Hz at 72$^{\circ}$ ZA. 
Efficient muon rejection cuts were developed [10] from comparing the measured muon event topology with MC 
simulated $\gamma$ candidates. 
Only a fraction of 1.8$\times$10$^{-3}$ survived the cuts while the majority of $\gamma$s
passed these cuts. 

For the search of a possible excess of localized $\gamma$ emission 
 around SN 1006 we applied the False Source Method (FSM). 
A grid of 0.1$^{\circ}$ step size was defined in the FOV and at each knot we searched for a pointlike 
$\gamma$ source. This method works also for slightly extended sources underestimating somewhat the $\gamma$ flux. 
At each knot we applied the usual Distance cut and applied a 12.5$^{\circ}$ cut in the ALPHA distribution. The
remaining sample is basically a coarse measurement of the hadron flux across the FOV 
(it should be noted that there is a twofold ambiguity because one is not able to distinguish the 
head/tail of the shower). Fig 1A shows the reconstructed hadronic background as well as a possible excess. 
Already at this step an excess around  RA=15:04:10 and DEC =-41:40:00 is visible. 
In the next step we tried to subtract the background by extrapolating in each ALPHA distribution the background 
in the 0-12.5$^{\circ}$ ALPHA range. 
For this procedure we fitted a second degree polynomial (without the linear term) to the data between 20 and 
80$^{\circ}$. Then we subtracted the extrapolated background into the low ALPHA regime from the actual data 
and determined the excess. The procedure has been extensively tested [9]. 
The excess map is shown in Fig 1B. A clear excess is found at the NE rim .
Significance is in part influenced by the background statistics.
We improved the background precision by averaging over a ring around the center of the FOV but leaving out a section, around the position under test , which might contain $\gamma$ events [10]. 
This method increases the background statistics by a factor $\approx$ 3 and reflects the true behavior 
at the low ALPHA range. 
The lower panel in Fig 1 shows the ALPHA distribution of the excess bin together with the higher statistics 
background distribution in the ring. 
An excess of 103 $\pm$17 over a background of 225 events has been determined using the
recommended procedure [7]. The significance of the excess is  5.6$\sigma$ . 
The excess location  is nearly coincident with the observation of the Cangaroo group. 
If we allow for an uncertainty 9 bins we obtain a reduced significance of 5.1$\sigma$.
We verified the correct positioning of the telescope at large ZA 
by measuring the position of two bright stars in the FOV. The positions of $\kappa$ Cen (mag=3.13)
and $\beta$ Lup (mag=2.68) allowed us to determine the excess position with 0.1$^{\circ}$ precision.
For a 5$\sigma$ excess it is not possible to determine a spectrum. 
From MC simulations including all above mentioned cuts we calculated a threshold of 18 TeV 
and the collection area assuming an unbroken powerlaw with spectral index of -2.3 [5]. 
Using the experimental data and the parameters from the simulation we calculated a preliminary 
integral flux of  $\Phi$($>$18 TeV) = (2.5$\pm$ 0.5$_{stat}$)$\times$10$^{-13} \frac{ph}{cm^{2} s}$.
The evaluation for other photon indices and of the systematic error (at least 35$\%$) is still ongoing.

\section {Summary and Conclusions}

From the analysis of the large zenith angle observation of SN 1006 we conclude:
\ a) that we see evidence (5.1 $\sigma$) for $\gamma$-emission at the NE rim of 
the shell coinciding within 0.1$^{\circ}$ with the Cangaroo observation at lower energies.
\ b) The threshold of CT1 at 72$^{\circ}$ ZA is 18 TeV assuming an unbroken powerlaw 
with a spectral coefficient of -2.3 .
\ c) A preliminary integral flux of $\Phi$(E$>$18 TeV) = (2.5 $\pm$ 0.5$_{stat}$)$\times$10$^{-13} \frac{ph}{cm^{2} s}$
 has been determined. The systematic error is still under study but is at least 35$\%$. 
\ d) We observe no excess at the SW rim and determined an upper limit of 1.6$\times$10$^{-13} \frac{ph}{cm^{2} s}$
(C.L. 99.5$\%$)
\ We conclude further that it is possible to carry out searches for $\gamma$ sources at large ZA 
although limitations from the pixel size and the large threshold are obvious.

\

REFERENCES: 1. Long \& al. 2003, ApJ 586 \ 2. Ghavamian \& al. 2002, ApJ 572 
3. Koyama \& al. 1995, Nature 378 \ 4. Willingale \& al 1996, MNRAS 278\ 
5. Tanimori \& al. 1998, ApJ 497L\ 6. Berezhko \& al. 2002, A\& A 395\ 
7. Li \& Ma 1983, ApJ 272\ 8. Kestel M. PhD th.\ 9. Kranich D. 2001, PhD th.\  
10. Vitale V. PhD th.,in prep.

\begin{figure}[t]
  \begin{center}
    \includegraphics[width=30pc,height=25pc]{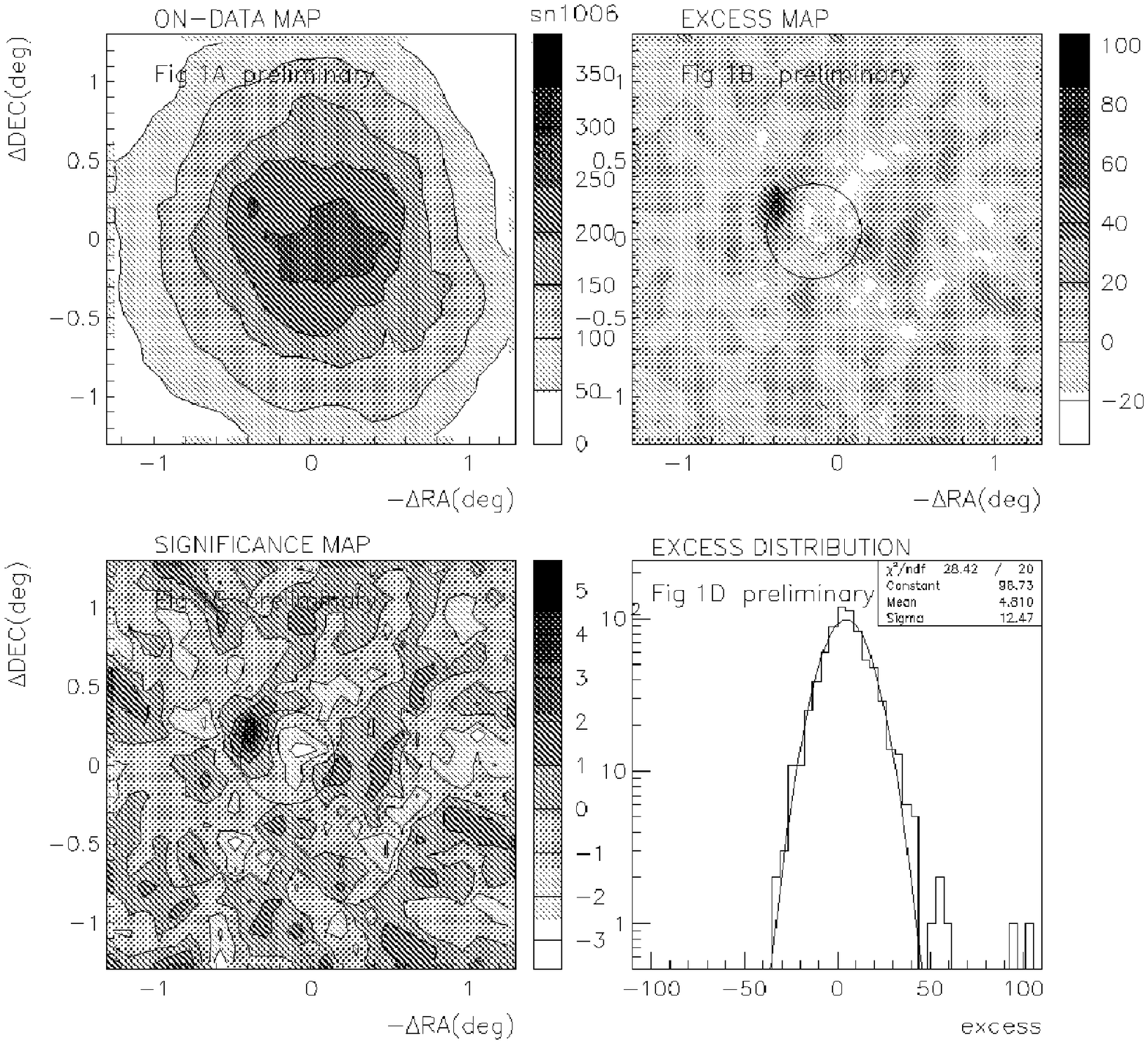} 
    \includegraphics[width=24pc,height=16pc]{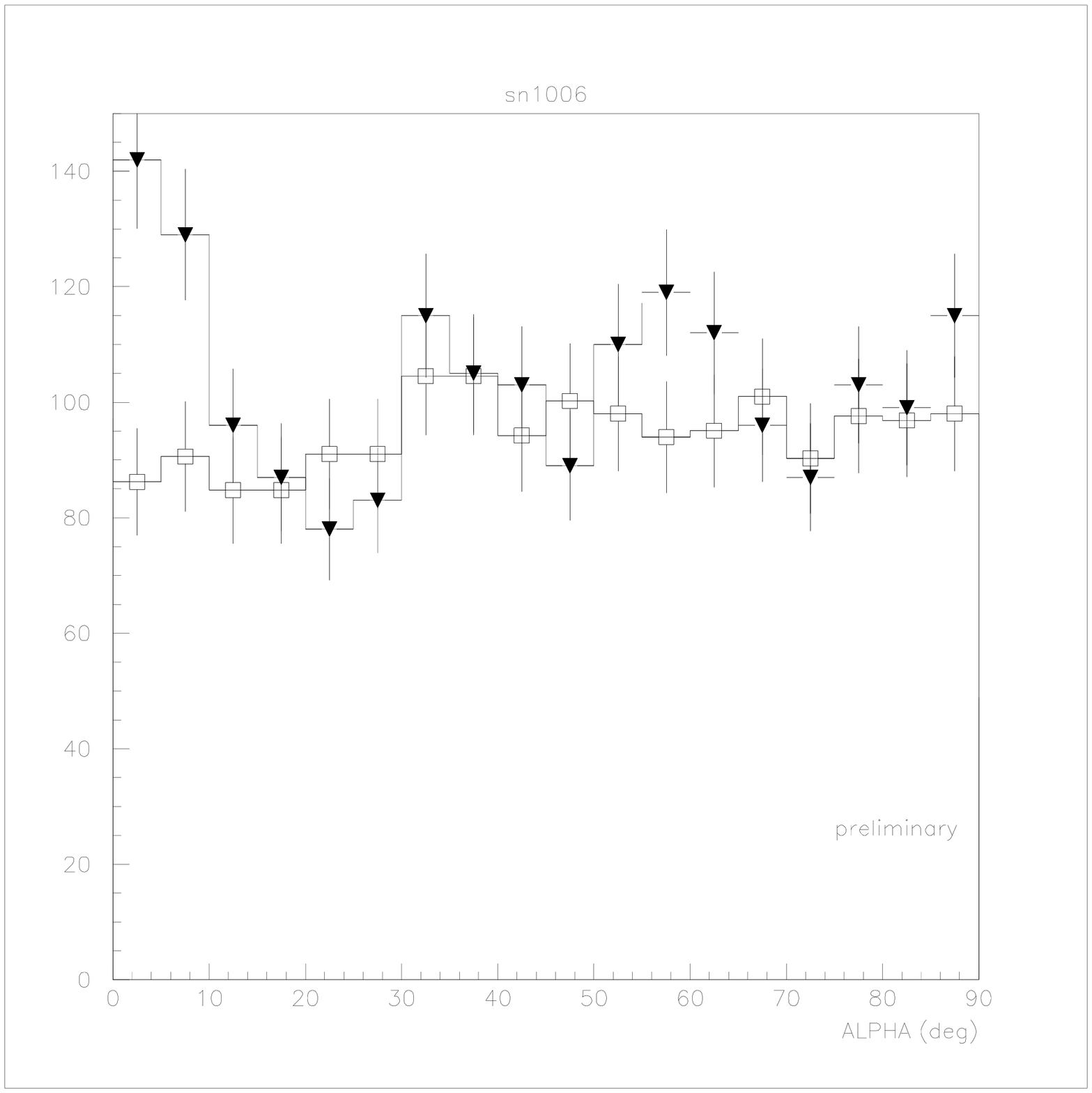}
  \end{center}
  
  \caption{Fig 1A is the sky map of ON-source data after $\gamma$ selection but before of subtraction of residual background; Fig 1B is the sky map of the excess (= ON-source - residual background); Fig 1C is the sky map of excess significance; Fig 1D is  the distribution of the excess. Fig 1 lower panel is alpha distribution for ON-source data and background (ring method) at the NE rim position. All maps show the FOV of CT1; RA increases from right to left. 
The black circle in 1B and 1C shows the position of the SNR shell. 
The NE part of the rim is around the upper-left part of this circle.}

\end{figure}

\endofpaper